\documentstyle[12pt]{article}
\input epsf

\def\SECTION#1{\begin{center}{\large\bf #1}\end{center}}
\def\SUBSECTION#1{\begin{center}{\bf #1}\end{center}}
\def\figsize{10cm}

\def\rulerheight{0.5pt}

\def\U1{$U(1)$}
\def\SU5{$SU(5)$}
\def\SO10{$SO(10)$}
\def\422{$SU(4)\otimes SU(2)_L \otimes SU(2)_R$}

\def\diag.{\hbox{diag.}}

\def\refeqn#1{(\ref{#1})}
\def\M_U{\hbox{$M_U$}\ }
\def\M_P{\hbox{$M_P$}\ }

\def\tanb{\hbox{$\tan \beta$}}
\def\GeV{{GeV}}
\def\MSSM+N{\hbox{MSSM+$\nu$}}
\def\etal{{\it et al.}}
\def\bigsim{{\>{\buildrel {\scriptstyle \sim} \over {\scriptstyle  >}}\>}}

\def\bigsml{{\>{\buildrel {\scriptstyle >} \over {\scriptstyle <}}\>}}

\def\ibid{{\it ibid.}}
\def\ie{{\it i.e.~}}

\def\FigGHFeyn{1}
\def\FigBFeyn{2}
\def\FigGaugeUnif{3}
\def\FigMtVsAlphasPic2{4}
\def\FigMtVsAlphas{5}
\def\FigTanbVsAlphas{6}
\def\FigMtVsYukX{7}
\def\FigMtVsMs{8}
\def\FigMtVsTanb{9}
\def\FigRunYuk{10}
\def\FigRunNeutMass{11}
\def\FigMtVsMnu{12}
\def\Figm0VsMgaug{13}
\def\FigMtopVsMgaug{14}
\def\FigMtopVsMg4{15}
\def\FigM2LVsM4{16}
\def\FigMtopVsM4M2L{17}
\def\FigMtopVsMFL{18}
\def\FigMtopVsMFLR{19}
\def\FigMtopVsMh{20}
\def\FigMtopVsMhMF{21}
\def\FigMFVsMh{22}
\def\FigMtVsMhScan{23}
\def\FigMtVsMuScan{24}
\def\FigMtVsGLNScan{25}
\def\FigMtVsMCHScan{26}
\def\FigMtVsNTLScan{27}
\def\FigMtVsStauScan{28}
\def\FigMtVsSNeuScan{29}
\def\FigMtVsSbottomScan{30}
\def\FigMtVsMaScan{31}
\def\FigBsgLoops{32}
\def\FigBsgChargLoops{33}
\def\FigBsgAmp{34}
\def\FigBsgBR{35}
\def\FigTmgChargLoop{36}
\def\FigTmgAmp{37}
\def\FigTmgBR{38}

\def\TabCasesDef{I}
\def\TabMainPred{II}
\def\TabMtTanbYukG{III}
\def\TabMnu{IV}
\def\TabBsgYuk{V}
\def\TabMsPred{VI}
\def\TabGluinoCorr{VII}
\def\TabHiggsinoCorr{VIII}
\def\TabBinoCorr{IX}
\def\TabSusyCorr{X}
\def\TabDtermCharges{XI}

\def\AppendixA{A}
\def\AppendixB{B}
\def\AppendixC{C}
\def\AppendixD{D}

\begin{document}
\baselineskip 24pt
\newcommand{\sheptitle}
{Yukawa Unification as a Window into
the Soft Supersymmetry Breaking Lagrangian}

\newcommand{\shepauthor}
{S. F. King
and
M.Oliveira}

\newcommand{\shepaddress}
{Department of Physics and Astronomy, University of Southampton \\
        Southampton, SO17 1BJ, U.K}

\newcommand{\shepabstract}
{We study Yukawa unification,
including the effects of a physical neutrino mass consistent
with the Superkamiokande observations,
in a string/$D$-brane inspired \422 model which 
allows the most general non-universal scalar and gaugino masses,
including the usual $D$-term contributions which arise in $SO(10)$.
We investigate how the tight constraints from
rare decays such as  $b \rightarrow s \gamma$
and $\tau \rightarrow \mu \gamma$
can provide information about the family dependent supersymmetry
breaking soft Lagrangian, for example the trilinears
associated with the second and third family.
Many of our results also apply to $SO(10)$ to which the model approximately
reduces in a limiting case. In both models
we find that Yukawa unification is perfectly viable providing the
non-universal
soft masses have particular patterns. In this sense Yukawa unification
acts as a window into the 
soft supersymmetry breaking Lagrangian.}

\begin{titlepage}
\begin{flushright}
hep-ph/0008183\\
\end{flushright}
\begin{center}
{\large{\bf \sheptitle}}
\\ \shepauthor \\ \mbox{} \\ {\it \shepaddress} \\
{\bf Abstract} \bigskip \end{center} \setcounter{page}{0}
\shepabstract
\begin{flushleft}
\today
\end{flushleft}
\end{titlepage}

\newpage

\SECTION{I. INTRODUCTION}

One of the earliest successes of the Minimal Supersymmetric Standard
Model (MSSM) \cite{MSSM} was that it could allow for
the unification of the gauge couplings at high energy 
($Q=M_X$) \cite{GaugeUnif},
thereby opening the door for Supersymmetric Grand Unified Theories (SUSY GUTs)
\cite{ElKeNa,RoRo,LaPo2,LaPo}. 
SUSY GUTs typically involve
some third family Yukawa unification 
\cite{LaPo,KeLoNa,CaPoWa,ArCaPiRa,AlKi2}.
For example minimal $SU(5)$ \cite{GeGl} predicts
bottom-tau Yukawa unification at $M_X$
($\lambda_b=\lambda_\tau$) and allows for the prediction of the top quark
mass ($m_t$) in terms of the ratio $m_b/m_\tau$.
In models like \SO10, and in the \422 (422) Pati-Salam model \cite{PaSa},
complete third family Yukawa unification
holds ($\lambda_t=\lambda_b=\lambda_\tau$).
Thus the ratio of the vacuum expectation values (VEVs) of
the up/down Higgs bosons doublets --  $\tan\beta=v_2/v_1$ -- 
can be predicted.
Since the top and bottom Yukawa couplings
start out equal at the GUT scale, and their renormalization group
(RG) evolution is very similar (they differ only by the \U1\
couplings at one loop) their low energy values are approximately
equal so we expect $\tan\beta \sim m_t /m_b \sim 40$.

In this paper we shall include the neutrino Yukawa coupling in the
analysis giving quadruple Yukawa unification 
($\lambda_t=\lambda_b=\lambda_\tau=\lambda_{\nu}$)
as was first done in Ref.~\cite{AlKi2}.
In the light of the Superkamiokande results \cite{SKamiokandeColl} 
(assuming hierarchical neutrino masses and the see-saw
mechanism) we regard the third family neutrino mass ($m_\nu$) as an input
which allows for an additional prediction -- 
the mass scale at which the heavy right-handed neutrinos
decouple -- $M_\nu$. 
However such predictions depend sensitively
on the parameters of the soft SUSY breaking Lagrangian, as we now
discuss.

Yukawa unification (assuming the MSSM and SUSY GUTs)
at first was thought to lead to an acceptable top mass prediction,
given the experimentally permitted range of top and bottom quark masses
and $\alpha_s$, with a unified
Yukawa coupling at $M_X$ of order one ($\lambda_X \sim 1$)
\cite{CaPoWa,AlKi2,AnLaSh,BaBeOh,CaPiRa,KaKoRoWe}.
However, when low energy SUSY corrections
to the running bottom quark mass
due to the decoupling of SUSY particles were included
\cite{CaOlPoWa,HaRaSa}, the initial good agreement was spoiled.
The SUSY corrections to the bottom quark mass are :
\begin{equation}
\Delta m_b = {\delta m_b \over m_b}
= \frac{\mu \tan \beta}{4 \pi}\left[
\frac{8}{3}\alpha_s m_{\tilde g}
I(m^2_{\tilde g},m^2_{\tilde b_1}, m^2_{\tilde b_2})+
{\lambda_t^2 \over 4\pi} A_t
I(\mu^2,m^2_{\tilde t_1}, m^2_{\tilde t_2},)\right]
\end{equation}
where the function $I(x,y,z)$ is given in \cite{CaOlPoWa}.
The corrections can be positive or negative depending on the sign of
$\mu$, and being proportional to $\tan \beta$, are rather large.

With universal soft masses and {\em negative} $\mu$,
the SUSY corrections to the bottom mass lead to large negative
$\Delta m_b = \delta m_b / m_b \sim -20 \% $ corrections
which implies a larger $b$ quark mass
before the SUSY corrections, which effectively
lowers the previously successful top mass prediction to
$m_t \sim 150 $ GeV which is too small \cite{CaOlPoWa}.
With universal soft masses and {\em positive} $\mu$ the top mass is
predicted to be too large, outside its perturbative upper limit.
It has already been pointed out in the literature that $\delta m_b$
can be made small by either assuming explicit
non-universal scalar soft masses \cite{OlPo2,BaDiFeTa,CaWa}
or by introducing approximate
Peccei-Quinn and R symmetries \cite{HaRaSa,RaSa}.
However in the framework of $SO(10)$ the degree of
non-universality one could assume until recently appeared rather
limited. For example, if both Higgs doublets arise from a single
$\underline{10}$ dimensional representation
then they will necessarily have a common soft scalar mass at
the GUT scale, and the same applies for the light Higgs bidoublet
of the Pati-Salam model. Such universality is also a
problem for electroweak symmetry breaking where one requires
a large hierarchy of vacuum expectation values (VEVs)
starting from very symmetrical
initial conditions where the two Higgs doublet soft masses are
equal, and where the approximately equal top and bottom Yukawa
couplings tend to drive both Higgs masses negative equally,
making large $\tan \beta$ rather difficult to achieve.

A large step forward for both these problems has been to realise
the importance of $D$-term contributions to scalar masses \cite{Dterms},
which naturally split the two Higgs doublet masses.
If the up-type Higgs doublet mass at the GUT scale is
smaller than the down-type Higgs doublet mass, then this
makes electroweak symmetry breaking with large $\tan \beta$ much more
natural. Assuming negative $\mu$, 
the $D$-terms also allow a choice of non-universal scalar masses which
reduces the correction to the $b$-quark mass and hence allow a
larger top quark mass. Although these problems appear to be resolved with
$D$-terms, one still faces difficulties with rare decays such as
$b \rightarrow s \gamma$, which is also enhanced for large 
$\tan\beta$, and negative $\mu$ \cite{bsgSUSY}. This problem can be avoided
either by increasing the masses of all the superpartners,
or by considering positive $\mu$ which will tend to cancel
the SUSY contributions, however both these procedures
lead to fine-tuning.
Another possibility which we pursue in this paper is to
consider the effect of non-universality in the family space,
which may lead to additional contributions to
$b \rightarrow s \gamma$ which can cancel those coming from
the family universal sources. We explicitly check that the
contributions that we introduce do not introduce problems
elsewhere such as with $\tau \rightarrow \mu \gamma$.

The main purpose of this paper is to provide a detailed analysis
of Yukawa unification, post-Superkamiokande, allowing the most
general non-universal soft SUSY breaking masses possible. Apart
from non-universality in the soft scalar masses arising from $D$-terms,
we shall also consider more general types of non-universality
which may arise in models such as the string/$D$-brane inspired \422
model \cite{AnLe},\cite{string422}, 
which reduces to the $SO(10)$ model in a limiting case.
This allows more general soft scalar mass non-universality
including similar $D$-term contributions as in $SO(10)$
(since the relevant broken $U(1)$ factors all arise from
the breaking of the \422 subgroup), and also permits
violations of gaugino mass universality which have not so far been
studied. The underlying theme of our approach is that
the sensitivity of Yukawa unification to soft SUSY breaking
parameters is to be welcomed, since it provides
a window into the soft SUSY breaking Lagrangian, both from the
point of view of family universal and family non-universal
soft SUSY breaking parameters.

The layout of the remainder of the paper is as follows.
In the section II we present our calculational approach,
and in section III we review the SUSY corrections to the
bottom and tau masses. Section IV deals with the effect of neutrino
Yukawa couplings on Yukawa unification and in section V we present
results for gauge and Yukawa unification, assuming minimal
supergravity (mSUGRA) with universal
soft SUSY breaking parameters and including the 
neutrino threshold in addition to the usual low energy
thresholds. In section VI we turn to Yukawa unification in the
\422 model, including the effect of physics above the GUT scale,
and non-universal soft scalar and gaugino masses permitted
in this model. We also consider the effect of $D$-terms in the
$SO(10)$ limit of the model, and then study
$b \rightarrow s \gamma$ and $\tau \rightarrow \mu \gamma$ including
non-universality in the family-dependent trilinear parameters.
Section VII concludes the paper.

\newpage

\SECTION{II. FRAMEWORK}

In this section we will summarize how we implemented the RGEs
and matching boundary conditions of the model.
In the region $Q < M_Z$ the effective theory is
$SU(3)_c \otimes U(1)_{em}$, thus 3-loop QCD
\cite{QCDRGEs}
plus 2-loop QED RGEs \cite{ArCaKeMiPiRaWr} apply.
Between $Q = M_Z$ and $Q = M_S$,
where $M_S$ is the scale that parameterises the energy at which
the theory effectively becomes supersymmetric
(see Appendix \AppendixB\ for details),
we considered 2-loop Standard Model (SM) RGEs
in the gauge and Yukawa couplings \cite{BaBeOh,ArCaKeMiPiRaWr,MaVa2}.
In the region $M_S < Q < M_X$
we evolved 2-loop (1-loop) gauge/Yukawa (all other parameters) MSSM RGEs
\cite{BaBeOh,MaVa},
properly adapted and extended to take into account the presence (and
decoupling) of right handed neutrinos $\nu^c$.

\SUBSECTION{A. Input}

The low energy input was :
the running electromagnetic coupling $\alpha_e^{-1}(M_Z)=127.8 $,
the pole tau lepton mass $M_\tau = 1.784$ \GeV\ and several values for
the pole bottom quark mass $M_b = 4.7,4.8,..,5.1$ GeV.
We have converted the pole bottom mass into ``running'' bottom mass in
the $\overline{\rm MS}$ scheme using two loop QCD perturbation
theory \cite{ArCaKeMiPiRaWr}~:
\begin{equation}
m_b(Q=M_b) = {M_b \over
              {\displaystyle
              1 + {4 \over 3}{\alpha_s(M_b) \over \pi} +
              K_b \, \biggl({\alpha_s(M_b) \over \pi}\biggr)^2 }
              \vrule width 0pt height 20pt }
\label{mbMb}
\end{equation}

\noindent
where $K_b=12.4$.
The corresponding running masses,
for $\alpha_s=0.120$, are
$m_b(M_b)=$ 4.06, 4.15, 4.24, 4.33, 4.42 GeV.
The experimental range for $m_b$ estimated from bottomonium and $B$
masses is 4.0--4.4 GeV \cite{CCasoEtal}.
The strong coupling $\alpha_s(M_Z)$ was taken to be in the range 0.110--0.130.

The low energy input was complemented by the Super-Kamiokande
atmospheric neutrino data that suggests
$m_{\nu_3}^2-m_{\nu_2}^2 = 10^{-2}$ to $10^{-3}$ eV$^2$
\cite{SKamiokandeColl}.
Assuming that the neutrino
masses are hierarchical we required that
$m_{\nu_3} \sim 0.05 $ eV.

The universal high energy inputs 
assumed initially are motivated by mSUGRA: spontaneously broken
supergravity in which the local N=1 supersymmetry
breaking occurs in a
``hidden'' sector and is only transmitted to the ``visible'' sector through
weak gravitational -- flavour blind -- interactions. Thus we assume
to begin with universal soft SUSY breaking masses (USBM) 
given by a common gaugino mass $M_{1/2}$ and soft
scalar masses $m_0$ at $Q=M_X$. The trilinear $A$-terms were set to
zero. In section VI we relax the universality assumption.

In the Yukawa sector we assumed that the third family Yukawa couplings
have their origin in a unified renormalizable interaction which fixes
their values to be $\lambda_X$.

\SUBSECTION{B. Running and matching boundary conditions}

The process by which the output was generated relied in the initial
estimation and successive iterative refinement of {\it a priori}
unknown parameters such as $M_t$ and $\tan\beta$.
The procedure is described in detail in section VII of Ref.~\cite{CaPiRa}.

Starting at $Q=M_Z$, $\alpha_s(M_Z)$
and $\alpha_e(M_Z)$ were firstly
run down to $Q=1$ GeV (where the up, down and strange quark running
masses were fixed \cite{ArCaKeMiPiRaWr}) and secondly up to $Q=M_Z$ using
$SU(3)_c \otimes U(1)_{em}$ RGEs.
In the ``running up'' process the
heavy (charm, tau and bottom) fermion pole masses
were converted to
running ones using the expressions in Ref.~\cite{ArCaKeMiPiRaWr}.
At $Q=M_Z$, $\alpha_e$ was corrected for the
decoupling of heavy gauge bosons, Higgs and Nambu-Goldstone bosons,
and ghosts \cite{LHall}. Afterwards the value of $s_\theta^2$
\cite{LaPo2} was used to obtain $g'$ and $g$.
The bottom and tau Yukawa couplings, in the Standard Model,
were evaluated using
$\lambda_{b,\tau}^{SM}(M_Z) =
  m_{b,\tau}(M_Z)/v$, where the VEV is $v$ = 174 GeV.
Next, the gauge and the Yukawa couplings were run from $Q=M_Z$ to
the estimated pole top mass $Q=M_t$ using 2-loop SM RGEs,
at which point $M_t$ was converted to $\overline{\rm MS}$
running mass $m_t(M_t)$, joining the list of
parameters to be integrated to $Q=M_S$. All along, threshold
corrections in the gauge couplings were included by changing
the 1-loop $\beta$-functions (using the ``step'' approximation.)
At $Q=M_S$ the $\overline{\rm MS}$ gauge couplings $\alpha_i$
were converted to the DR ones \cite{LaPo2,CaPoWa,BDWright}
and corrected for the cumulative
effect of decoupling of all the SUSY particles \cite{BDWright}.
The top, bottom and tau Yukawa couplings were then converted from the SM to
the MSSM normalization and corrected for the SUSY corrections
\cite{BDWright} :
\begin{eqnarray}
\lambda_t^{MSSM}    &=& {\lambda_t^{SM} / \sin\beta} \\
\lambda_b^{MSSM}    &=& {\lambda_b^{SM} / \cos\beta} -
                        {\delta m_b / v_1} \\\
\lambda_\tau^{MSSM} &=& {\lambda_\tau^{SM} / \cos\beta} -
                        {\delta m_\tau / v_1}
\label{YukMatch}
\end{eqnarray}

Afterwards we run up all the above couplings, together with an
estimate for $\lambda_{\nu}^{MSSM}(M_S)$, in two stages, firstly
from $Q=M_S$ to $Q=M_\nu$ and afterwards from
$Q=M_\nu$ to $Q=M_X$,
properly excluding (including) $\nu^c_\tau$ in the former (latter) stage.
In our model $M_X$ was fixed to be the scale at which only the
$U(1)$ and $SU(2)$ gauge couplings unify :
$\alpha_1(M_X)=\alpha_2(M_X)\ne\alpha_3(M_X)$.

At $M_X$ gauge and Yukawa unification were tested leading to eventual
wiser choices to the next estimates for $M_t$, $\tan\beta$,
$M_X$ and $M_\nu$.

The iteration cycle was completed by setting the USBM to
their unification values after which all the couplings and masses
of the model were run down from $Q=M_X$ to $Q=M_S$
using the inverse ``running up'' procedure described above.
Finally at $Q=M_S$ the 1-loop effective Higgs potential was
minimized \cite{GaRiZw,PHChankowski}
and the SUSY Higgs mixing parameter $\mu^2$ and the
corresponding soft term $m_3^2$ were determined using
\cite{CaPiRa,DeLaTa,BaBeOh2} :
\begin{equation}
\mu^2 = \frac{m_1^2-m_2^2 \> \tan^2\beta}{\tan^2\beta - 1} -
{1 \over 2} \> M_Z^2
\label{mu2}
\end{equation}
\begin{equation}
m_3^2 = (\mu_2^2+\mu_1^2) \> {\tan\beta \over \tan^2\beta+1}
\label{m32}
\end{equation}
where $\mu^2_{2,1} = m^2_{2,1}+\mu^2$ and the up/down soft Higgs boson
masses are $m^2_{2,1} = m^{2tree}_{2,1}+\Sigma_{2,1}$.
\footnote{The $\Sigma_{2,1}$ parameterise
the 1-loop corrections to the tree level Higgs potential
\cite{ArNaALL}.}

We note that the number of independent parameters the model
can predict is four -- as many as the constraints imposed
(one gauge and three Yukawa unification conditions.)
We took them to be $M_X$, $M_t$, $\tan\beta$ and $M_\nu$.
The latter was fixed by requiring that $m_{\nu_3} = 0.05$ eV.

\SECTION{III. SUPERSYMMETRIC BOTTOM \\ AND TAU MASS CORRECTIONS}

In the Standard Model it is frequent to work in off-shell mass schemes
in which the running masses $m_f(Q)$ differ from their physical masses
$M_f$ (defined as the real part of the complex pole position of its
propagator) by some finite correction \cite{HeKn}.
For the quarks the most important corrections (arising from gluon loops)
are well known and particularly affect bottom quark mass
(see Eq. \refeqn{mbMb}.)
In supersymmetric models additional corrections are also present.
For large values of $\tan\beta$ some of these SUSY corrections
can indeed affect the running bottom mass $m_b$ by 20 \%
\cite{CaOlPoWa,HaRaSa},
thus their consideration is crucial for the prediction of $M_t$.
In this section we review the origin of the SUSY corrections to the
bottom quark ($\delta m_b$) and tau lepton mass ($\delta m_\tau$.)

\bigskip\bigskip

\vbox{
\vbox{
\hfil
\vbox{
\hbox{
\hbox{
\epsfxsize=5cm
\epsffile[0 0 346 226]{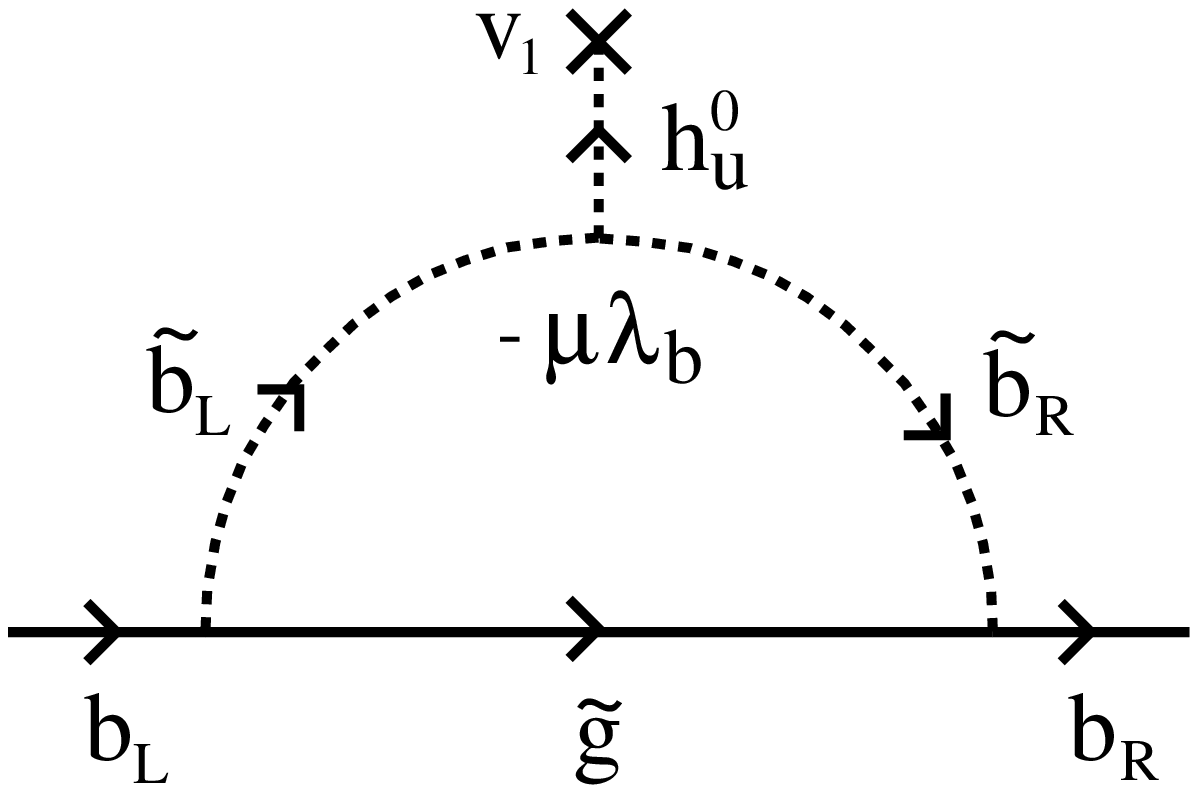}}
\hbox{
\epsfxsize=5cm
\epsffile[0 6 346 180]{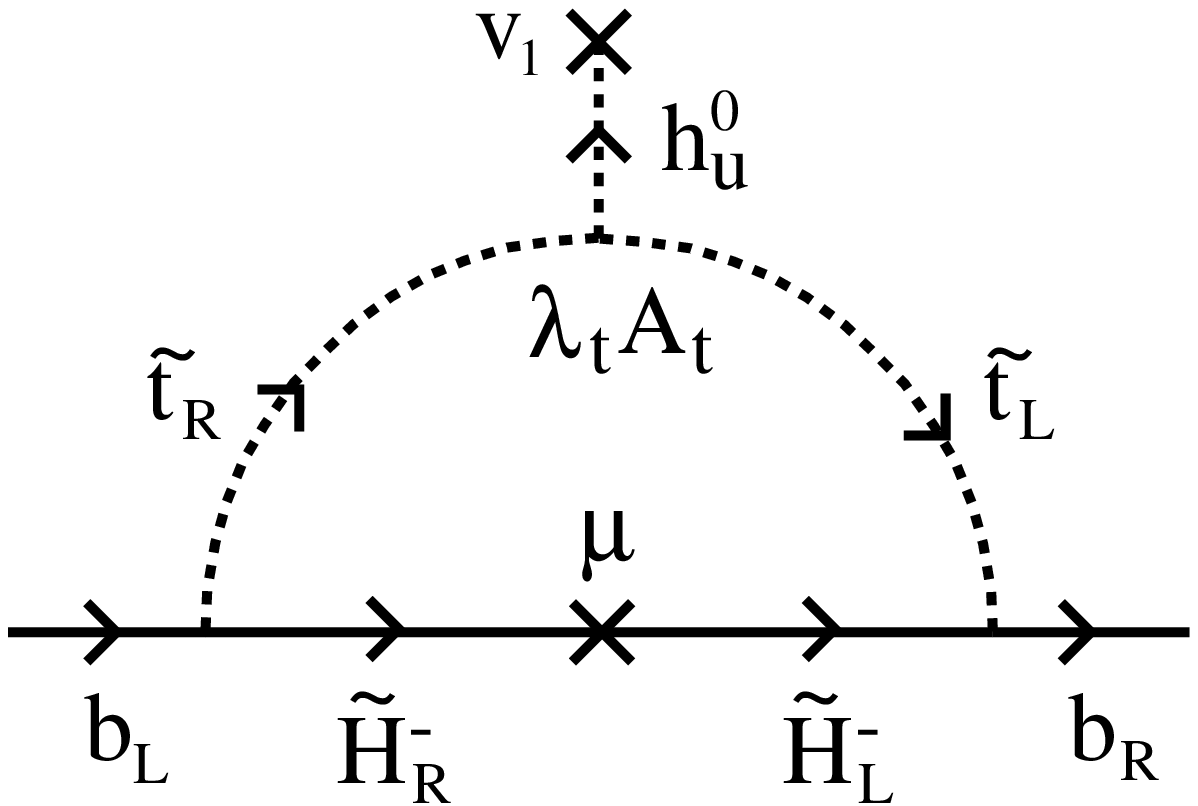}}
\hskip 0.5cm}}

\medskip

{\narrower\narrower\footnotesize\noindent
{FIG. \FigGHFeyn}
Gluino and higgsino loop diagrams corresponding to the bottom
mass corrections $\delta^{\tilde g} m_b$ and $\delta^{\tilde H} m_b$
of Eq.~\refeqn{gluinoCorr} and Eq.~\refeqn{HiggsinoCorr}.\par
\bigskip}}

\vbox{
\hfil
\hbox{
\vbox{
\epsfxsize=5cm
\epsffile[0 0 346 226]{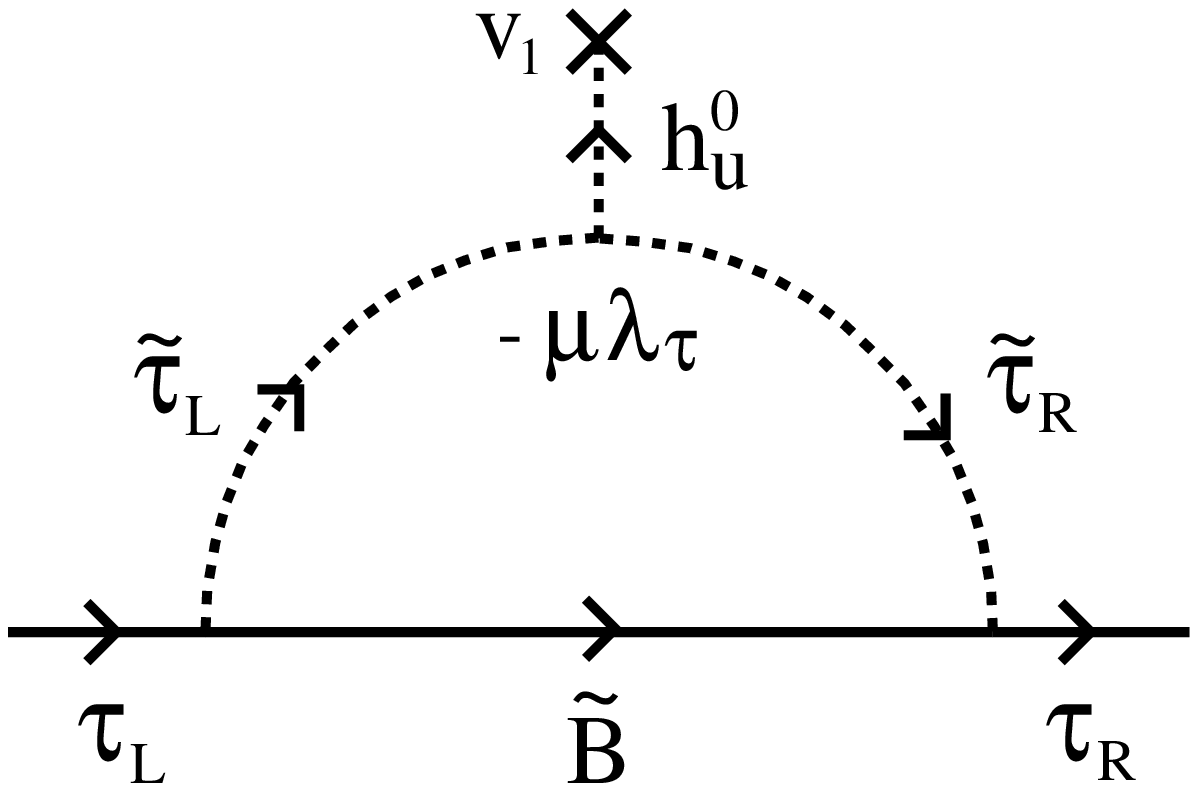}}}

\medskip

{\narrower\narrower\footnotesize\noindent
{FIG. \FigBFeyn}
Bino diagram corresponding to the tau mass correction
$\delta^{\tilde B} m_\tau$ of Eq.~\refeqn{BinoCorr}.\par
\bigskip}}}

In the MSSM, at tree level, the bottom quark acquires a mass
exclusively through the non-vanishing VEV of $\langle h_d^0 \rangle = v_1$.
However, at 1-loop level, the bottom quark also receives a ``small''
mass from $\langle h_u^0 \rangle = v_2$. The dominant processes
responsible for the additional corrections to the bottom (tau) mass
are characterized by gluino/sbottom and higgsino/stop (bino/stau)
diagrams which we illustrate in Fig.~\FigGHFeyn\ (and Fig.~\FigBFeyn.)
\noindent
The corresponding
expressions for $\delta^{\tilde g} m_b$, $\delta^{\tilde H}m_b$
and $\delta^{\tilde B} m_\tau$ are given
by~\cite{CaOlPoWa,HaRaSa,PiBaMaZh}~:
\begin{eqnarray}
\Delta^{\tilde g} m_b =
\frac{\delta^{\tilde g}m_b}{m_b} &=& \frac{2\alpha_3}{3\pi} \>
m_{\tilde g} \> \mu \> \tan\beta \> I(m_{\tilde g}^2, m_{\tilde b_1}^2,
m_{\tilde b_2}^2 ) \label{gluinoCorr} \\
\Delta^{\tilde H} m_b =
\frac{\delta^{\tilde H}m_b}{m_b} &=& \frac{\alpha_t}{4\pi} \>
A_t \> \mu \> \tan\beta \> I(\mu^2, m_{\tilde t_1}^2,
m_{\tilde t_2}^2 ) \label{HiggsinoCorr} \\
\Delta^{\tilde B} m_\tau =
\frac{\delta^{\tilde B}m_\tau}{m_\tau} &=& \frac{\alpha'}{4\pi} \>
m_{\tilde B} \> \mu \> \tan\beta \> I(m_{\tilde B}^2, m_{\tilde\tau_1}^2,
m_{\tilde\tau_2}^2) \label{BinoCorr}
\end{eqnarray}
where $m_{\tilde g}$ ($m_{\tilde B}$)
is the gluino (bino) mass, $\alpha_t = \lambda_t^2/4\pi$,
$\alpha' = g'^2/4\pi$, $A_t$ the top soft trilinear term and
$m_{\tilde b,\tilde t,\tilde\tau}$ the sbottom, stop and stau masses.
The function $I$ is positive, symmetric,
smallest for degenerate masses and approximately scales with
the inverse of its biggest argument.
\footnote{See footnote in appendix \AppendixC.}
We also find convenient to define the
total absolute and relative bottom corrections :
\begin{equation}
\delta m_b = \delta^{\tilde g} m_b+\delta^{\tilde H} m_b \quad\quad
\Delta m_b = \Delta^{\tilde g} m_b+\Delta^{\tilde H} m_b.
\end{equation}
\noindent
The bottom mass before the SUSY corrections are included --
$m_b=\lambda_b v_1$ --
is related to the bottom mass after the SUSY corrections are included
-- $m_b^{SUSY}$ -- through :
\begin{equation}
m_b^{SUSY} = m_b+ \delta m_b = m_b (1+\Delta m_b)
\label{mbsusy}
\end{equation}
The pole mass after the SUSY corrections are included
is given by Eq.~\refeqn{mbMb} using $m_b^{SUSY}$
in Eq.~\refeqn{mbsusy}.

\newpage

\SECTION{IV. DECOUPLING OF THE HEAVY \\ RIGHT HANDED NEUTRINO}

In this section we briefly discuss the decoupling of
the heavy SM singlet right-handed neutrino
with mass $M_\nu$.
We assume that the neutrino Yukawa matrix is dominated by 
a single entry in the 33 position, although in realistic
models of neutrino mixing one would expect that 
there are at two entries which may have similar magnitude.
We have checked that our results are insensitive to 
the presence of a large off-diagonal entry
in the neutrino Yukawa matrix, assuming that the charged lepton Yukawa
matrix does not have any large off-diagonal entries.
\footnote{We focus on the third
          family only, thus we simplify our notation by replacing
          $\nu_\tau \to \nu$ and $\nu^c_\tau \to \nu^c$.}
The decoupling is achieved via see-saw mechanism \cite{seesaw} 
which generates a small mass for the left-handed neutrino through
the presence of a right-left Dirac Yukawa coupling.
The part of the superpotential of interest is~:
\begin{equation}
{\cal W} = \nu^c \> \lambda_\nu \> \nu \> h_u^0 +
           {\textstyle {1 \over 2}} \> M_\nu \> \nu^c \> \nu^c.
\end{equation}
Thus the light neutrino tau acquires a mass
$m_\nu = \lambda_\nu^2 v_2^2 / (4 M_\nu)$. In our model we used
$m_\nu = 0.05$ eV suggested by Ref.~\cite{SKamiokandeColl}
to fix $M_\nu \sim 10^{13}$ GeV
(see results in Table~\TabMnu\ in section V.C.)

\SECTION{V. mSUGRA RESULTS}

We now proceed to discuss the results generated by the model described
in the previous sections.
Although many of the results presented here appear elsewhere,
we find it useful to compile and review them here for the purposes
of comparison to the new situations we discuss later such as the
effect of neutrinos and non-universal soft masses.
These have been organized in three categories which are
suitable to expose their variation with $\alpha_s(M_Z)$ in the range
0.110--0.130, selected pole bottom masses $M_b=4.7,4.8,..,5.1$ GeV and
universal gaugino/soft scalar masses $M_{1/2}$, $m_0$ $ < 1$ TeV.
We plotted graphs showing the dependence of the results
with $\alpha_s$ for various $M_b$ and
fixed $M_{1/2}=400$, $m_0 = 200$ GeV;
graphs scanning the $M_{1/2}$--$m_0$
parameter space for illustrative fixed $\alpha_s$ and $M_b$; and
a set of numerical tables corresponding to the results obtained
from nine models for which the input is listed in Table \TabCasesDef.

\vbox{
\begin{center}
\begin{tabular}{cccccc}
\multicolumn{6}{c}{TABLE \TabCasesDef.} \cr
\noalign{\medskip}
\noalign{\hrule height\rulerheight}
\noalign{\smallskip}
\noalign{\hrule height\rulerheight}
\noalign{\medskip}
Case & $\alpha_s(M_Z)$ & $M_b$ & $m_b(M_b)$ & $M_{1/2}$ & $m_0$ \cr
\noalign{\medskip}
\noalign{\hrule height\rulerheight}
\noalign{\medskip}
A & 0.1150 & 4.70 & 4.12 & 400 & 200 \cr
B & 0.1150 & 4.70 & 4.12 & 800 & 400 \cr
C & 0.1150 & 5.10 & 4.49 & 400 & 200 \cr
D & 0.1150 & 5.10 & 4.49 & 800 & 400 \cr
E & 0.1250 & 4.70 & 3.99 & 400 & 200 \cr
F & 0.1250 & 4.70 & 3.99 & 800 & 400 \cr
G & 0.1250 & 5.10 & 4.36 & 400 & 200 \cr
H & 0.1250 & 5.10 & 4.36 & 800 & 400 \cr
I & 0.1250 & 4.80 & 4.08 & 600 & 400 \cr
\noalign{\medskip}
\noalign{\hrule height\rulerheight}
\noalign{\smallskip}
\noalign{\hrule height\rulerheight}
\end{tabular}
\end{center}

{\narrower\narrower\footnotesize\noindent
{TABLE \TabCasesDef.}
Input values of $\alpha_s(M_Z)$, the pole bottom mass $M_b$,
the running bottom mass $m_b(M_b)$ obtained from $M_b$ using
two loop QCD corrections only (see Eq.~\refeqn{mbMb}),
and the universal gaugino and soft scalar masses $M_{1/2}$, $m_0$
(given in GeV units) at the
unification scale $Q=M_X$ for a list of models we will refer
in the main text as Case A,B,..,I. \par
\bigskip}}

\SUBSECTION{A. Gauge unification}

In Fig.~\FigGaugeUnif\ we plot the unified $U(1)$ and $SU(2)$ gauge couplings
$\alpha_1^{-1}(M_X)$=$\alpha_2^{-1}(M_X)$= $\alpha_X^{-1}$ together with
$\alpha_3^{-1}(M_X)$ against $\alpha_s(M_Z)$ for
$M_{1/2}=400$, $m_0=200$ GeV and several values of $M_b$.
The mismatch $\Delta\alpha_X^{-1}=\alpha_3^{-1}(M_X)-\alpha_X^{-1}$
in gauge unification is small and decreasing for increasing $\alpha_s$.
In fact, for large values of $\alpha_s \sim 0.128$ complete gauge unification
occurs. The sensitiveness of $\Delta\alpha_X^{-1}$ to $M_b$ is quite
small.
Complete gauge unification can also be present for lower values of
$\alpha_s \sim 0.125$ as is shown by case H in Table~\TabMainPred\
where we find $\alpha_X^{-1} \sim \alpha_3 \sim 25.18$
($M_{1/2} = 800$, $m_0=400$ GeV.)

\vbox{
\noindent
\hfil
\vbox{
\epsfxsize=\figsize
\epsffile[130 380 510 735]{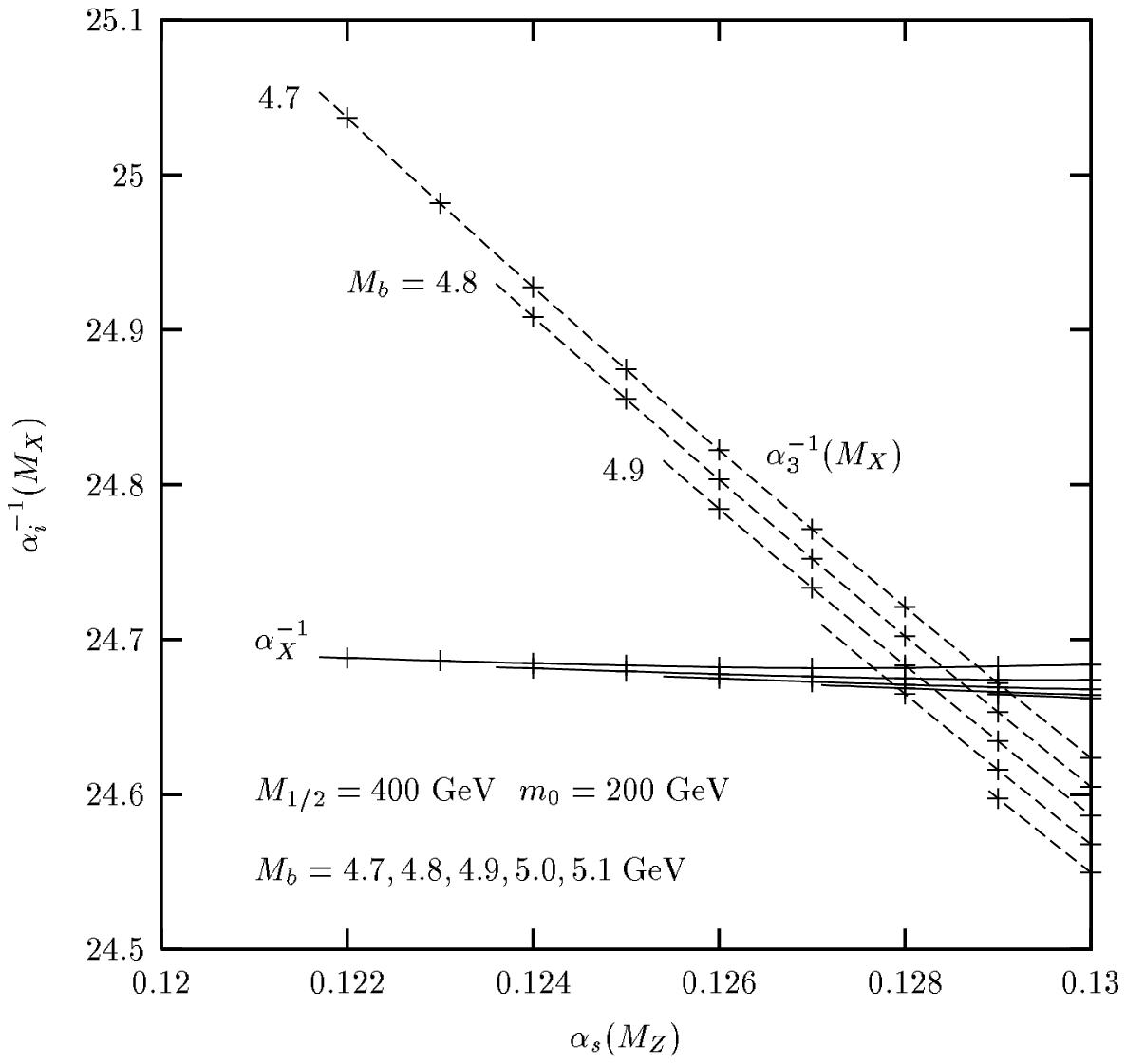}}

{\narrower\narrower\footnotesize\noindent
{FIG. \FigGaugeUnif.}
Dependence of unified gauge couplings
$\alpha_1^{-1}(M_X) = \alpha_2^{-1}(M_X) = \alpha_X^{-1}$ and
$\alpha_3^{-1}(M_X)$ on $\alpha_s(M_Z)$ for
several bottom masses $M_b$ and fixed gaugino and soft scalar masses
$M_{1/2}(M_X)$, $m_0(M_X)$. \par
\bigskip}}

\vbox{
\begin{center}
\begin{tabular}{cccc}
\multicolumn{4}{c}{TABLE \TabMainPred.} \cr
\noalign{\medskip}
\noalign{\hrule height\rulerheight}
\noalign{\smallskip}
\noalign{\hrule height\rulerheight}
\noalign{\medskip}
Case &
$\alpha_X^{-1} $ &
$\alpha_3^{-1}(M_X) $ &
$M_X / 10^{16}$ \\
\noalign{\medskip}
\noalign{\hrule height\rulerheight}
\noalign{\medskip}
A & 24.70 & 25.46 & 1.64 \\
B & 25.21 & 25.83 & 1.37 \\
C & 24.70 & 25.38 & 1.52 \\
D & 25.18 & 25.75 & 1.26 \\
E & 24.68 & 24.87 & 1.83 \\
F & 25.20 & 25.25 & 1.54 \\
G & 24.68 & 24.80 & 1.70 \\
H & 25.18 & 25.17 & 1.42 \\
I & 24.98 & 25.09 & 1.66 \\
\noalign{\medskip}
\noalign{\hrule height\rulerheight}
\noalign{\smallskip}
\noalign{\hrule height\rulerheight}
\end{tabular}
\end{center}

{\narrower\narrower\footnotesize\noindent
{TABLE \TabMainPred.}
Predicted values for the unified gauge coupling
$\alpha_X^{-1}=\alpha_1^{-1}(M_X)=\alpha_2^{-1}(M_X)$,
for the strong coupling $\alpha_3^{-1}(M_X)$ and
for the unification scale $M_X$ (given in GeV.) \par
\bigskip}}

Analysing Table~\TabMainPred\ we find that
$\delta\alpha_X^{-1}=\Delta\alpha_X^{-1} / \alpha_X^{-1} < 3 \%$.
Gauge unification also depends on the SUSY spectrum.
Generally, increasing $M_{1/2}$ and/or $m_0$ decreases $M_X$ and
$\delta\alpha_X^{-1}$.
In short partial gauge unification is achieved for
$M_X/10^{16} \sim 1-2$ GeV with $\alpha_X^{-1} \sim 25.0 \pm 0.5$
while complete unification $\alpha^{-1}_X = \alpha^{-1}_3(M_X)$
is favoured by large $\alpha_s(M_Z)$, $M_{1/2}$ and $M_b$.

\SUBSECTION{B. Top mass prediction \\
               with SUSY bottom and tau mass corrections}

We now turn to the predictions for the top mass. 
Experimentally the top mass has been measured to be \cite{TopAvGroup} :
\begin{equation}
M_t \sim 174.3\pm 3.2{\rm (stat.)}\pm 4.0{\rm (syst.)} {\rm\ GeV} 
\end{equation}

\vbox{
\noindent
\hfil
\vbox{
\epsfxsize=\figsize
\epsffile[130 380 510 735]{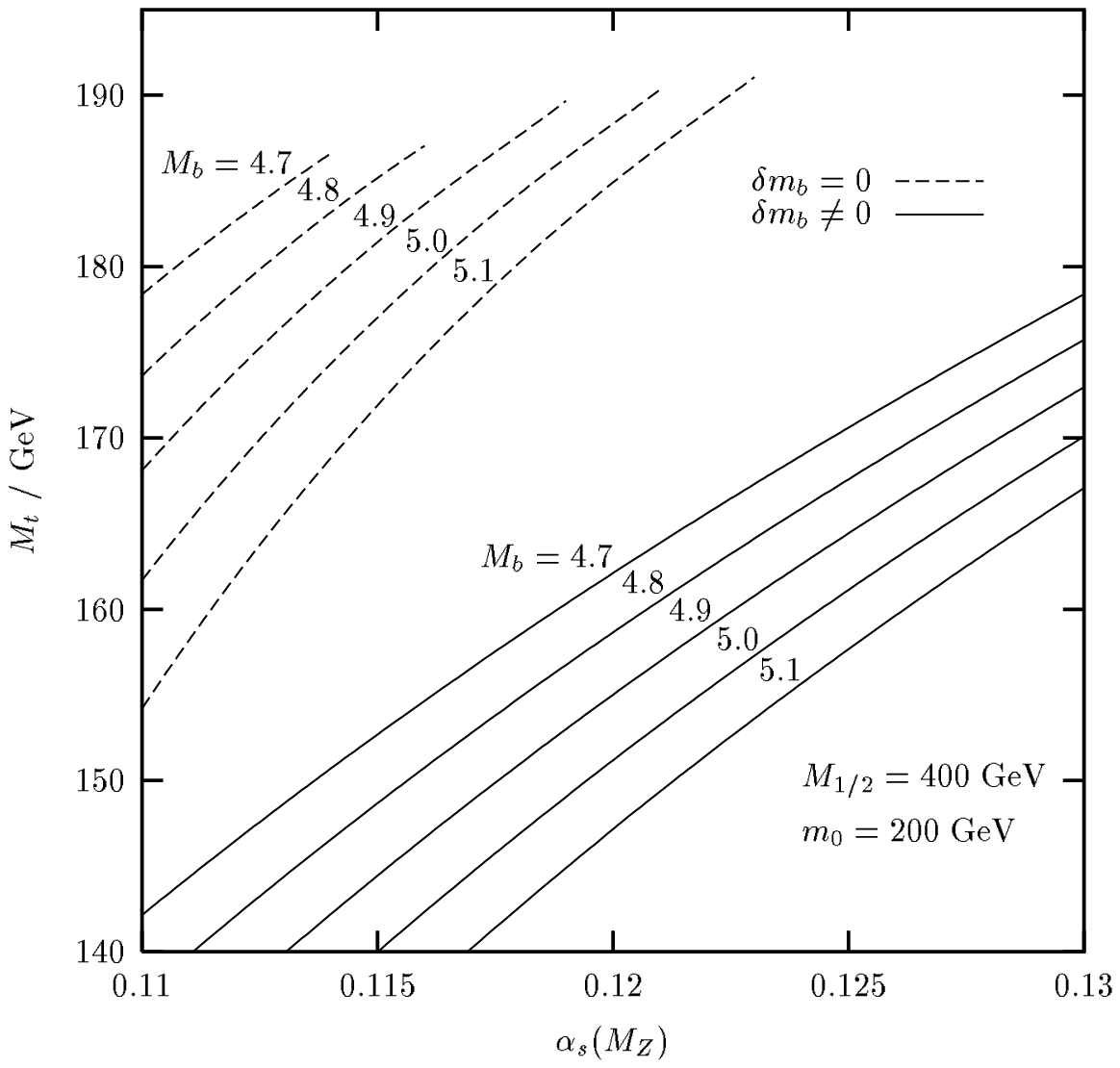}}

{\narrower\narrower\footnotesize\noindent
{FIG. \FigMtVsAlphasPic2.}
Predicted values for the pole top mass $M_t$ against
$\alpha_s(M_Z)$ for several values of the pole bottom mass
in the range $M_b = 4.7-5.1$ GeV. The dashed line refers to the top
mass prediction when no SUSY correction to the bottom mass is
considered ($\delta m_b = 0$).
The prediction for the top mass when negative
corrections to the bottom mass
are included ($\mu < 0$) is indicated by the solid line
($\delta m_b \ne 0$.)
\par\bigskip}}

In Fig.~\FigMtVsAlphasPic2\ we compare the values for the pole top
mass $M_t$ prediction obtained when no SUSY correction to the
bottom mass is considered ($\delta m_b = 0$), indicated with a dashed
line, and when they are included ($\delta m_b \ne 0$ and negative)
plotted with a solid line. We see that the SUSY
bottom corrections (for our choice of negative $\mu$) decrease the
top mass considerably.

In fact, the magnitude of the SUSY correction is so large that
it excludes the possibility of an eventual positive sign for $\mu$.
The reason is because if we have a positive $\mu$ the SUSY bottom
corrections are positive. Thus, in order to keep the bottom mass,
after the SUSY corrections are included $m^{SUSY}_b$, in the allowed
experimental range, the bottom mass before the SUSY corrections are
included must decrease.
In this case, third family unification leads to an
increase in the top Yukawa coupling.
However, it turns out that the required increase in the top Yukawa
coupling is so large that it drives it, at high energy,
well beyond perturbation theory.
Numerically, we find that the RGEs fail to converge.
\footnote{One can turn the argument around by saying that, if the top
quark mass is fixed to be around 175 GeV then third family
Yukawa unification at $M_X$ requires,
for positive $\mu$, a very large bottom mass prediction
after the SUSY corrections are included.}

At this point it is also interesting to comment on the effect
the bino correction to the tau mass has on the prediction for the
top mass. Again, a negative $\mu$ means that $\delta^{\tilde B} m_\tau$
is negative (see Eq.~\refeqn{BinoCorr}). Thus, in order to keep the tau mass
after the SUSY correction is included unchanged, the tau mass before
the SUSY correction is included must increase, implying that,
third family unification predicts a larger top mass.
Numerically we find that the top mass increases by 3--4 GeV.

Returning to Fig.~\FigMtVsAlphasPic2 we observe that when
the bottom corrections are included the top mass prediction is
only acceptable for large values of $\alpha_s(M_Z)$.
For this reason, in the analysis that will follow,
we will study the implications of including the SUSY bottom corrections
in our model by taking $\alpha_s(M_Z)$ to be in the range 0.120--0.130.

\newpage

\vbox{
\vbox{
\noindent
\hfil
\vbox{
\epsfxsize=\figsize
\epsffile[130 380 510 735]{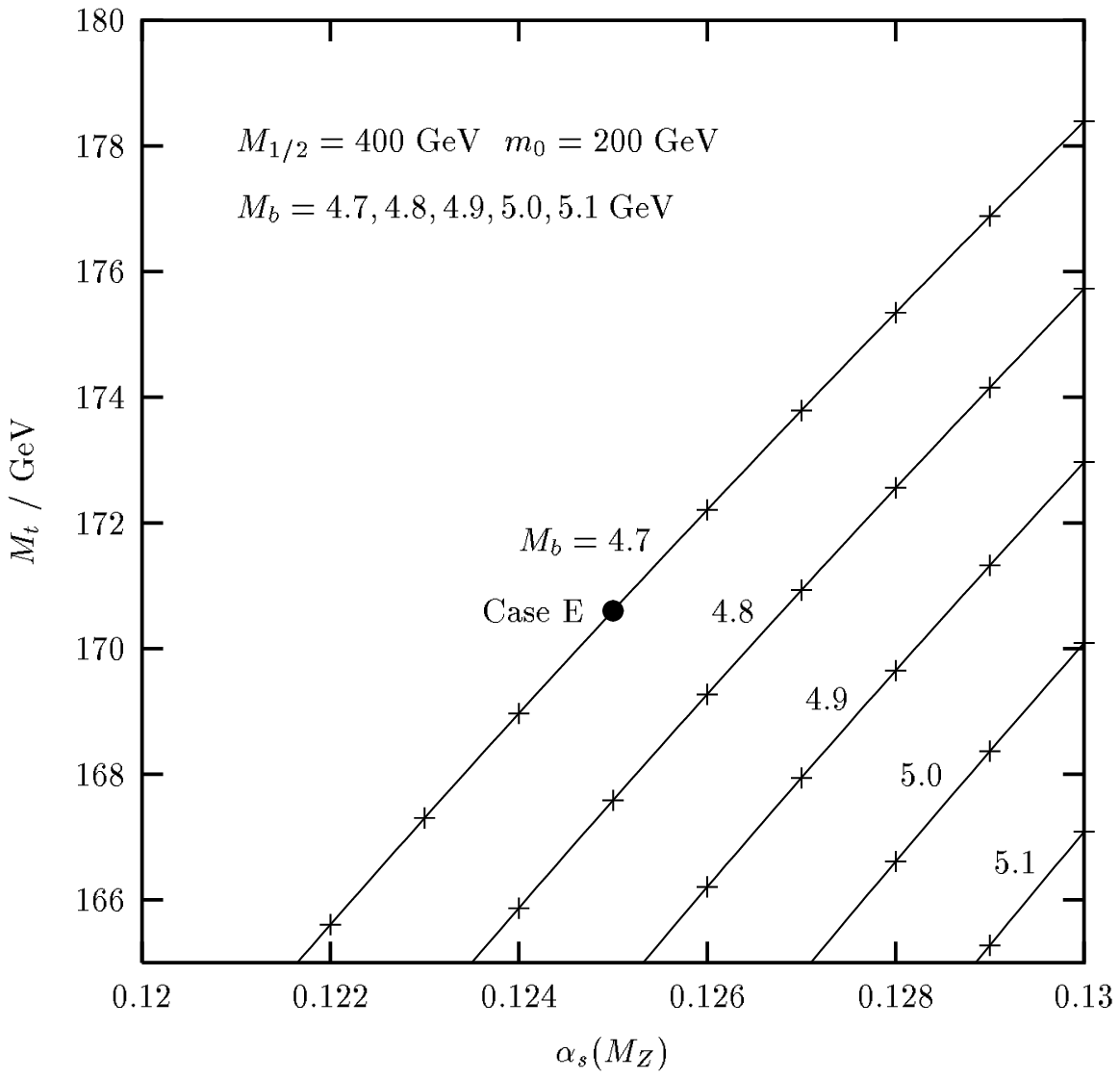}}

{\narrower\narrower\footnotesize\noindent
{FIG. \FigMtVsAlphas.}
Pole top mass prediction $M_t$ against $\alpha_s(M_Z)$ for
several values of the pole bottom mass $M_b$ and fixed $M_{1/2}$, $m_0$. \par
\bigskip}}

\vbox{
\noindent
\hfil
\vbox{
\epsfxsize=\figsize
\epsffile[130 380 510 735]{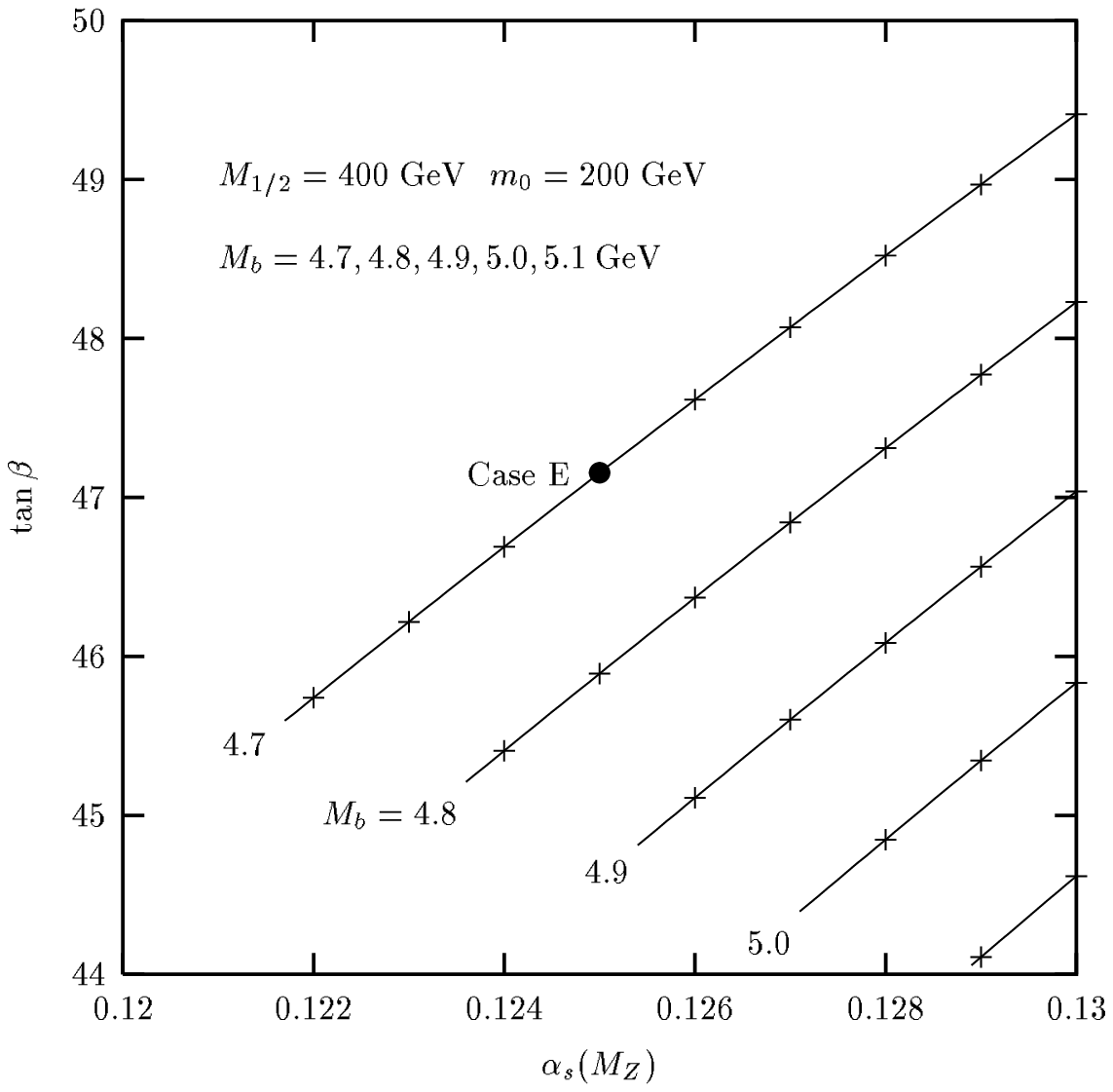}}

{\narrower\narrower\footnotesize\noindent
{FIG. \FigTanbVsAlphas.}
Prediction for $\tan \beta$ against $\alpha_s(M_Z)$ for
several values of the pole bottom mass $M_b$ and fixed $M_{1/2}$, $m_0$. \par
\bigskip}}}

In Fig.~\FigMtVsAlphas\ and \FigTanbVsAlphas\
the pole top mass $M_t$ and $\tan \beta$ are plotted against $\alpha_s$
for several values of the bottom mass $M_b$.
We can see that both are significantly sensitive to $\alpha_s$
and to $M_b$. For $M_b = 4.7$ (the first upper
line) the top mass takes values within $165.6 < M_t < 178.4$ GeV and
$45.73 < \tan\beta < 49.41$ for $0.122 < \alpha_s < 0.130$.
Additionally for a fixed value of $\alpha_s = 0.125$ the top mass
increases from 157.7 to 170.6 GeV and $\tan\beta$ from 41.99 to 47.15 when
$M_b$ decreases from 5.1 to 4.7 GeV.
In Table \TabMtTanbYukG\ we list the numerical predictions for the top
mass, $\tan\beta$ and the value of the unified Yukawa coupling
$\lambda_X$.

\vbox{
\begin{center}
\begin{tabular}{cccc}
\multicolumn{4}{c}{TABLE \TabMtTanbYukG.} \cr
\noalign{\medskip}
\noalign{\hrule height\rulerheight}
\noalign{\smallskip}
\noalign{\hrule height\rulerheight}
\noalign{\medskip}
Case &
$M_t$ &
$\tan\beta$ &
$\lambda_X$ \\
\noalign{\medskip}
\noalign{\hrule height\rulerheight}
\noalign{\medskip}
A & 152.7 & 42.16 & 0.426 \\
B & 158.1 & 44.16 & 0.469 \\
C & 135.3 & 35.94 & 0.317 \\
D & 139.7 & 37.51 & 0.340 \\
E & 170.6 & 47.15 & 0.569 \\
F & 176.2 & 49.40 & 0.643 \\
G & 157.7 & 41.99 & 0.428 \\
H & 163.3 & 44.06 & 0.473 \\
I & 170.7 & 47.34 & 0.561 \\
\noalign{\medskip}
\noalign{\hrule height\rulerheight}
\noalign{\smallskip}
\noalign{\hrule height\rulerheight}
\end{tabular}
\end{center}

{\narrower\narrower\footnotesize\noindent
{TABLE \TabMtTanbYukG.}
Predicted values for the top mass (in GeV), $\tan\beta$ and for
the value of the third family unified Yukawa coupling
$\lambda_X$.
\par}}

In Fig.~\FigMtVsYukX\ the strong correlation between
$M_t$ and $\lambda_X$ is
exposed. It is clear that since $\lambda_X$ is not large the top
quark mass prediction is far from its infrared (IR) fixed point $\sim 200$ GeV
\cite{BaBeOhPh,TopIRFixPnt}.

In Fig.~\FigMtVsMs\ we show that $M_t$ increases with 
increasing SUSY particle masses.
The reason for such dependence can be traced to the $SU(3)$ gauge
group factors in the SM and MSSM RGEs :
d$\lambda_t^{SM,MSSM}$/dt $\sim -b_t^{SM,MSSM} g_3^2$.
Numerically we have $b_t^{SM}=8 > b_t^{MSSM}=16/3$ thus increasing
$M_S$ allows for a wider $M_Z < Q < M_S$ range of
integration for the SM which favours an enhancement for the top mass.

\vbox{
\vbox{
\noindent
\hfil
\vbox{
\epsfxsize=\figsize
\epsffile[130 380 510 735]{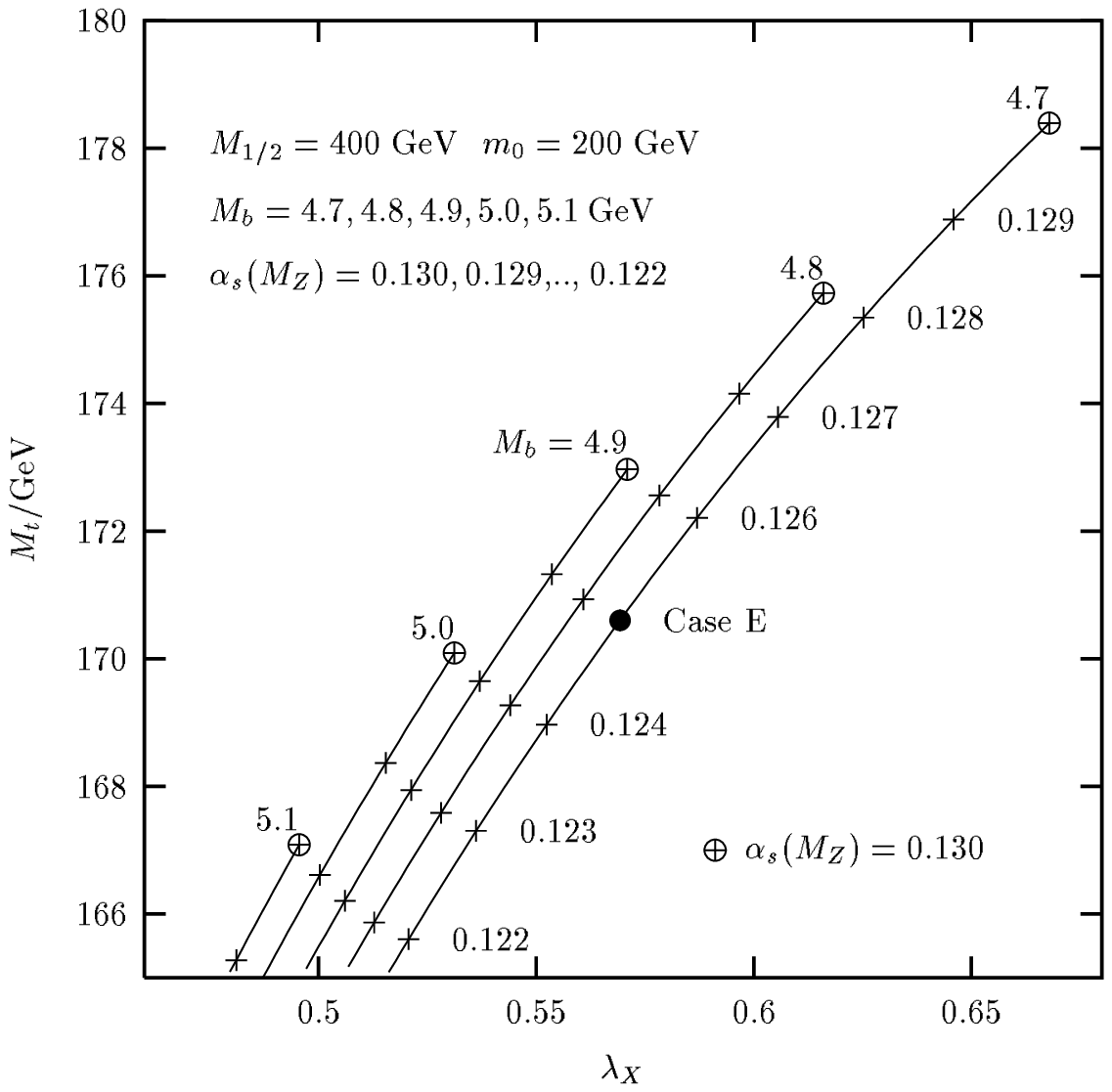}}

{\narrower\narrower\footnotesize\noindent
{FIG. \FigMtVsYukX.}
Correlation between the top pole mass prediction $M_t$ and the value
of the unified gauge coupling at $M_X$, $\lambda_X$.
Each line corresponds to a fixed choice of $M_b$ and along it
$\alpha_s(M_Z)$ decreases from a maximum value of 0.130
(indicated with a crossed circle) to lower values at 0.001 intervals
marked with crosses.
\par\bigskip}}

\vbox{
\noindent
\hfil
\vbox{
\epsfxsize=\figsize
\epsffile[130 380 510 735]{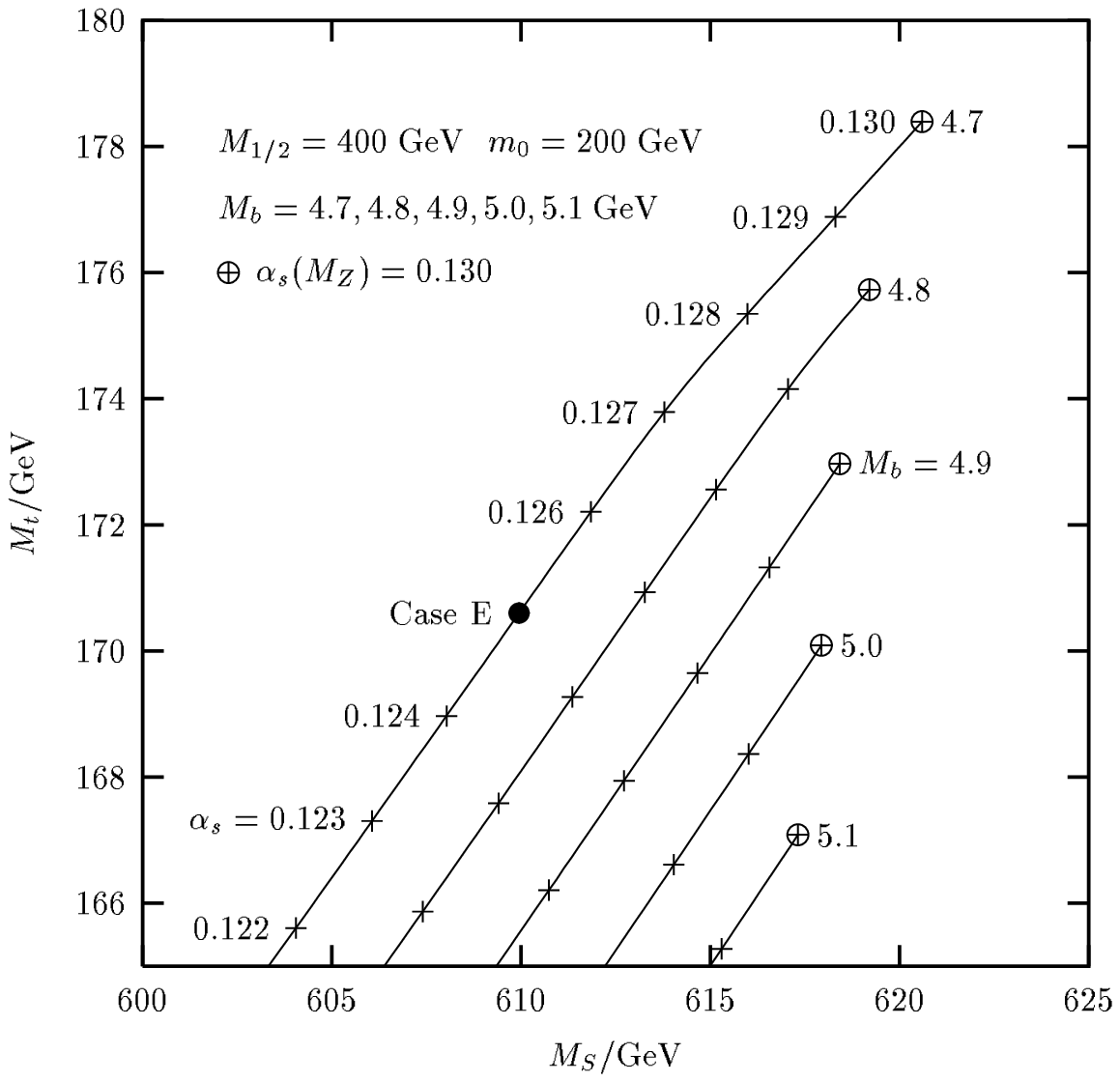}}

{\narrower\narrower\footnotesize\noindent
{FIG. \FigMtVsMs.}
Dependence of the top mass prediction $M_t$
on the effective supersymmetry scale $M_S$.
Each line corresponds to a
fixed choice of $M_b$ and along it $\alpha_s(M_Z)$ decreases
from a maximum value of 0.130 (indicated with a crossed circle)
to lower values at 0.001 intervals marked with crosses. \par
\bigskip}}}

\vbox{
\noindent
\hfil
\vbox{
\epsfxsize=\figsize
\epsffile[130 380 510 735]{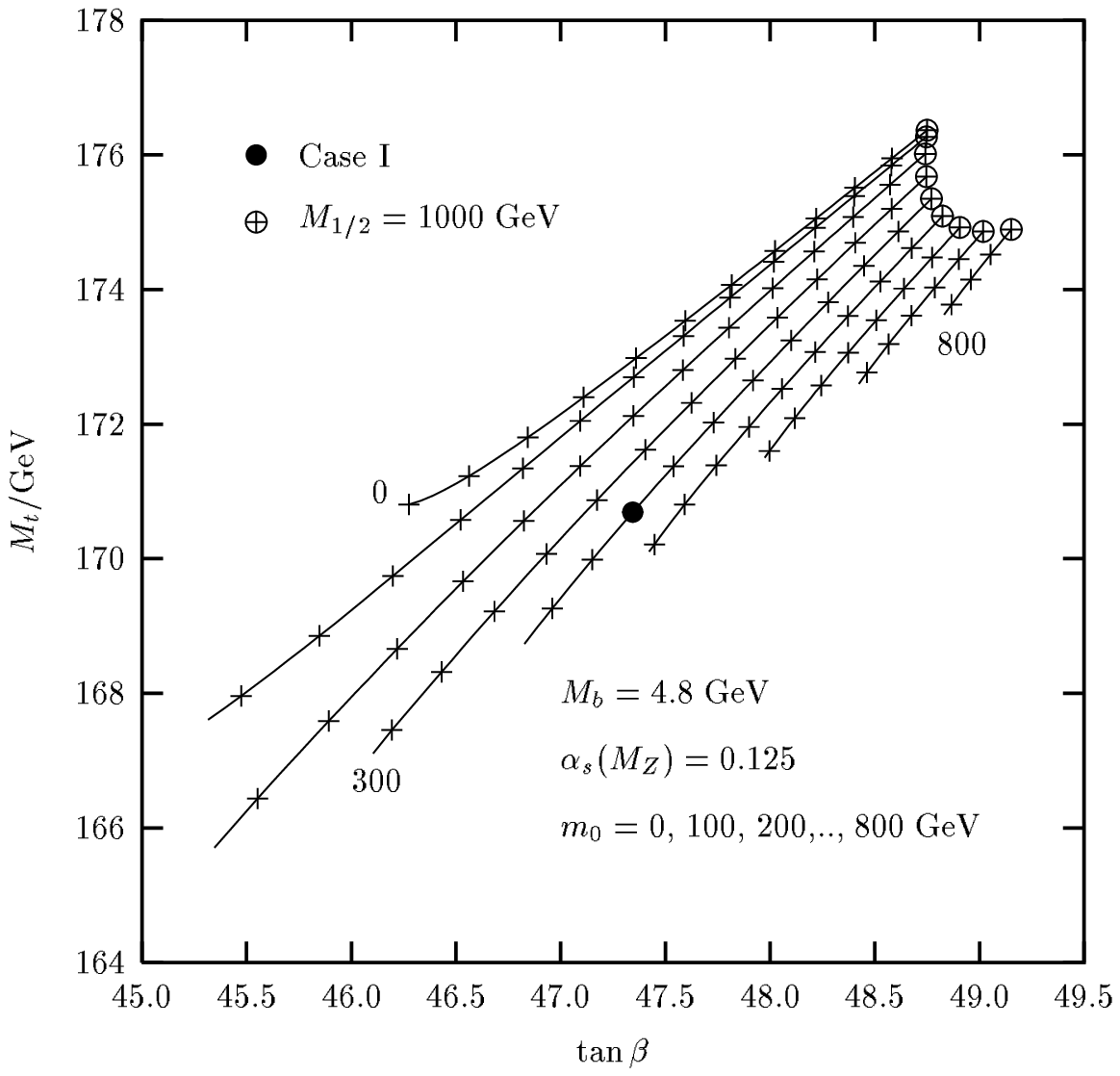}}

{\narrower\narrower\footnotesize\noindent
{FIG. \FigMtVsTanb.}
Correlation between the top mass prediction $M_t$
and the prediction for $\tan \beta$.
Each line corresponds to a
fixed choice of $m_0$ and along it $M_{1/2}$ decreases
from a maximum value of 1 TeV (indicated with a crossed circle)
to lower values at 50 GeV intervals marked with crosses. \par
\bigskip}}

Perhaps more interesting than Fig.~\FigMtVsMs\ where variations in $M_S$ are
indirectly induced by $\alpha_s$ is Fig.~\FigMtVsTanb\
where $\alpha_s = 0.125$ is keep fixed but $M_{1/2}$ and $m_0$
are allowed to vary.
The results are displayed in the $M_t$--$\tan \beta$ plane.

A quick estimate of the effect of including a consistent
supersymmetric scale $M_S$ on the top mass prediction can easily
be computed.
Taking, for example, $m_0=400$ GeV, varying $M_{1/2}$ from 600 to 700 GeV
will increase $M_t$ from 170.7 to 172.0 GeV (see Fig.~\FigMtVsTanb)
and $M_S$ increases from 915 to 1047 GeV.
Thus if we had considered a ``rigid'' $M_S \sim M_Z \sim 100$ GeV
we would find a value for $M_t$ decreased by $\Delta M_t \sim -8$ GeV.

\newpage

\SUBSECTION{C. Decoupling of the heavy tau neutrino}

In Fig.~\FigRunYuk\ we show an example of third family Yukawa unification
generated for case I. We note that the inclusion of the SUSY
correction $\delta m_b$ leads to low values for the unified Yukawa
coupling $\lambda_X \sim 0.55$.
Consulting Table~\TabMtTanbYukG\ we find
$0.32 < \lambda_X < 0.64$ in contrast with typical
$\lambda_X \sim 1$ predictions when no SUSY corrections are considered.
A low $\lambda_X$ implies that the RGEs which govern
the Yukawa evolution are dominated by the gauge terms,
thus the effect of decoupling the right-handed tau neutrino is small.

\vbox{
\noindent
\hfil
\vbox{
\epsfxsize=\figsize
\epsffile[130 380 510 735]{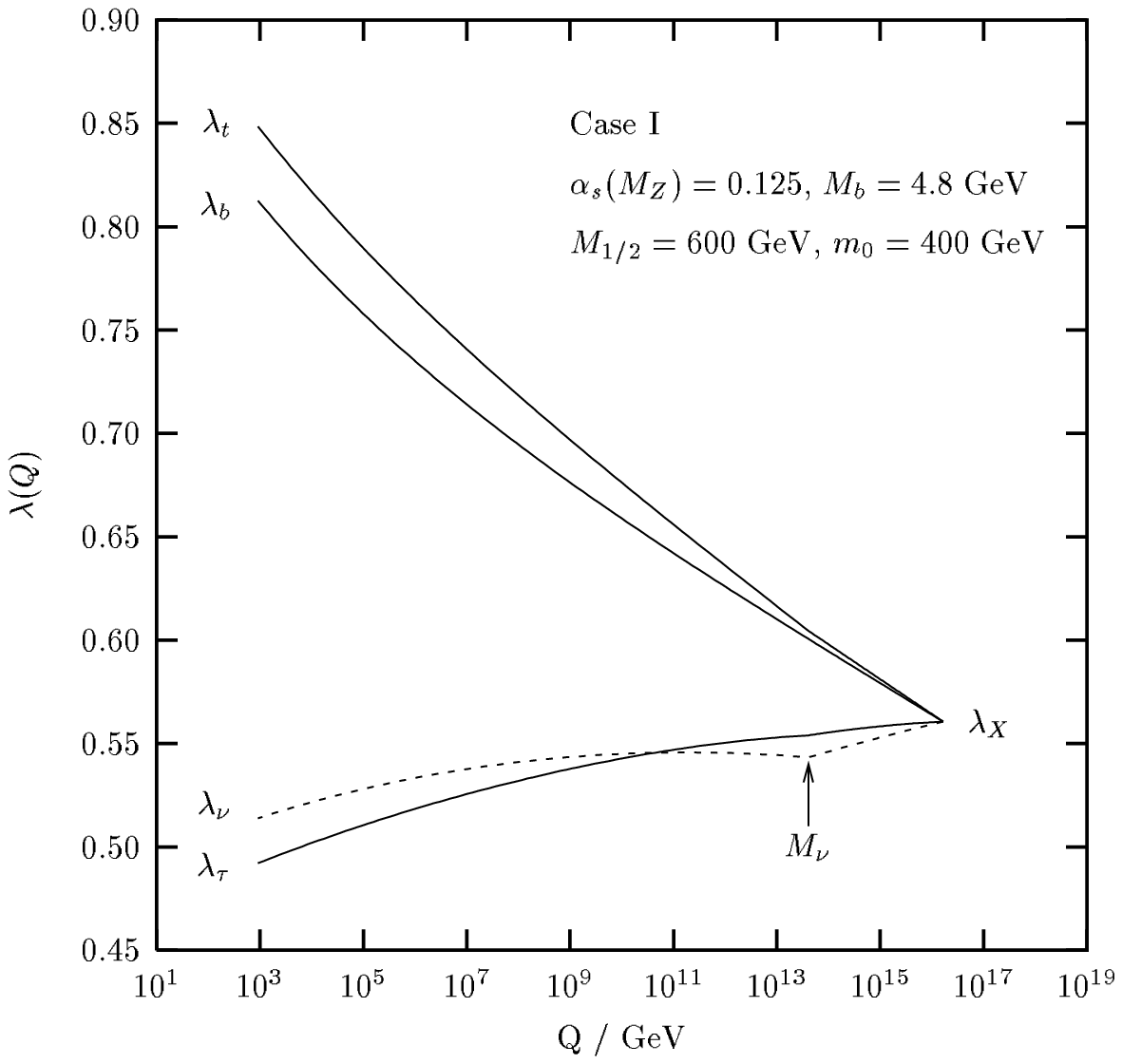}}

{\narrower\narrower\footnotesize\noindent
{FIG. \FigRunYuk.}
Running of the third family $t-b-\tau-\nu_\tau$ Yukawa couplings between
the unification scale $Q=M_X$ and the effective supersymmetry
breaking scale $Q=M_S$. The decoupling of the tau right-handed
neutrino at $Q=M_\nu$ mainly affects the running of
$\lambda_\nu$ (dashed line.) \par
\bigskip}}

In Table~\TabMnu\ we list the predicted values for the heavy
right-handed neutrino mass scale $M_\nu$ which in our model is fixed
by the requirement that the light left-handed neutrino tau has a mass
of $m_\nu = 0.05$ eV.
We see that $M_\nu$ is in the range $2-5 \times 10^{13}$ GeV.

\vbox{
\begin{center}
\begin{tabular}{cccccccccc}
\multicolumn{10}{c}{TABLE \TabMnu.} \cr
\noalign{\medskip}
\noalign{\hrule height\rulerheight}
\noalign{\smallskip}
\noalign{\hrule height\rulerheight}
\noalign{\medskip}
Case & A & B & C & D & E & F & G & H & I \\
\noalign{\medskip}
\noalign{\hrule height\rulerheight}
\noalign{\medskip}
$M_{\nu}/ 10^{13}$ &
 2.97 & 3.35 & 1.98 & 2.18 & 4.16 & 4.77 & 2.96 & 3.35 & 4.09 \\
\noalign{\medskip}
\noalign{\hrule height\rulerheight}
\noalign{\smallskip}
\noalign{\hrule height\rulerheight}
\end{tabular}
\end{center}

{\narrower\narrower\footnotesize\noindent
{TABLE \TabMnu.}
Predicted values for the mass of the heavy right-handed neutrino
$M_\nu$ (given in GeV) required to generate a light left-handed neutrino
mass $m_{\nu_3} = 0.05 {\rm\ eV}$.
\par\bigskip}}

\vbox{
\noindent
\hfil
\vbox{
\epsfxsize=\figsize
\epsffile[130 380 510 735]{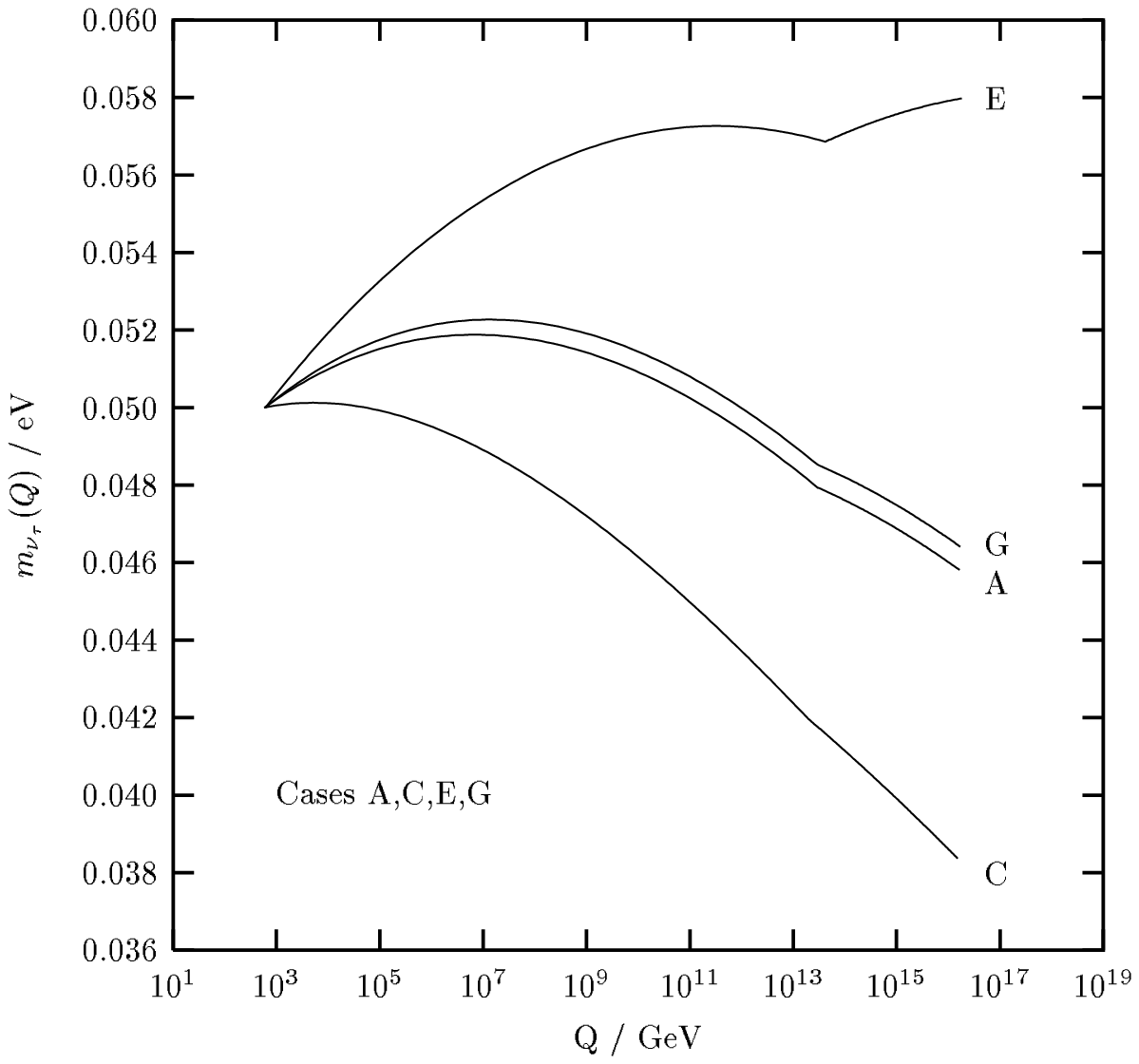}}

{\narrower\narrower\footnotesize\noindent
{FIG. \FigRunNeutMass.}
Running of the tau neutrino mass from the unification scale to low
energy for Cases A,C,E,G in Table~\TabCasesDef.\par
\bigskip}}

In Fig.~\FigRunNeutMass\ the evolution of
$m_{\nu_3}$ from $M_X$ to $M_S$
is shown for cases A,C,E,G. For each line a different $M_\nu(Q=M_X)$
was chosen in order to assure the same $m_\nu = 0.05$ eV at low energy.
For the purpose of illustration we have in Fig.~\FigMtVsMnu\
(and {\it only} in this plot)
relaxed the constraint upon $M_\nu$ and set it to vary at ten log
intervals from $M_X$ to $10^{-5} M_X$.
The effect of changing $M_\nu$ is shown in the $M_t$--$\lambda_X$ plane.
Comparing this plot with the tables throughout this article
we can conclude that the uncertainties attached to
$\alpha_s$, $M_b$ and $M_S$ are far more important than the ones
which might affect the determination of $M_\nu$. The predictions for
$M_t$ and $\lambda_X$ increase for decreasing $M_\nu$.
The variation in $M_t$ is quite small (typically less than 1 \%).
On the other hand variations in $\lambda_X$ are larger
and of about 3--4 \%.

\vbox{
\noindent
\hfil
\vbox{
\epsfxsize=\figsize
\epsffile[130 380 510 735]{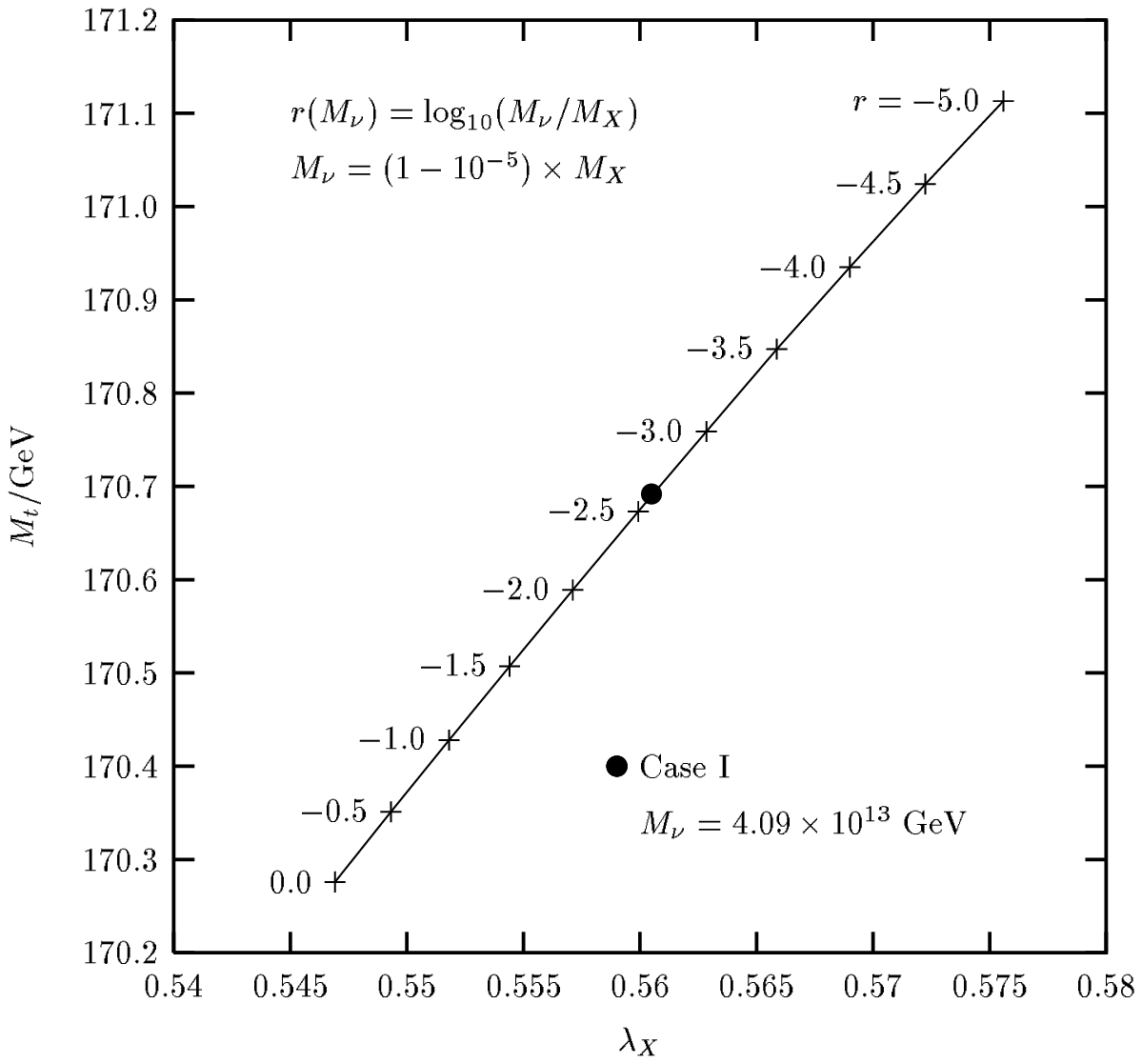}}

{\narrower\narrower\footnotesize\noindent
{FIG. \FigMtVsMnu.}
Variations in the predictions for the top mass $M_t$ and in the unified
Yukawa coupling $\lambda_X$ induced by changing $M_\nu$
in the range $(1-10^{-5}) \times  M_X$.
The dark circle refers to the choice of $M_\nu$ which for Case I
gives $m_{\nu_3} = 0.05$ eV.\par
\bigskip}}

\SECTION{VI. YUKAWA UNIFICATION IN THE \422 MODEL}

In this section we will extend the analysis of the top quark mass,
so far considered in the context of the MSSM with a right-handed tau neutrino,
by assuming that above the unification scale, at which the gauge couplings
meet, the gauge group is \422 \cite{PaSa}. We will proceed in three
steps. Firstly 
we review the model constrained by universal soft SUSY breaking masses (USBM).
Secondly, the model is considered under non-USBM conditions and thirdly
we will allow for the presence of $D$-terms which generally arise from the
reduction of the gauge group rank~\cite{Dterms}.
Finally, 
we conclude by showing that the 422 model with non-universal family
dependent $A$-terms, is compatible with third family Yukawa unification
(which requires a negative sign for $\mu$) and with the
experimental branch ratio for the $b\to s\gamma$ and 
$\tau\to\mu\gamma$ decays.

\SUBSECTION{A. The model}

Here we briefly summarize the parts of the model that
are relevant for our analysis.
For a more complete discussion see Ref.~\cite{AnLe}.
The SM fermions, together with the right-handed tau neutrino, are conveniently
accommodated in the following $F^c=(\bar 4,1,\bar 2)$ and $F=(4,2,1)$
representations :
\begin{equation}
F^c_A = \left(\matrix{
            d^c & d^c & d^c & e^c \cr
            u^c & u^c & u^c & \nu^c \cr}
            \right)_A
\qquad
F_B   = \left(\matrix{
            u & u & u & \nu \cr
            d & d & d & e \cr}
            \right)_B
\label{FcFields}
\end{equation}
The MSSM Higgs bosons fields are contained in $h=(1,\bar 2,2)$ :
\begin{equation}
h = \left(\matrix{h_d^- & h_u^0 \cr
                  h_d^0 & h_u^+ \cr} \right)
\label{SMHiggs}
\end{equation}
whereas the heavy Higgs bosons $\bar H=(\bar 4,1,\bar 2)$ and $H=(4,1,2)$
are denoted by :
\begin{equation}
\bar H = \left(\matrix{ \bar H_d & \bar H_d & \bar H_d & \bar H_e \cr
                        \bar H_u & \bar H_u & \bar H_u & \bar H_\nu
                        \cr}\right)
\qquad
H  = \left(\matrix{ H_d & H_d & H_d & H_e \cr
                    H_u & H_u & H_u & H_\nu  \cr}\right).
\label{RHiggs}
\end{equation}
In addition to the Higgs fields in Eqs.~\refeqn{SMHiggs} and \refeqn{RHiggs}
the model also involves an $SU(4)$ sextet field $D=(6,1,1)=(D_3,D^c_3)$.
\footnote{In fact, since we wish to keep the gauge couplings unified above
$M_X$ we also trivially include another pair of heavy Higgs boson fields
in the $(4,2,1)$ and $(\bar 4,\bar 2,1)$ representations
and six more replicas of $D$'s.
This is sufficient to assure that, above the unification scale,
the one loop RGEs beta functions of the gauge couplings are equal.
\cite{AlKi} }

The superpotential of the minimal 422 model is :
\footnote{Note that $FFD$ and $F^c F^c D$ terms, associated with
baryon number violating processes, can be forbidden by imposing a
global R-symmetry. \cite{KiSh}}
\begin{eqnarray}
{\cal W} &=& F^c_A \lambda_{AB} F_B h + \lambda_h S h h + \nonumber \\
         & & \lambda_S S (\bar H H - M^2_H) +
             \lambda_H H H D + \lambda_{\bar H} {\bar H} {\bar H} D +
             F^c_A \lambda'_{AB} F^c_B \,{H H \over M_P}
\label{W3}
\end{eqnarray}
where $S$ denotes a gauge singlet superfield, the $\lambda$'s
are real dimensionless parameters and $M_H \sim M_X$.
Additionally, the Planck mass is denoted by
$M_P \sim 2.4 \times 10^{18}$ GeV.
As a result of the superpotential terms involving the singlet $S$,
the Higgs fields develop VEVs
$\langle H \> \rangle = \langle H_\nu \rangle \sim M_X$ and
$\langle \bar H \rangle = \langle \bar H_\nu \rangle \sim M_X$
which lead to the symmetry breaking~:
\begin{equation}
SU(4) \otimes SU(2)_L \otimes SU(2)_R \to
SU(3)_c \otimes SU(2)_L \otimes U(1)_Y.
\label{422:321}
\end{equation}
The singlet $S$ itself also naturally develops a small VEV of the order
of the SUSY breaking scale \cite{KiSh} so that the $\lambda_h S$ term in Eq.~\refeqn{W3}
gives an effective $\mu$ parameter of the correct order of magnitude.
Under Eq.~\refeqn{422:321} the Higgs field $h$ in Eq.~\refeqn{SMHiggs}
splits into the familiar MSSM doublets $h_u$ and $h_d$ whose neutral
components subsequently develop weak scale VEVs
$\langle h_u^0 \rangle = v_2$ and $\langle h_d^0 \rangle = v_1$ with
$\tanb = v_2/v_1$.
The neutrino field $\nu^c$ acquires a large mass
$M_{\nu} \sim \lambda' \langle HH \rangle / M_P$ through the
non-renormalizable term in ${\cal W}$ which, together
with the Dirac $\nu^c$ -- $\nu$ interaction
(proportional to $\lambda \langle h_u^0 \rangle$),
gives rise to a 2 $\times$ 2 matrix that generates, via see-saw mechanism,
a suppressed mass for the left-handed neutrino state \cite{AnLe}.
The $D$ field does not develop a VEV but the terms $HHD$ and $\bar H \bar H D$
combine the colour triplet parts of $H$, $\bar H$ and $D$ into acceptable
GUT scale mass terms \cite{AnLe}.

In addition to the terms generated by the superpotential of Eq.~\refeqn{W3}
the lagrangian of the 422 model also includes the corresponding
trilinear soft terms
$\tilde F^c \tilde A \tilde F h +
 \tilde\lambda_h S h h$,
\footnote{Often re-parametrised by $\tilde A = A \lambda$ and
      $m_3^2 = \tilde\lambda_h \langle S \rangle$.}
masses for the $SU(4)$, $SU(2)_L$, $SU(2)_R$ gauginos
$M_4$, $M_{2L}$, $M_{2R}$ and explicit soft masses for the
scalar fields
$\tilde m^2_F |\tilde F|^2 +
 \tilde m^2_{F^c} |\tilde F^c|^2 + {\tilde m}^2_h |\;\! h|^2$.

Finally we remind that the symmetry breaking in Eq~\refeqn{422:321}
leads to specific relations between the
$SU(4)$, $SU(2)_{2R}$ and $U(1)_Y$ gauge couplings and gaugino
masses at $M_X$ given by \cite{KaMuYa} :
\footnote{See appendix \AppendixD\
for a comprehensive derivation of these relations.}
\begin{equation}
{5 \over \alpha_1} = {2\over \alpha_4}+{3\over \alpha_{2R}}
\hskip 2cm
{5 M_1\over \alpha_1} = {2 M_4\over \alpha_4}+
                        {3 M_{2R}\over \alpha_{2R}}
\label{Gauge422}
\end{equation}

\SUBSECTION{B. The universal model}

In this section we briefly review the 422 model with USBM.
The main motivation for USBM is the smallness of flavour changing
neutral currents (FCNC) and it's simplicity
--- few input parameters are needed to specify a
otherwise complex and largely unconstrained parameter space.

The model is very similar to the one presented in the previous section
except that we have imposed the high energy boundary conditions at the
Planck scale instead of at $M_X$, thus we allowed for the
running of the parameters between these two scales.
The relevant input parameters were sign $\mu < 0$, $A_0=0$ and :
\begin{equation}
0 < M_{1/2} < 2 {\rm\ TeV} \hskip 2cm
0 < m_0 < M_{1/2}
\end{equation}
 where
$M_{1/2}=M_4=M_{2L}=M_{2R}$ is the
common gaugino mass and
$m^2_0=\tilde m^2_F=\tilde m^2_{F^c}={\tilde m}^2_h$
is the universal scalar mass ($\alpha_s = 0.120$, $M_b = 4.8$ GeV).
The results are presented in scattered plots
in Fig.~\Figm0VsMgaug\ and Fig.~\FigMtopVsMgaug.

In Fig.~\Figm0VsMgaug we plot $m_0$ against $M_{1/2}$.
The black (white) circles indicate points that do (not) satisfy EWSB,
\ie $m^2_1-m^2_2 \bigsml M^2_Z$.
\footnote{This condition can be obtained from
Eqs.~\refeqn{mu2} and \refeqn{m32} in the limit of large $\tan\beta$.}
We observe that increasing $m_0$ is disfavoured by EWSB
whereas increasing $M_{1/2}$ makes EWSB easier to occur.
The interplay between these two dependencies dictates that only the
region with $M_{1/2} \bigsim  m_0$, with a threshold for
$M_{1/2} \sim 500-600$ GeV, is allowed.
\footnote{Note that this value corresponds to the gaugino
mass at the Planck scale which in our model decreases
between $M_P$ and $M_X$. At the GUT scale we obtained
$M_{1/2} > 400$ GeV.}
The large gaugino mass (required in models with USBM and large \tanb)
has important implications, it leads to a large negative $m_2$
parameter at low energy. Thus the EWSB condition in Eq.~\refeqn{mu2}
can only be satisfied if $|m^2_2|-\mu^2 \sim M^2_Z/2$ with
$|m^2_2| \sim \mu^2 \gg M^2_Z/2$ --- fine tuning ${\cal O}$(500)
is inevitable \cite{OlPo2,RHempfling2}.
Furthermore large values for $\mu$, which is correlated with $M_{1/2}$,
increase the SUSY correction to the bottom mass in
Eq.~\refeqn{gluinoCorr} and Eq.~\refeqn{HiggsinoCorr} pushing the
prediction for the top mass below the experimental lower bound.

\vbox{
\noindent
\hfil
\vbox{
\epsfxsize=\figsize
\epsffile[130 380 510 735]{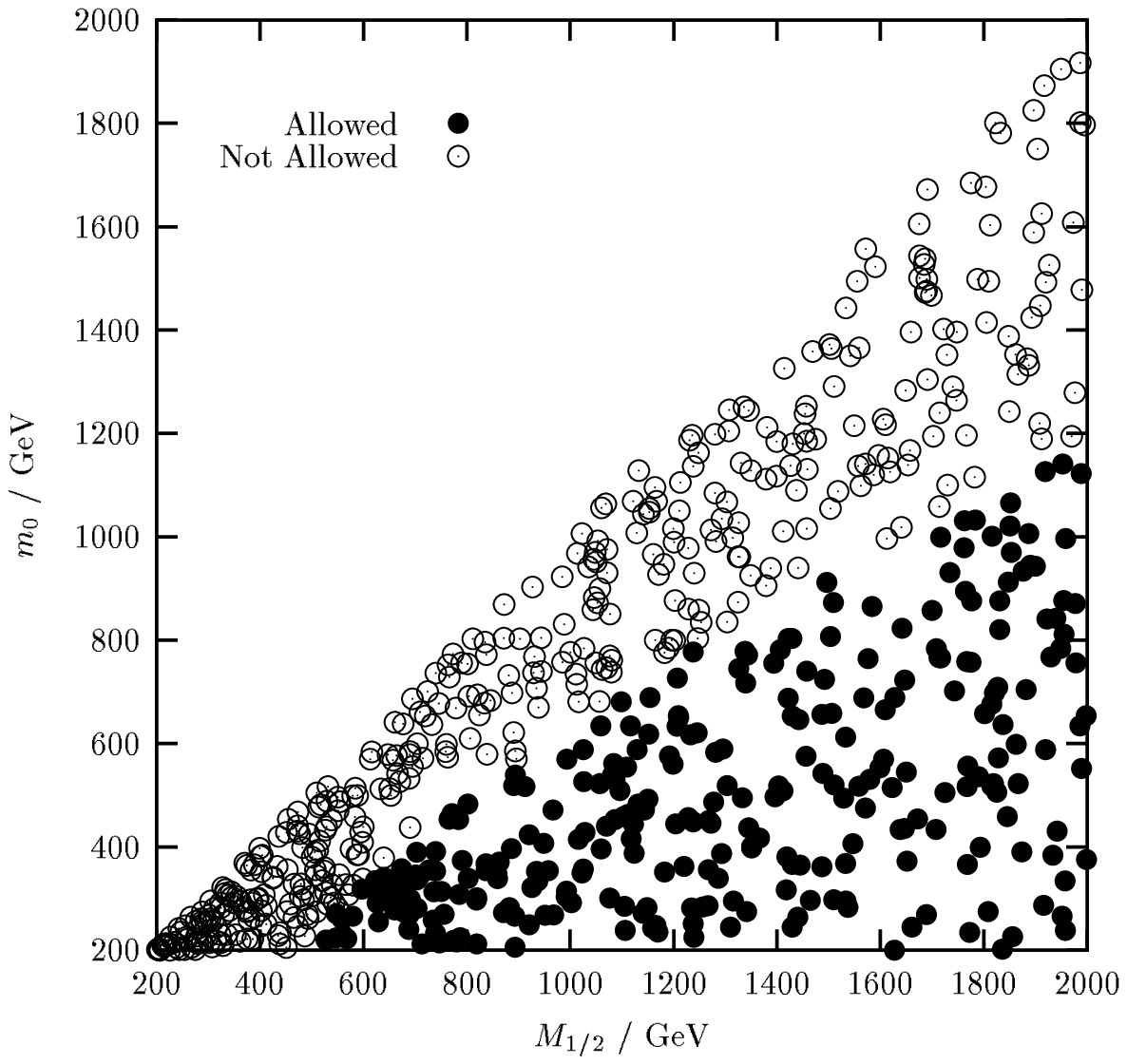}}

{\narrower\narrower\footnotesize\noindent
{FIG.~\Figm0VsMgaug.}
Common scalar mass $m_0$ against unified gaugino mass $M_{1/2}$ fixed
at the Planck scale. The white circles denote points that fail to
satisfy the electro-weak symmetry breaking conditions.
\par}}

In Fig.~\FigMtopVsMgaug\ we show the top mass prediction $M_t$ against $M_{1/2}$.
We can see that not only is $M_t$ smaller than 170 GeV
(even for $M_{1/2}$ as large as 2 TeV) but also that for fixed $M_{1/2}$
its dependence with $m_0$ is small. It is relevant to note that,
in the attractive scenario of lowest $M_{1/2}$, corresponding to
lighter sparticles and smallest fine tuning, the largest top mass
prediction is unacceptable. Indeed, we found, by pushing the stability
of the Higgs potential to the extreme, that for a low gaugino mass
$M_{1/2} \sim 600$ GeV
\footnote{Corresponding to $M_{1/2}(M_X)$ = 485 GeV and
gluino masses $m_{\tilde g} (M_Z) = 1145$ GeV.}
we obtain $M_t \sim 157$ GeV (\tanb = 43).
Nevertheless, the fact that in Fig.~\FigMtopVsMgaug\ the white circles
are above of the blacks suggests that, if the scalar masses
$\tilde m_F$, $\tilde m_{F^c}$, $\tilde m_h$
are allowed to increase
(and/or split) and be bigger than $M_{1/2}$, then $M_t$ may increase.
Thus, the main problem is how to conciliate EWSB with scalar masses
bigger than gaugino masses.

\vbox{
\noindent
\hfil
\vbox{
\epsfxsize=\figsize
\epsffile[130 380 510 735]{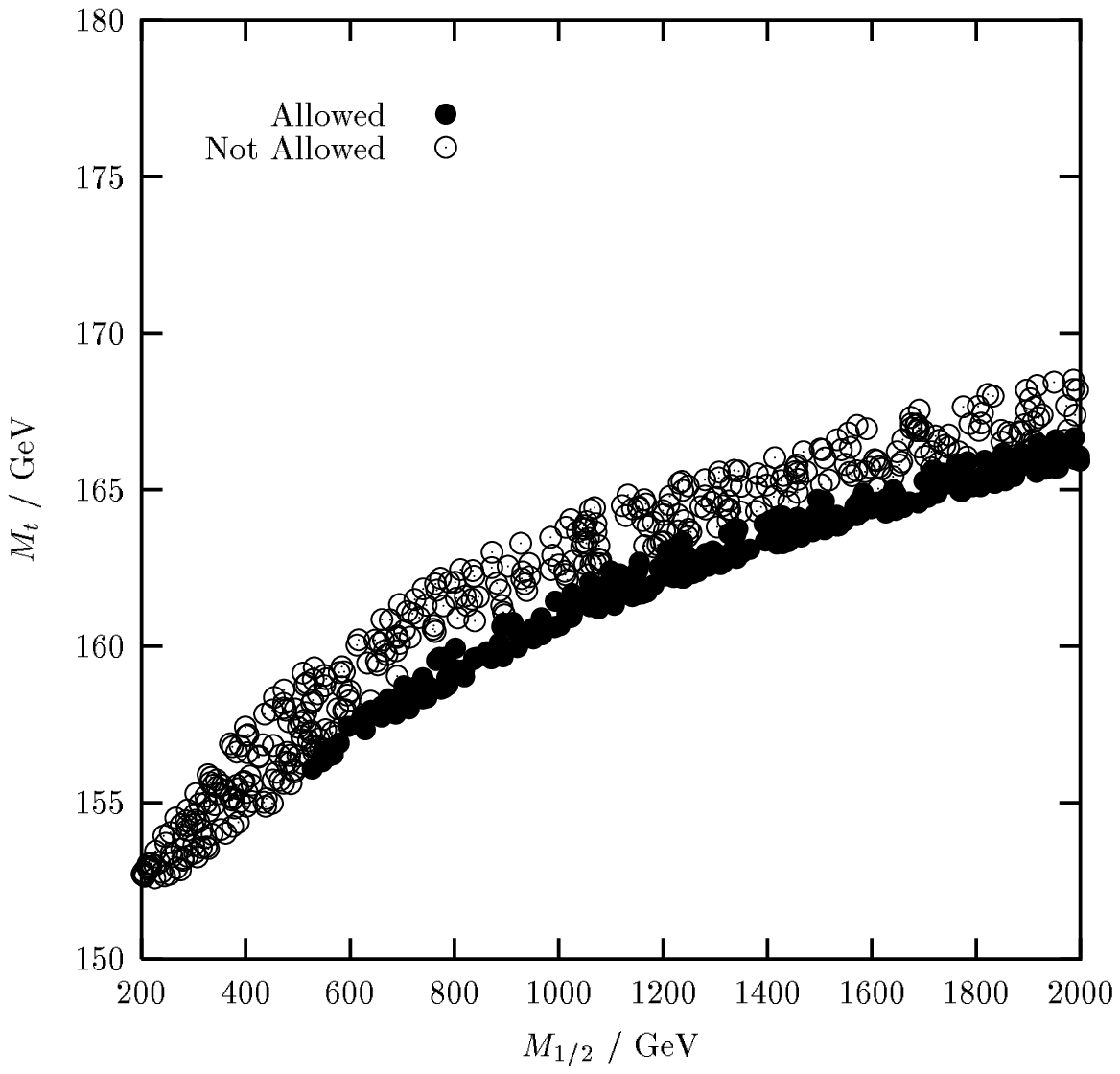}}

{\narrower\narrower\footnotesize\noindent
{FIG.~\FigMtopVsMgaug.}
Top mass prediction $M_t$ against the common gaugino mass $M_{1/2}$
fixed at the Planck scale ($\alpha_s = 0.120$.)
\par}}

\SUBSECTION{C. The non-universal model}

Over the past decade the numerous studies of models with USBM
\cite{AnLaSh,KaKoRoWe,CaOlPoWa,OlPo1,DrNo,MaMo,BaBeOh3}
at the GUT scale showed that they often face recurrent difficulties
with low energy phenomenology.
The unsatisfactory results have stimulated the interest in
non-universal models
\cite{HaRaSa,OlPo2,BaDiFeTa,KaMuYa,RaSaHa,PoPo2,PoPo3,LlMu}
which are well motivated from a theoretical
point of view. As was emphasized in \cite{PoPo2,PoPo3}, even when
universality is imposed at the Planck scale, radiative corrections
between $M_P$ and $M_X$ and heavy threshold effects
lead to non-universal parameters at the GUT scale.
Moreover, the most general SUGRA models
(with non-canonical kinetic terms) in which supersymmetry is
broken in a hidden sector, and/or superstring theories in which supersymmetry
is broken by the $F$ component of the moduli fields with different
weights, show that non-universality can be generated
at the Planck or string scale.

In this section we consider the 422 model with non-universal
boundary conditions at $M_P$. The independent input parameters were
$\tilde m^2_F$, $\tilde m^2_{F^c}$ (proportional to the unit matrix),
${\tilde m}^2_h$ and $M_4=M_{2R}\ne M_{2L}$.
\footnote{Note that, for the sake of simplicity,
we have taken the $SU(4)$ and $SU(2)_{R}$
gaugino masses equal at $M_P$. Since their one-loop beta functions are
identical, and the gauge couplings are roughly unified above $M_X$
we also have $M_4(M_X) = M_{2R}(M_X)$.}
These parameters were made to vary at random in the ranges :
\vskip -1cm
\begin{equation}
200 {\rm\ GeV} < M_4=M_{2R} < 2 {\rm\ TeV}
\end{equation}
\begin{equation}
{\textstyle {1 \over 2}} M_4 < M_{2L} < 2 M_4
\end{equation}
\begin{equation}
200 {\rm\ GeV} < \tilde m_F\ne \tilde m_{F^c}\ne {\tilde m}_h < M_4.
\end{equation}
\noindent
The results are presented in Figs.~\FigMtopVsMg4--\FigMtopVsMhMF\
which main purpose was
not to exhaustively scan all the parameter space but to be a guide
of the configuration of the parameter space which most
effectively increased $M_t$

In Fig.~\FigMtopVsMg4\ we plot the top mass against $M_4$.
Comparing with Fig.~\FigMtopVsMgaug\
we observe that while the average increase of $M_t$ with $M_4$ is
similar, the strict correlation present in the universal case is
replaced, in the non-universal model, by a dispersed region of
enhanced/suppressed $M_t$ predictions. This plot also illustrates that
one can expect $M_t \sim 170$ GeV (for $M_4 > 1000$ GeV) and
shows that EWSB can occur under more relaxed conditions, which is
obvious from the way the white and black circles distribute evenly.
\footnote{A similar graph is obtained when $M_t$ is
plotted against $M_{2L}$ thus we spared from including it here.}

In Fig.~\FigM2LVsM4 $M_{2L}$ is plotted against $M_4$. We see that
$M_{2L} > M_4$ is disfavoured by the condition on the Higgs
potential to be bounded. The majority of black circles is concentrated in the
$M_4 > M_{2L}$ region with $M_4 > 600$ GeV. This preference is
illustrated by the left-right asymmetry around the $x$-axis
of Fig.~\FigMtopVsM4M2L which displays $M_t$ against $M_4-M_{2L}$.

\vbox{
\vbox{
\noindent
\hfil
\vbox{
\epsfxsize=\figsize
\epsffile[130 380 510 735]{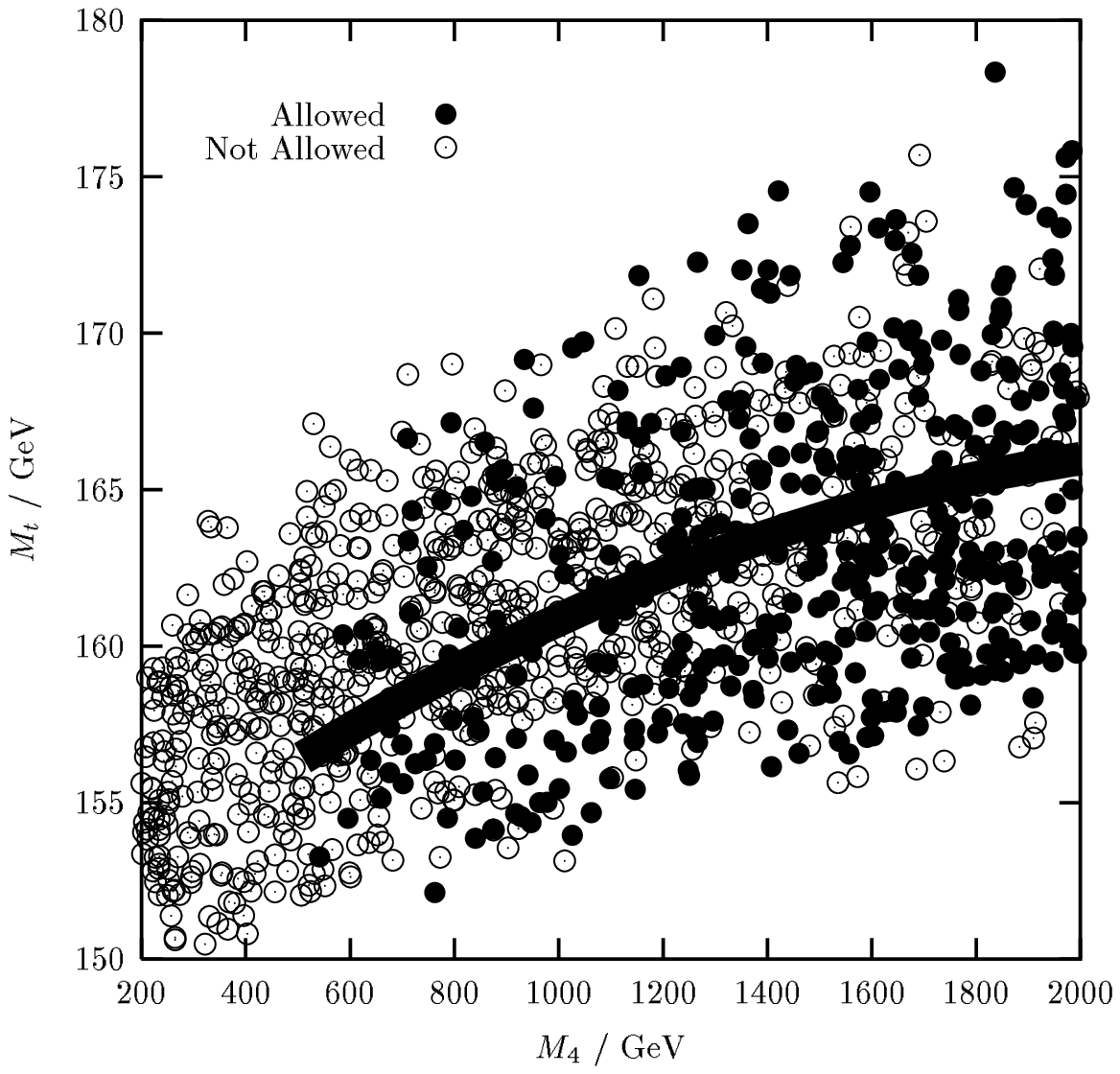}}

{\narrower\narrower\footnotesize\noindent
{FIG.~\FigMtopVsMg4.}
Top mass prediction $M_t$ against the $SU(4)$ gaugino mass $M_4$
fixed at the Planck scale in the non-universal model.
The thick central line indicates the prediction for $M_t$
in the universal model of Fig.~\FigMtopVsMgaug\ ($\alpha_s = 0.120$.)
\par\bigskip}}

\vbox{
\noindent
\hfil
\vbox{
\epsfxsize=\figsize
\epsffile[130 380 510 735]{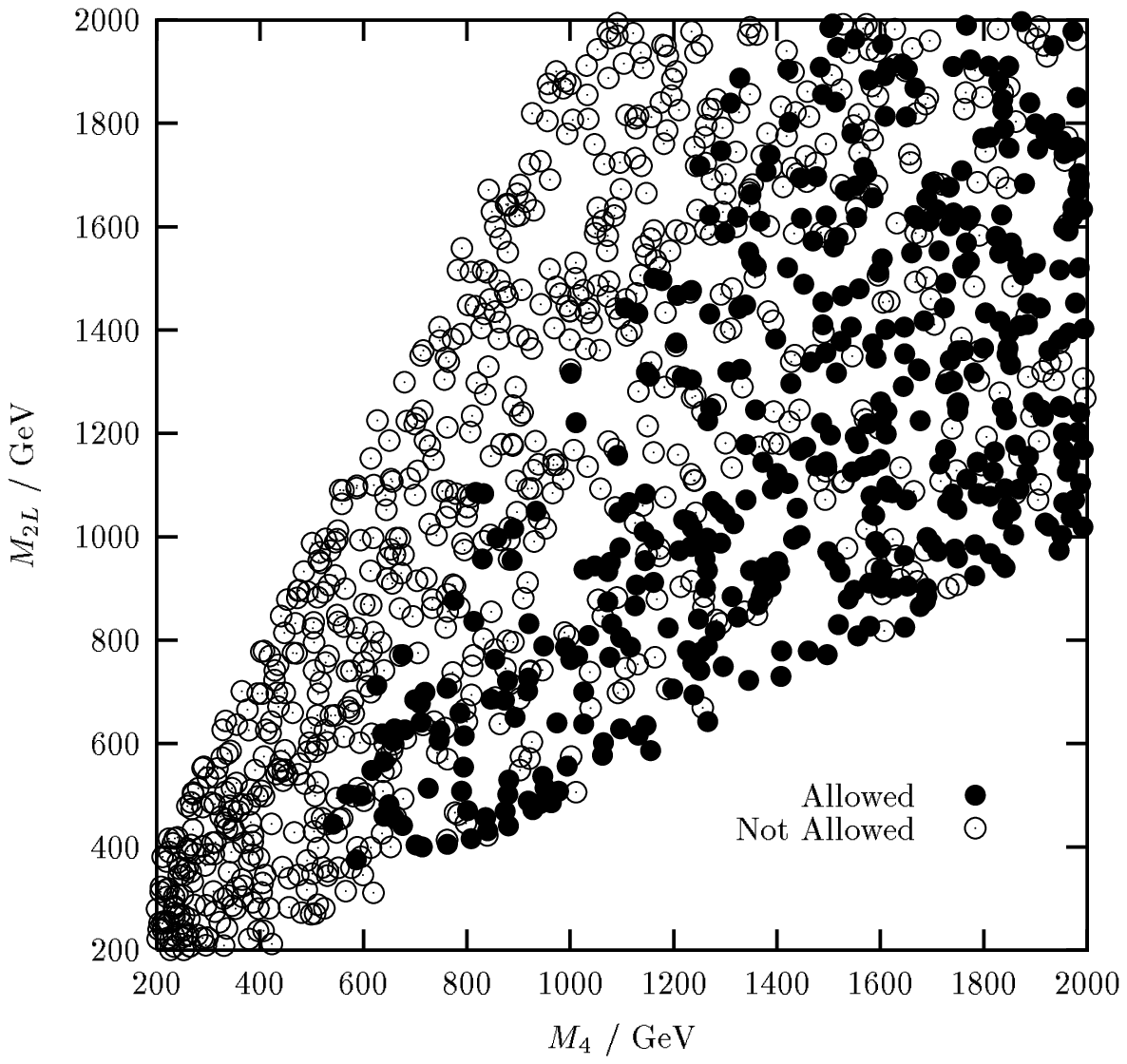}}

{\narrower\narrower\footnotesize\noindent
{FIG.~\FigM2LVsM4.}
The $SU(2)_L$ gaugino mass $M_{2L}$ against
the $SU(4)$ gaugino mass $M_4$.\par
\bigskip}}}

\vbox{
\noindent
\hfil
\vbox{
\epsfxsize=\figsize
\epsffile[130 380 510 735]{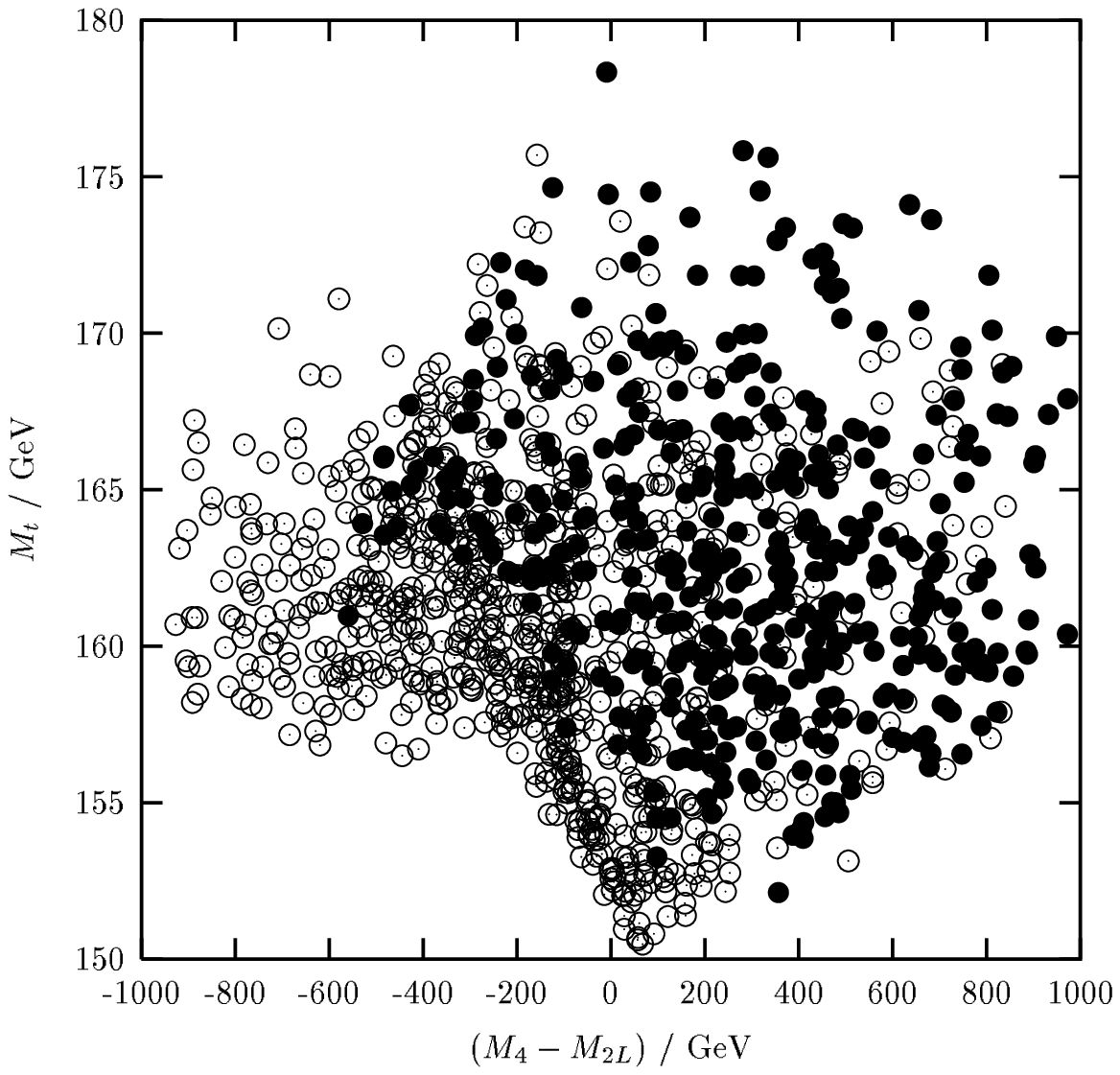}}

{\narrower\narrower\footnotesize\noindent
{FIG.~\FigMtopVsM4M2L.}
Top mass prediction $M_t$ against the difference
between the $SU(4)$ and $SU(2)_L$ gaugino masses $M_4-M_{2L}$.
The stronger concentration of black circles on the
right-half part of this figure indicates that EWSB favours $M_4$ to be
bigger than $M_{2L}$.\par
\bigskip}}

\vbox{
\noindent
\hfil
\vbox{
\epsfxsize=\figsize
\epsffile[130 380 510 735]{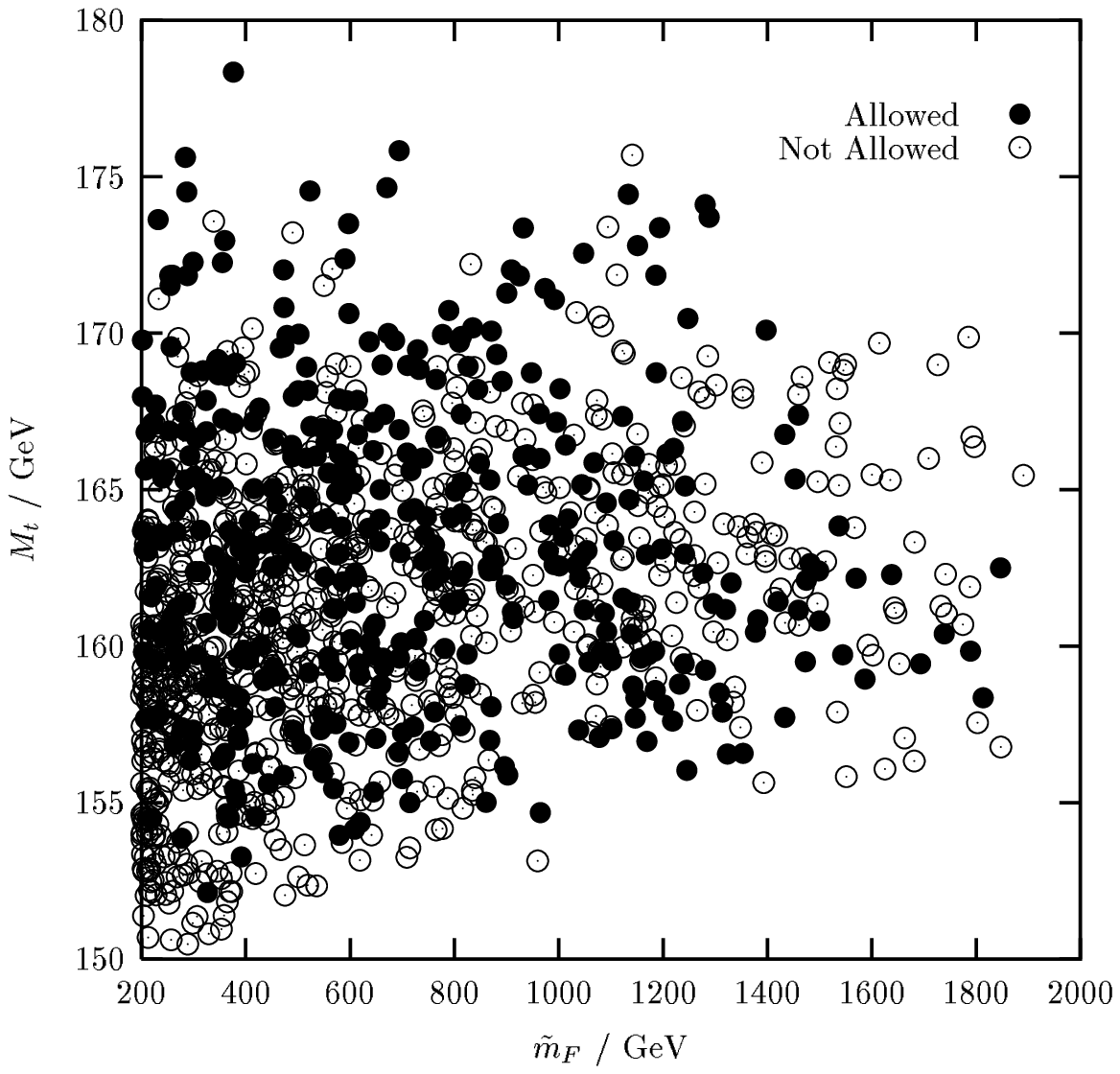}}

{\narrower\narrower\footnotesize\noindent
{FIG.~\FigMtopVsMFL.}
The top mass prediction $M_t$ against
the left-handed SUSY scalar mass $\tilde m_F$.
The lower $\tilde m_F$ the bigger the spread in $M_t$ ($\alpha_s = 0.120$.)
\par\bigskip}}

\vbox{
\noindent
\hfil
\vbox{
\epsfxsize=\figsize
\epsffile[130 380 510 735]{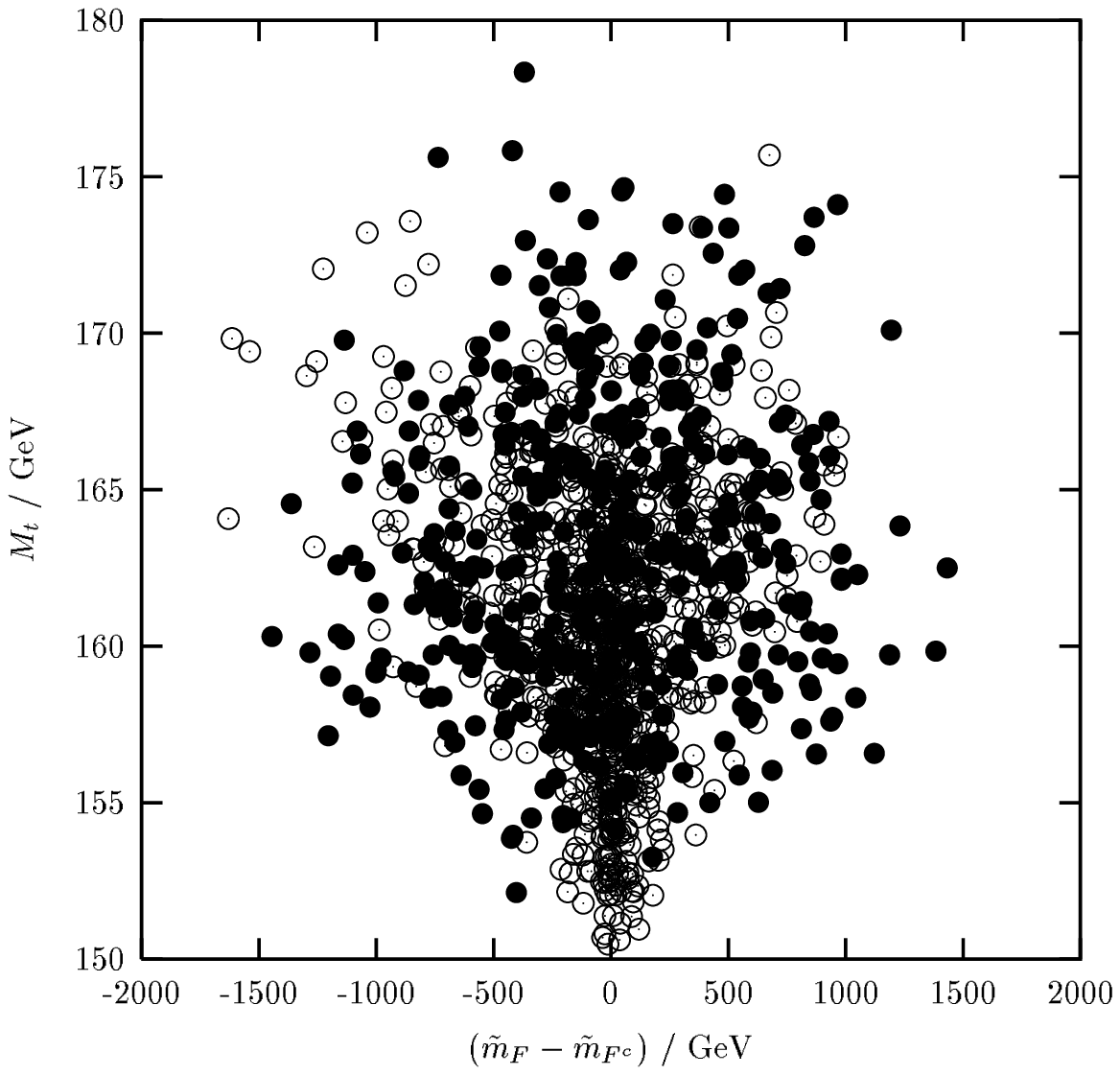}}

{\narrower\narrower\footnotesize\noindent
{FIG.~\FigMtopVsMFLR.}
Top mass prediction $M_t$ against the difference between
the left/right SUSY scalar masses $\tilde m_F - \tilde m_{F^c}$.
This symmetric plot shows that the top mass does not favour an
hierarchy between
$\tilde m_F$ and $\tilde m_{F^c}$ ($\alpha_s = 0.120$.)
\par\bigskip}}

\vbox{
\noindent
\hfil
\vbox{
\epsfxsize=\figsize
\epsffile[130 380 510 735]{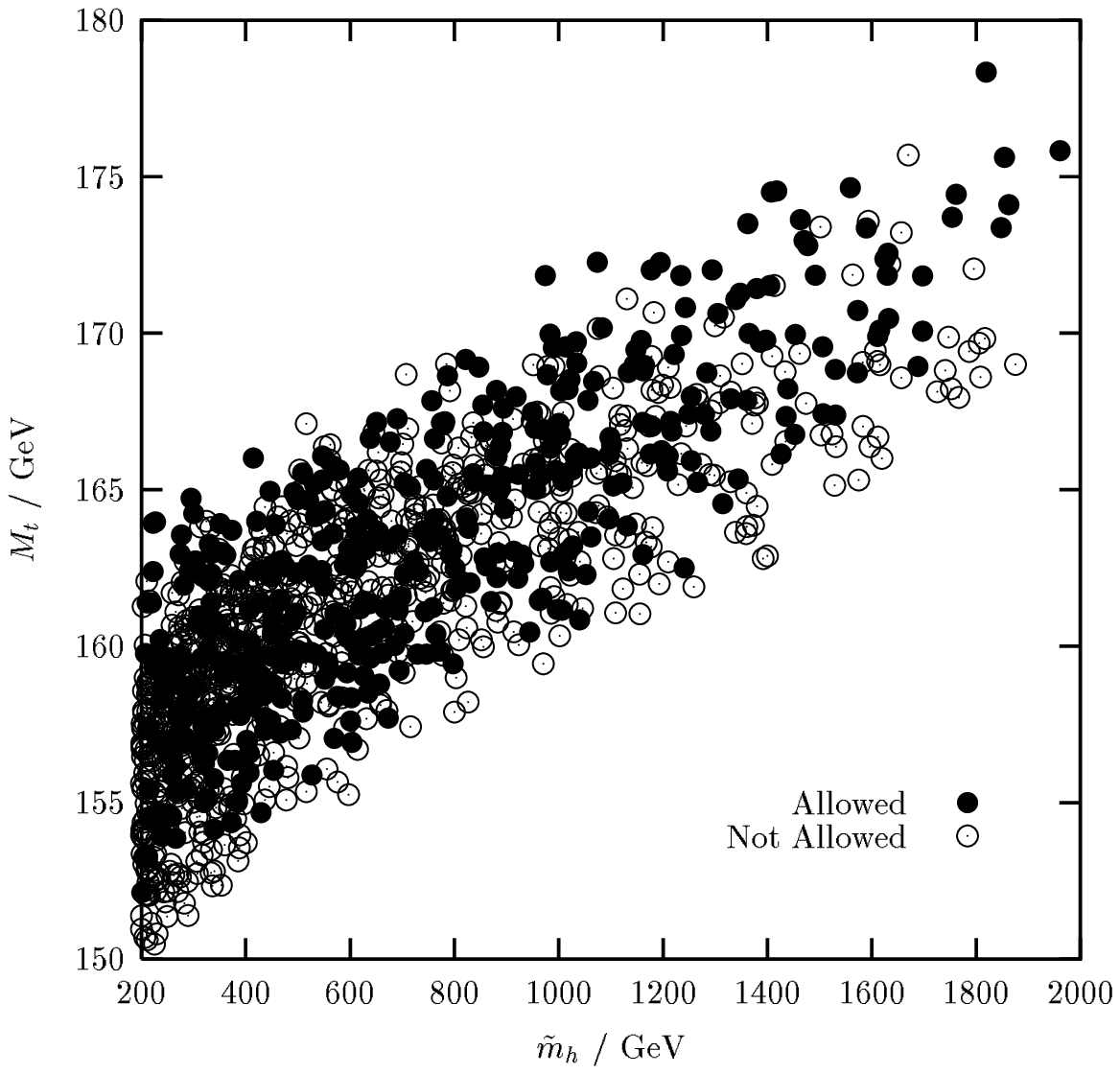}}

{\narrower\narrower\footnotesize\noindent
{FIG.~\FigMtopVsMh.}
Top mass prediction $M_t$ against the soft Higgs mass $\tilde m_h$.
We can see that a large top mass is favoured by a large Higgs
parameter ${\tilde m}_h$ ($\alpha_s = 0.120$.)
\par\bigskip}}

\vbox{
\noindent
\hfil
\vbox{
\epsfxsize=\figsize
\epsffile[130 380 510 735]{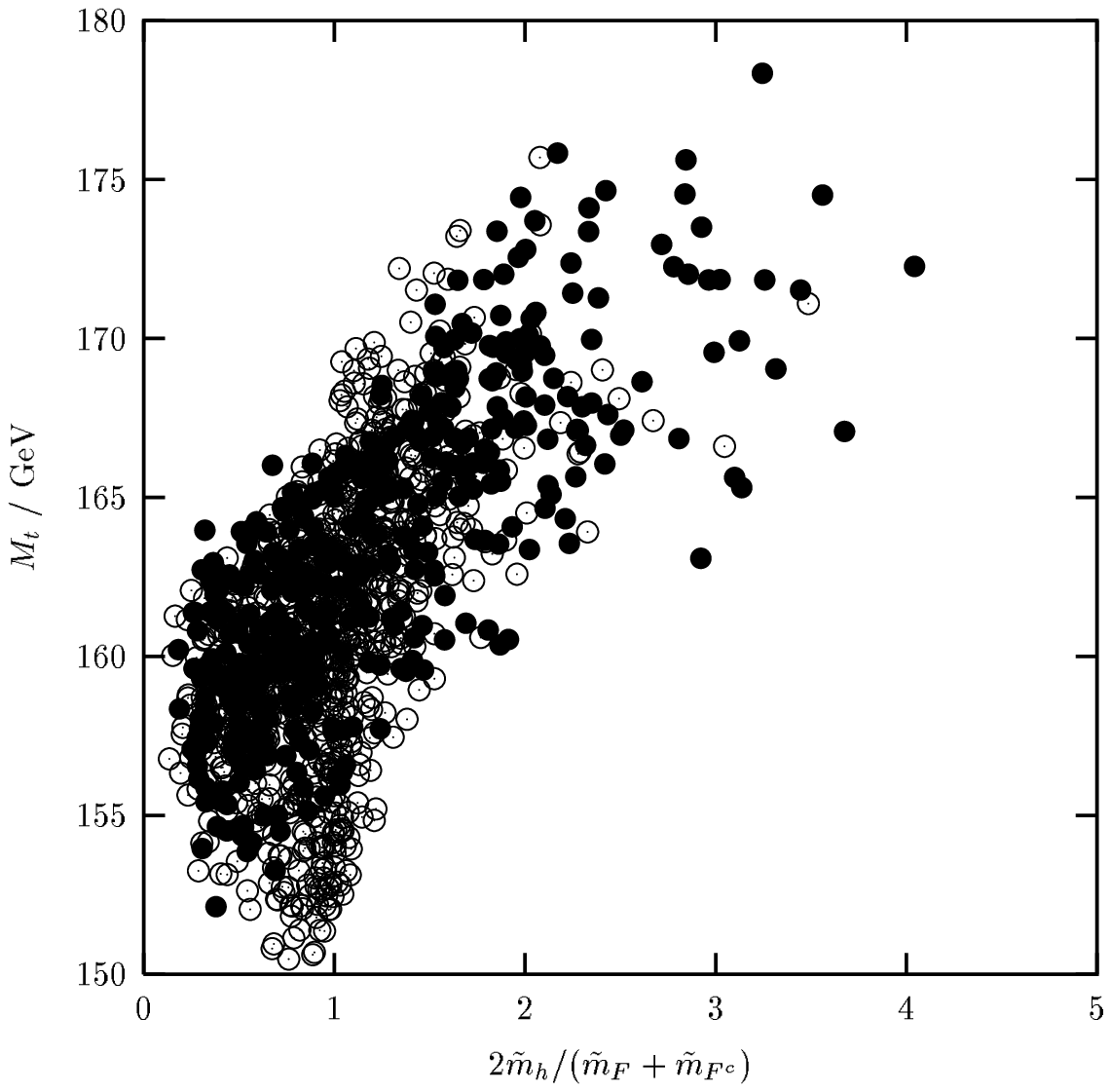}}

{\narrower\narrower\footnotesize\noindent
{FIG.~\FigMtopVsMhMF.}
Top mass prediction $M_t$ against the ratio between
the soft Higgs mass $\tilde m_h$ and
the average of the SUSY scalar masses
${1\over 2}({\tilde m_F}+\tilde m_{F^c})$.
The bigger this ratio the bigger $M_t$ is likely to be.
\par\bigskip}}

The Fig.~\FigMtopVsMFL\ shows that, generally, as $\tilde m_F$ decreases $M_t$
predictions are allowed to increase. A similar dependence of $M_t$ on
$\tilde m_{F^c}$ was found. Indeed, when $M_t$ is plotted against the
difference $\Delta \tilde m_F = \tilde m_F-\tilde m_{F^c}$ such as
in Fig.~\FigMtopVsMFLR\ a symmetric left-right graph emerges around the $x$-axis.
On the other hand, from Fig.~\FigMtopVsMh\
we can conclude that $M_t$ increases with
increasing soft Higgs mass ${\tilde m}_h$. These figures suggest that the
top mass may be increased by increasing the splitting between the
sfermions and Higgs soft mass. We therefore find interesting to
plot in Fig.~\FigMtopVsMhMF\ $M_t$ against
the ratio ${\tilde m}_h / \langle \tilde m_F \rangle$
where $\langle \tilde m_F \rangle = (\tilde m_F+\tilde m_{F^c})/2$ is
the average sfermion mass. Our suspictions are confirmed.

After combining the top mass dependencies on the input parameters
suggested by the previous figures with numerous case-by-case analyses
of ``outputs'' from our numerical model we arrived at the following
conclusions. The top mass prediction is strongly dependent on $M_4$.
\footnote{The reason is because, at the unification scale,
the gluino masses are set to be $M_3(M_X)=M_4(M_X)$. Thus the sensitiviness
of the results with $M_4$ is in fact a sensitiviness
to the masses of the coloured sparticles.}
Once $M_4$ is fixed, decreasing $\tilde m_F$ and $\tilde m_{F^c}$
increases both $M_t$ and the Higgs potential stability parameter
(that is positive for a ``bounded-from-below'' Higgs potential) :~
\footnote{In the large $\tan\beta$ limit $S^2$ is given by
$S^2 \sim m_1^2-m_2^2-M^2_Z \sim m_3^2 / \tan\beta$.}
\begin{equation}
S^2 = \mu^2_2+\mu^2_1-2|m_3^2| > 0
\end{equation}
However $\tilde m_{F,F^c}$ cannot be arbitrarily small since if
$\tilde m_{F,F^c}$ are very small the sleptons became too light.
Increasing $M_{2L}$ and ${\tilde m}_h$ increases the top mass, however they
also affect $S$. We found that the most efficient mechanism to
increase $M_t$ relied on decreasing $M_{2L}$ (which decreased $M_t$
moderately but increased $S$ substantially) and increasing ${\tilde m}_h$
(which increased $M_t$ significantly and decreased $S$ moderately).
For the sake of illustration, we found that, by pushing the EWSB
conditions to the extreme ($S \sim 0$, \ie tuning $m_3$),
it was possible to get $M_t \sim 175$ GeV only
if $M_4(M_P) \bigsim 800$ GeV. In this case we got $M_{2L}=500$ GeV,
$\tilde m_F=350$ GeV, $\tilde m_{F^c}=450$ GeV and ${\tilde m}_h=1000$ GeV.
\footnote{The corresponding values at $M_X$ are $M_4=653$ GeV,
$M_{2L}=408$ GeV, $\tilde m_F=456$ GeV, $\tilde m_{F^c}=567$ GeV,
and ${\tilde m}_h=972$ GeV.}
Naturally, for a larger $M_4$ mass, $S$ increases and an acceptable top mass
is easier to obtain with less tuning in $m_3$.

The main conclusion is that in the 422 model with non-universal soft
masses a top mass of around $M_t \sim 175$ GeV is only possible to
obtain in the context of a large gluino mass
$m_{\tilde g}(M_Z) \sim 1520$ GeV and
large $\tanb \sim 50$, by implicitly tuning the EWSB conditions,
and by choosing squark/slepton masses considerably smaller than the
soft Higgs mass $\tilde m_F, \tilde m_{F^c} < {\tilde m}_h$.

\newpage

\SUBSECTION{D. D-term contributions}

The symmetry breaking of the Pati-Salam to the SM gauge group given in
Eq.~\refeqn{422:321} reduces the group rank from five to
four. Although the GUT symmetry is broken at a very high energy it
nevertheless has important consequences to low energy TeV
phenomenology via $D$-term contributions to the scalar masses
\cite{BaDiFeTa,KaMuYa,FaHaKeNa}.
In the 422 model the GUT boundary conditions for the
scalar masses are \cite{KaMuYa}~:
\footnote{See appendix E for a detailed derivation of the
$D$-terms in the 422 model. Note that in the limit of unified gauge
couplings the $D$-terms give identical corrections to the fields
in the $\underline{16}$ 
dimensional representation of $SO(10)$.}
\begin{eqnarray}
\tilde m^2_q &=& \tilde m^2_F+g^2_4 D^2 \label{mQgD}\\
\tilde m^2_{u^c} &=& \tilde m^2_{F^c}-(g^2_4-2 g^2_{2R}) D^2 \\
\tilde m^2_{d^c} &=& \tilde m^2_{F^c}-(g^2_4+2 g^2_{2R}) D^2
\label {md-3gD} \\
\tilde m^2_l &=& \tilde m^2_F - 3 g^2_4 D^2 \label{ML-3gD}\\
\tilde m^2_{e^c} &=& \tilde m^2_{F^c} + (3 g^2_4 - 2 g^2_{2R}) D^2
\label{me+gD} \\
\tilde m^2_\nu &=& \tilde m^2_{F^c} + (3 g^2_4 + 2 g^2_{2R} ) D^2 \\
\tilde m^2_2 &=& {\tilde m}^2_h - 2 g^2_{2R} D^2 \\
\tilde m^2_1 &=& {\tilde m}^2_h + 2 g^2_{2R} D^2 \label{m1gD}
\end{eqnarray}
The $D$-term corrections, If $D$ is sufficiently large, are important
to consider because, firstly they leave an imprint in the scalar
masses of the charges carried by the broken GUT generator
(these charges determine the coefficients of the $g^2$ terms above),
therefore the analysis of the sparticle spectra \cite{FaHaKeNa} might reveal the
nature of the GUT symmetry breaking pattern; Secondly, they split the
soft Higgs masses by $m^2_2 - m^2_1 \sim -4 g^2_X D^2$, which for
positive $D^2$, makes radiative EWSB much easier to occur.
Indeed, we found that
once $D \bigsim 150-200$ GeV then EWSB no longer requires the large
gluino/squark masses characteristic of models with USBM.

\vbox{
\noindent
\hfil
\vbox{
\epsfxsize=\figsize
\epsffile[130 380 510 735]{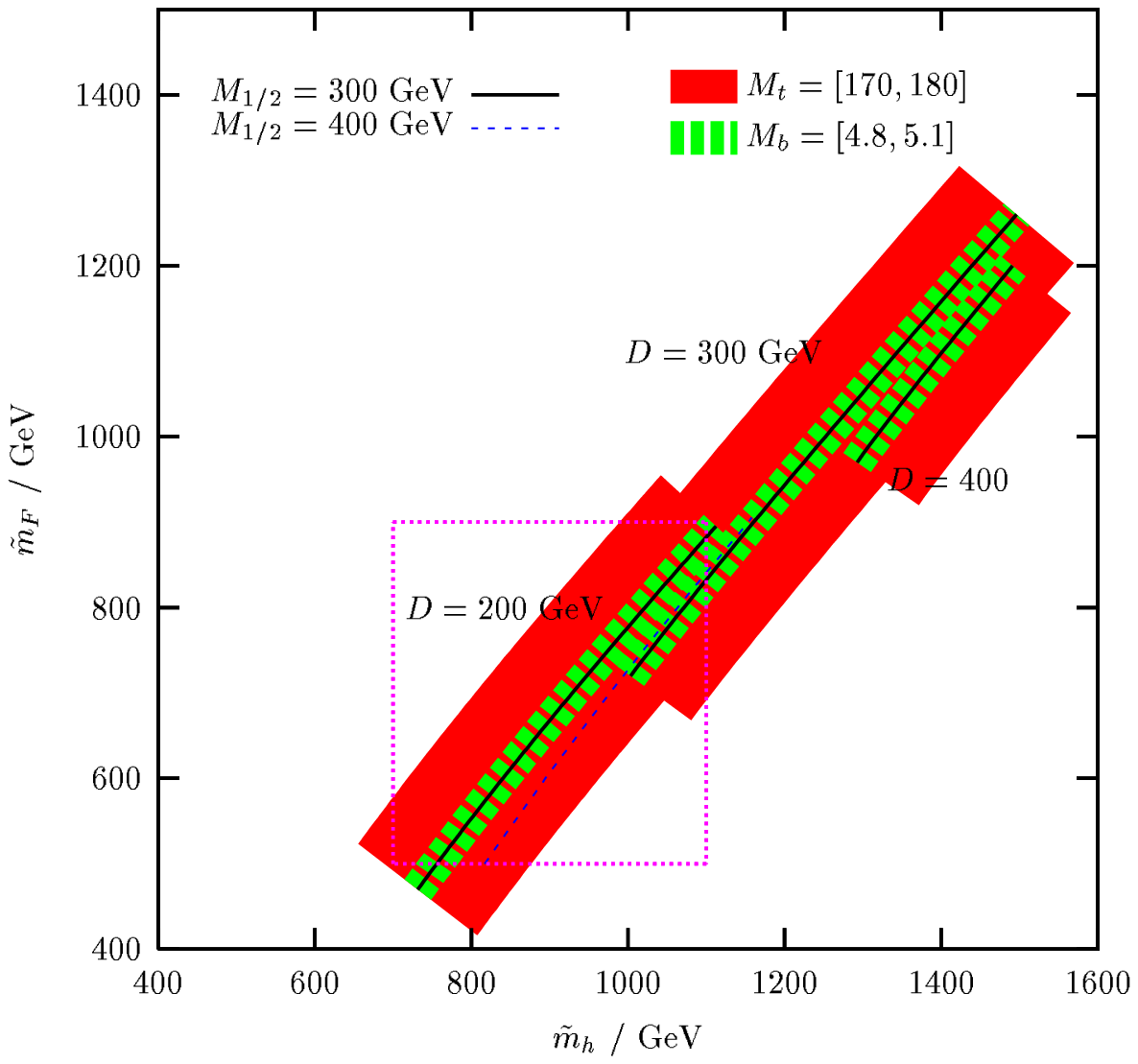}}

{\narrower\narrower\footnotesize\noindent
{FIG.~\FigMFVsMh.}
Correlation between the sfermion masses
${\tilde m}_F={\tilde m}_{F^c}$ and the Higgs soft
mass ${\tilde m}_h$ required to
obtain a top mass around 175 GeV ($\alpha_s = 0.120$.)
Three illustrative choices for the value of the $D$-term are taken.
The effect of experimental uncertainties in the top and bottom quark
masses is indicated by the toned areas.
The box around the $D=200 {\rm\ GeV}$ label indicates the scanned
region used to generate Figs.~\FigMtVsMhScan--\FigMtVsMaScan\
(see main text for details.) 
\par}}

In Fig.~\FigMFVsMh\ we summarize how an acceptable top mass prediction can be
achieved in the 422 model in the $SO(10)$ limit of universal
gaugino masses $M_{1/2}$ and universal squark and slepton mass parameters
${\tilde m}_F={\tilde m}_{F^c}$ defined at the Planck scale, but
with $D$-term corrections arising at the symmetry breaking scale.
\footnote{Although we refer to this as the $SO(10)$ limit,
in fact the two theories differ in the region between the GUT scale
and the Planck scale.} 
The three solid
lines establish the correlation between the sfermion masses
$\tilde m_F = \tilde m_{F^c}$ (plotted in the $y$-axis) and the Higgs
soft mass ${\tilde m}_h$ (plotted in the $x$-axis) required to obtain
a top mass $M_t = 175$ GeV, for $D=200,300,400$ GeV and
$M_4=M_{2R}=M_{2L}=300$ GeV.
For the $D=200$ GeV line, EWSB failed to occur beyond the point where
the line is discontinued in her way upward.
\footnote{The $D=300,400$ GeV lines continue beyond
the border of the figure.}
In all cases the lines were cut at the bottom edge
at points where the left-handed stau mass became smaller than 100 GeV.

The general behaviour in Fig.~\FigMFVsMh\ with increasing $D$-term values
--- lines shifting to the top-right corner --- results from the decrease
of $\tilde m^2_L \sim \tilde m^2_F - 3 g^2_X D^2$ in Eq.~\refeqn{ML-3gD} which
demands an increasing $\tilde m^2_F$ parameter. The broader darker
areas around the solid lines correspond to the uncertainty in the top
mass $175 \pm 5$ GeV, while the thinner patches result from
varying the pole bottom mass $M_b$ in the range 4.8--5.1 GeV.
We also analysed how the gaugino masses affect the solid lines.
For example, for $D=200$ GeV,
taking $M_{1/2}=400$ GeV generates the dashed line.
If $M_{2R}$ or $M_{2L}$ are taken different from $M_4$
the prediction for the top mass is very similar to the one with
universal gaugino masses with $M_{1/2}=M_4$.

In summary, Fig.~\FigMFVsMh\ shows that if $D \bigsim 200$ GeV then a
successful prediction for $M_t$ can be achieved, even for small
gluino/squark masses --- no fine tuning in EWSB conditions --- as long
as the soft Higgs mass ${\tilde m}_h$ is bigger than the sfermion masses
$\tilde m_F$, $\tilde m_{F^c}$. As an example, setting
$D=200$ GeV and $M_{1/2}=300$ GeV we found that $M_t \sim 175$ GeV could be obtained
by fixing $\tilde m_F=\tilde m_{F^c}=500$ GeV and ${\tilde m}_h=750$ GeV.
\footnote{
The corresponding values at $M_X$ are $M_{1/2}=247$ GeV,
$\tilde m_{F,F^c} = 474 {\rm\ GeV}$ and ${\tilde m}_h = 691 {\rm\ GeV}$.
After the $D$-term corrections the scalar masses are :
$\tilde m_q=\tilde m_{u^c}=\tilde m_{e^c}=496$ GeV,
$\tilde m_l=\tilde m_{d^c}=404$ GeV,
$\tilde m_{\nu^c}=572$ GeV (for the third family) and
$m_2 = 660$ GeV, $m_1=720$ GeV.}
This input lead to the following low energy predictions :
\tanb=52, $\lambda_X=0.70$ a gluino mass $m_{\tilde g}=618$ GeV,
a lightest neutralino mass $m_{\chi^0_1}=88$ GeV
(bino like, but with substantial higgsino component),
a lightest chargino mass $m_{\chi^-_1}=125$ GeV (higgsino like)
and masses for the lightest sfermions
$m_{\tilde\nu_\tau}=179$ GeV,
$m_{\tilde\tau_1}=189$ GeV and
$m_{\tilde b_2}=377$ GeV.
The lightest CP-even Higgs mass, computed using the one-loop
expressions of Ref.~\cite{ElRiZw} (that include the stop/sbottom corrections
only) was found to be $m_h=114 {\rm\ GeV}$ (note that this value should be
read with some caution, we estimate an error of
about 10 GeV in $m_h$.)

\newpage

In the last part of this section we present a series of figures
that show in detail how an acceptable top mass prediction can
be achieved and how it is correlated with the sparticle spectrum.
For the sake of illustration we take, as in the previous sections
$\alpha_s = 0.120$, $M_b=4.8 {\rm\ GeV}$ and the following
input at the Planck scale~:
\medskip
\begin{equation}
\hskip 26mm
\matrix{
\hfill M_{1/2} = 300 {\rm\ GeV}        & & &
       D  = 200 {\rm\ GeV}      \hfill \cr
& & & \cr
{\tilde m}_h = 700-1100 {\rm\ GeV} \hfill & & &
{\tilde m}_F = {\tilde m}_{F^c} = 500,550,..,900 {\rm\ GeV} \hfill
}
\label{RANGES422D}
\end{equation}
\medskip
These values correspond to a scan of the
${\tilde m}_h$--$\tilde m_{F,F^c}$ parameter space that is indicated
in Fig.~\FigMFVsMh\ with a box.

In Fig.~\FigMtVsMhScan\ we plot the pole top mass prediction
$M_t$ against ${\tilde m}_h$
for several choices of the sfermion masses
$\tilde m_{F,F^c}=500,550,..,900$ GeV.
We observe that $M_t$ increases with increasing ${\tilde m}_h$
and decreasing $\tilde m_{F,F^c}$.
The reason for such dependence is directly related with the
value of the Higgs mixing parameter $\mu$.

In Fig.~\FigMtVsMuScan\ we show how the top mass
prediction $M_t$ correlates with $\mu$.
In this plot each line
corresponds to a fixed $\tilde m_{F,F^c}$ mass
(labeled by the number beside it)
and along it ${\tilde m}_h$ is varying in the range
indicated by Eq.~\refeqn{RANGES422D}.
We see that decreasing $\tilde m_{F,F^c}$ shifts the lines to the
right of the graph thus making $\mu$ less negative.
Furthermore, as ${\tilde m}_h$ increases,
from the bottom to the top of the graph, $|\mu|$ decreases.
These dependencies lead to a small
value for $\mu$ (at the top-right corner of this graph)
which in turn lead to small SUSY bottom corrections that raise
the top mass prediction to an acceptable value.

In Fig.~\FigMtVsGLNScan\ we show the correlation between the top mass
$M_t$ and the gluino mass $m_{\tilde g}$ predictions.
We see that, for the choice of
$M_{1/2} = 300 {\rm\ GeV}$ in Eq.~\refeqn{RANGES422D},
if the top mass is in the 170--180 GeV range then the gluino mass is
in the 605--620 GeV range.
It is worth stressing that the 422 model
with $D$-terms offers the possibility of predicting
light gluinos. This is exciting from the experimental point of view
and theoretically desirable since it reduces the
fine-tuning in the Higgs potential parameters.

\vbox{
\vbox{
\noindent
\hfil
\vbox{
\epsfxsize=\figsize
\epsffile[130 380 510 735]{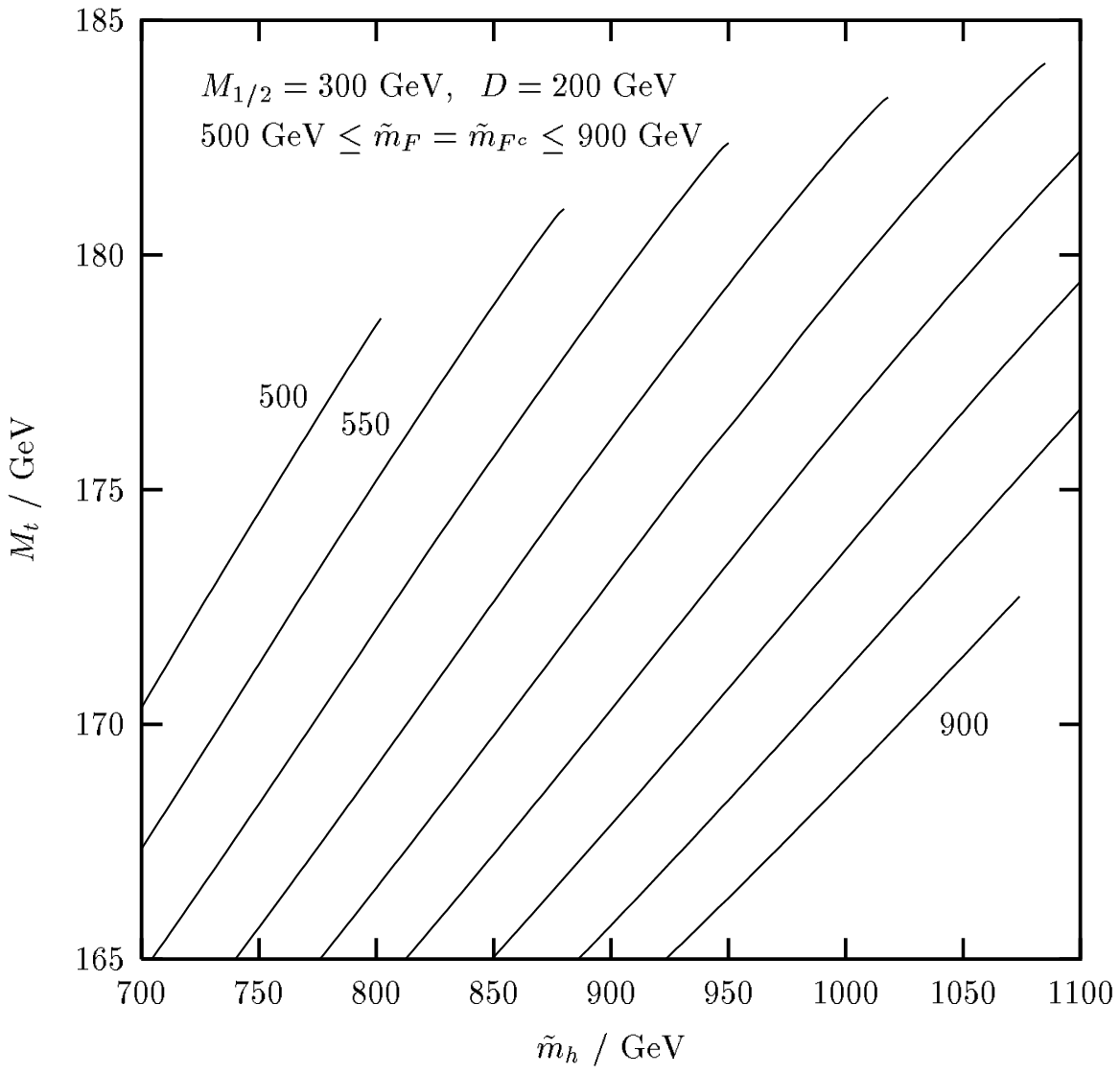}}

{\narrower\narrower\footnotesize\noindent
{FIG.~\FigMtVsMhScan}
Prediction for the top mass $M_t$ against
the soft mass for the unified Higgs bosons ${\tilde m}_h$
in Eq.~\refeqn{SMHiggs}.
The numerical labels, for each line, indicate the value for the
soft masses of the scalar fields $\tilde m_F=\tilde m_{F^c}$
used to generate the corresponding line ($\alpha_s = 0.120$.)
\par}}

\vbox{
\noindent
\hfil
\vbox{
\epsfxsize=\figsize
\epsffile[130 380 510 735]{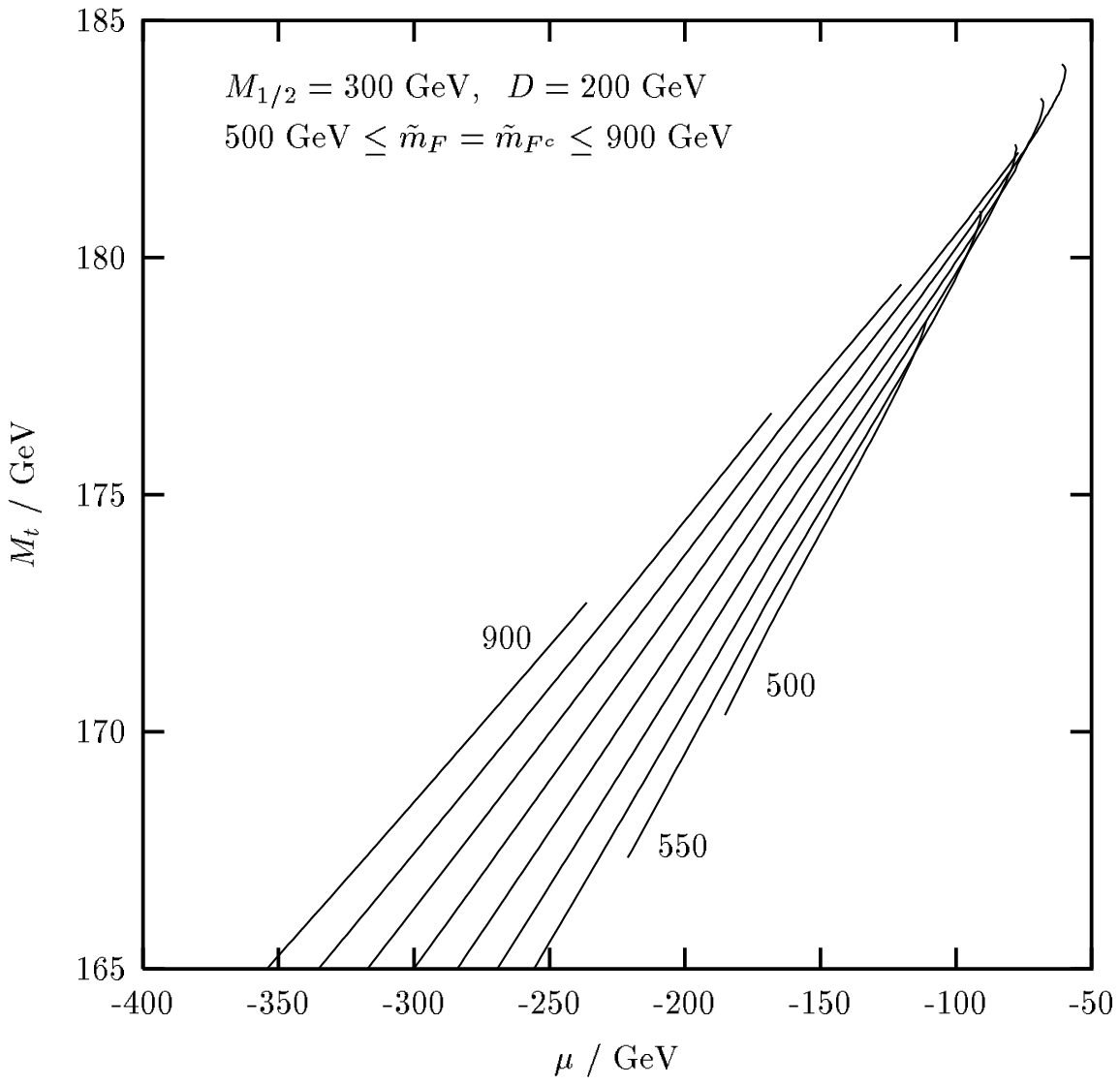}}

{\narrower\narrower\footnotesize\noindent
{FIG.~\FigMtVsMuScan}
Prediction for the top mass $M_t$ against
the Higgs mixing parameter $\mu$.
Each line corresponds to a fixed choice for the
soft mass of the scalar fields $\tilde m_F=\tilde m_{F^c}$
and along it the soft mass for the unified Higgs bosons ${\tilde m}_h$
is varying (increasing from the bottom to the top of the plot.)
We observe that $M_t$ increases with decreasing $|\mu|$ ($\alpha_s = 0.120$.)
\par}}}

\vbox{
\vbox{
\noindent
\hfil
\vbox{
\epsfxsize=\figsize
\epsffile[130 380 510 735]{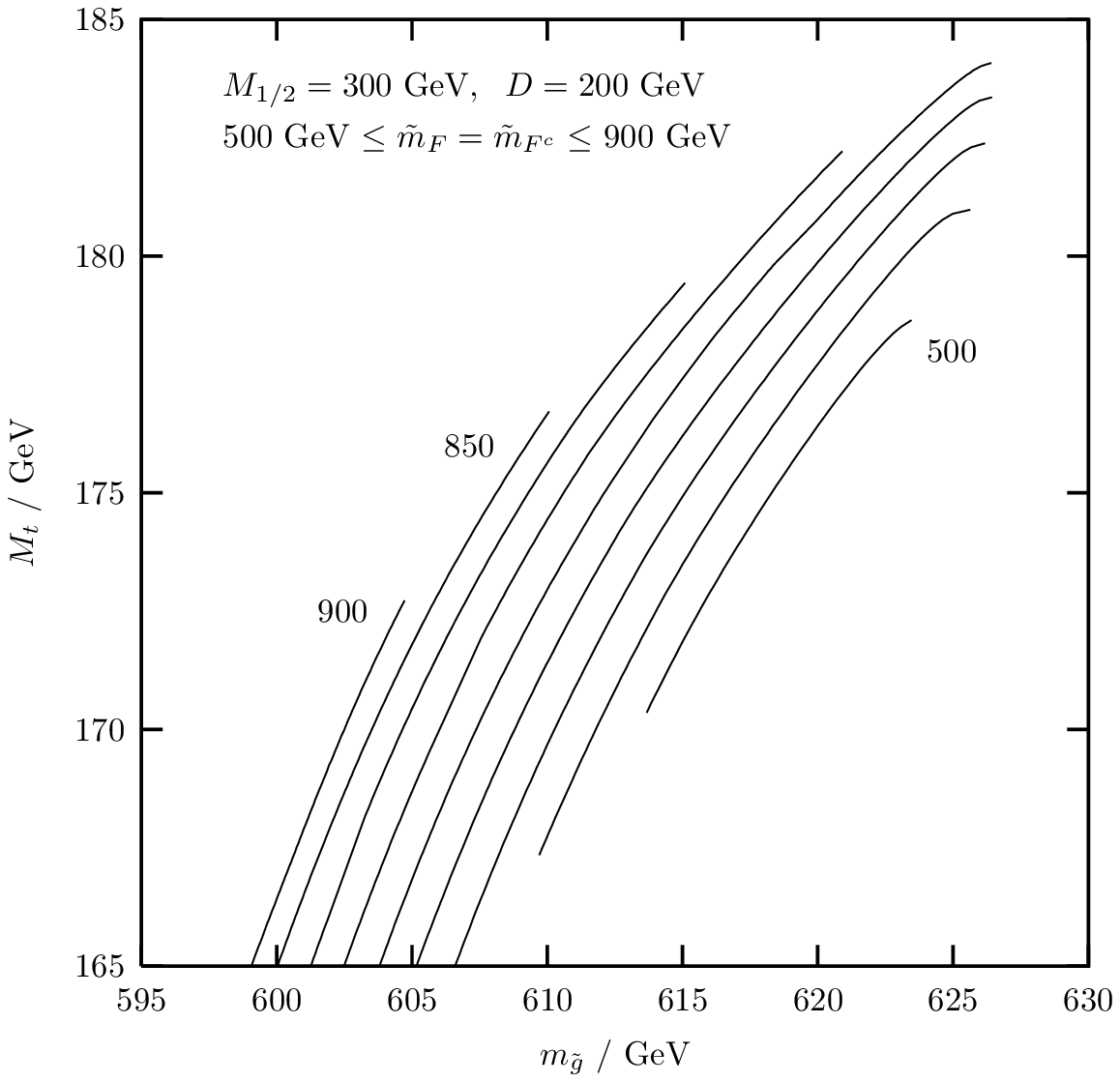}}

{\narrower\narrower\footnotesize\noindent
{FIG.~\FigMtVsGLNScan}
Correlation between the top mass prediction $M_t$ and
the gluino mass prediction $m_{\tilde g}$.
Each line corresponds to a fixed choice for the
soft mass of the scalar fields $\tilde m_F=\tilde m_{F^c}$
(labeled with a number)
and along it the soft mass for the unified Higgs bosons ${\tilde m}_h$
is varying
(increasing from the bottom to the top of the plot.)
\par}}

\vbox{
\noindent
\hfil
\vbox{
\epsfxsize=\figsize
\epsffile[130 380 510 735]{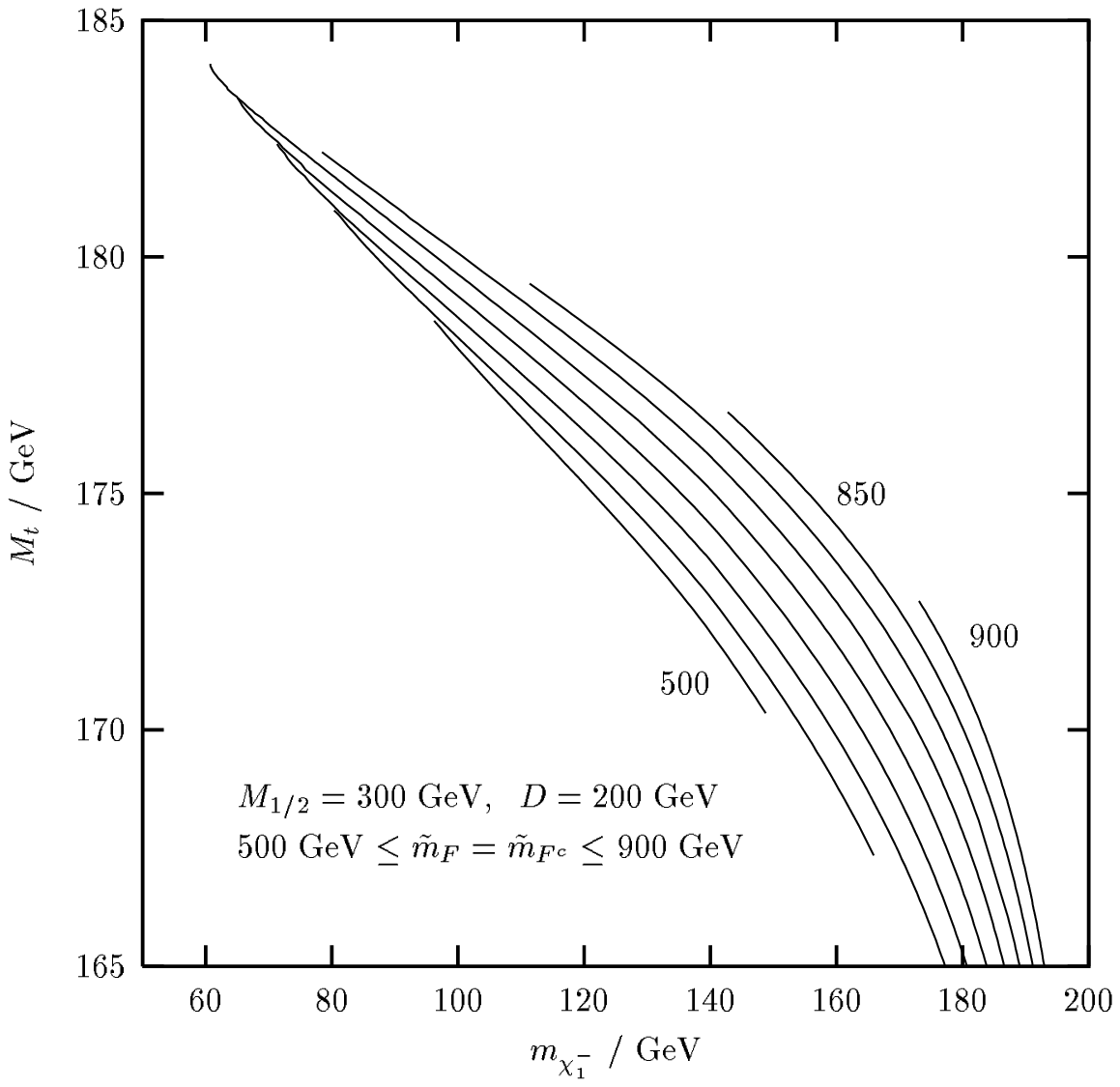}}

{\narrower\narrower\footnotesize\noindent
{FIG.~\FigMtVsMCHScan}
Correlation between the top mass prediction $M_t$ and
the lightest chargino mass prediction $m_{\chi^-_1}$.
Each line corresponds to a fixed choice for the
soft mass of the scalar fields $\tilde m_F=\tilde m_{F^c}$
(labeled with a number)
and along it the soft mass for the unified Higgs bosons ${\tilde m}_h$
is varying
(increasing from the bottom to the top of the plot.)
\par}}}

In Fig.~\FigMtVsMCHScan\ we plot the top mass $M_t$ against
the lightest chargino mass $m_{\chi^-_1}$ prediction.
In this model ${\chi^-_1}$ is roughly the charged higgsino which
is lighter than the charged wino because $\mu < m_{\tilde W}$ at low energy.
From this graph we read that $80 {\rm\ GeV} < m_{\chi^-_1} < 180 {\rm\ GeV}$
which should be compared with experimental bound from LEP2 :
$m^{exp}_{\chi^-_1} > 89$ GeV \cite{CCasoEtal}.

In Fig.~\FigMtVsNTLScan\ we show the correlation between
the top mass $M_t$ and the lightest neutralino mass $m_{\chi^0_1}$ predictions.
We note that since $\mu$ can be comparable with
of the bino mass, $\chi^0_1$ can have a substantial higgsino component.
From this plot we read that $60 {\rm\ GeV} < m_{\chi^0_1} < 105 {\rm\ GeV}$
which should be compared with the experimental bound :
$m^{exp}_{\chi^0_1} > 40$ GeV \cite{CCasoEtal}.

In Fig.~\FigMtVsStauScan\ we plot $M_t$ against the lightest charged
slepton mass $m_{\tilde\tau_1}$ prediction.
It is interesting to observe that the $D$-term correction to the
left-handed charged sleptons in Eq.~\refeqn{ML-3gD} is negative
while for the right-handed charged sleptons in Eq.~\refeqn{me+gD}
it is positive.
Thus the lightest stau is not right-handed
($\tilde\tau_2$) but left-handed ($\tilde\tau_1$).
From this graph we see that the prediction for $m_{\tilde\tau_1}$
can vary significantly, roughly we find that only an
upper bound can be imposed $m_{\tilde\tau_1} < 600 {\rm\ GeV}$
(the experimental lower bound is
$m^{exp}_{\tilde\tau_1} > 81 {\rm\ GeV}$. \cite{CCasoEtal})

In Fig.~\FigMtVsSNeuScan\ the prediction for the top mass $M_t$ is
plotted against the lightest sneutrino mass $m_{\tilde\nu_\tau}$.
We note that for the choices of
$\tilde m_{F,F^c}= 500,550,600,650,700$ GeV the $\tilde\nu_\tau$
mass is driven to zero as ${\tilde m}_h$ increases from the bottom to the top
of the graph (thus, it is possible that $\tilde\nu_\tau$
could be the lightest SUSY particle (LSP).)
Comparing this figure with Fig.~\FigMtVsStauScan\
we see that the prediction for the $\tilde\tau_1$ and $\tilde\nu_\tau$
masses are similar. We find that the predicted upper bound for
$m_{\tilde\nu_\tau} < 650 {\rm\ GeV}$ is compatible with the
experimental lower bound
$m^{exp}_{\tilde\nu_\tau} > 43 {\rm\ GeV}$ \cite{CCasoEtal}.

\newpage

\vbox{
\vbox{
\noindent
\hfil
\vbox{
\epsfxsize=\figsize
\epsffile[130 380 510 735]{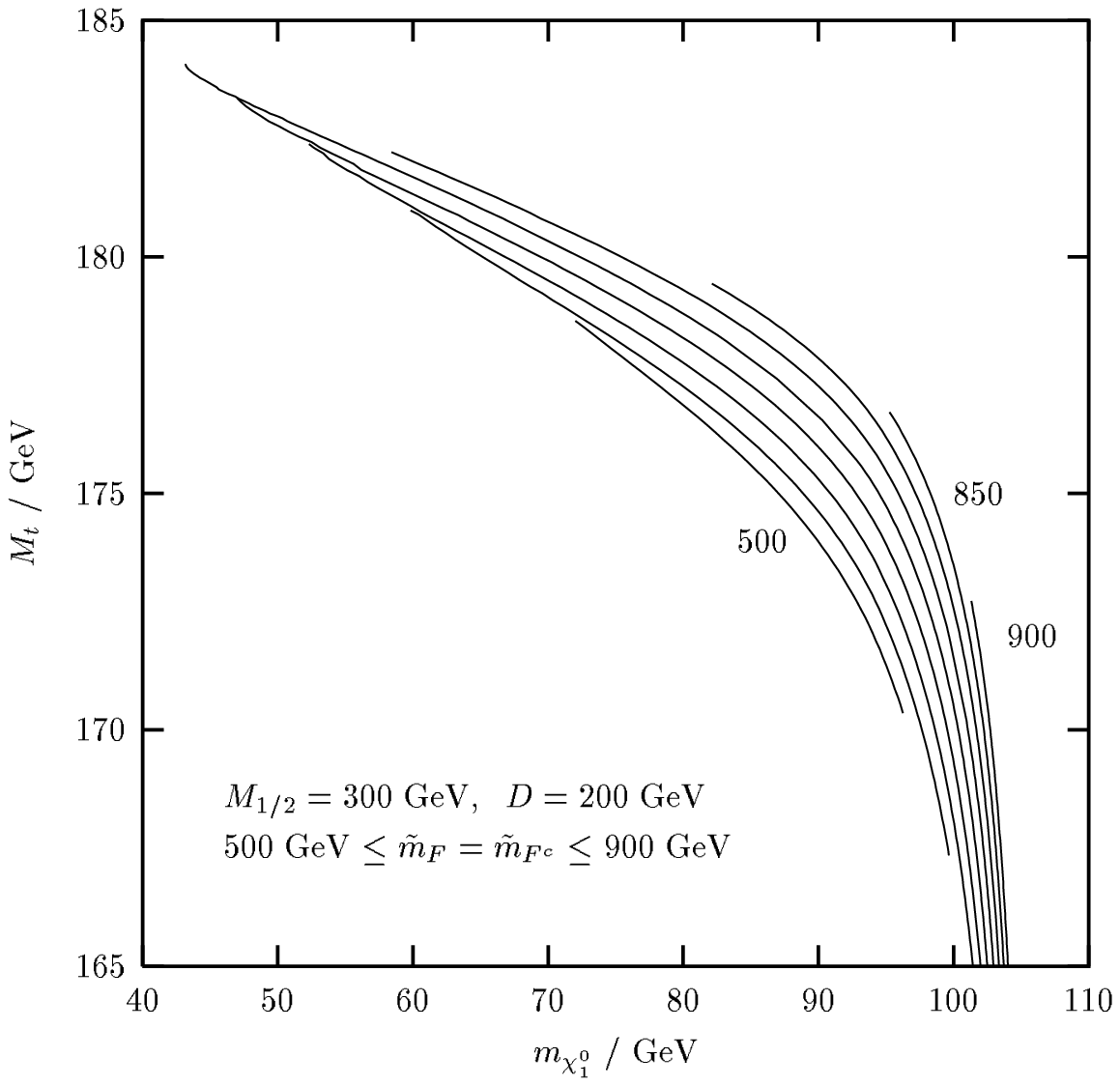}}

{\narrower\narrower\footnotesize\noindent
{FIG.~\FigMtVsNTLScan}
Correlation between the top mass prediction $M_t$ and
the lightest neutralino mass prediction $m_{\chi^0_1}$.
Each line corresponds to a fixed choice for the
soft mass of the scalar fields $\tilde m_F=\tilde m_{F^c}$
(labeled with a number)
and along it the soft mass for the unified Higgs bosons ${\tilde m}_h$
is varying
(increasing from the bottom to the top of the plot.)
\par}}

\vbox{
\noindent
\hfil
\vbox{
\epsfxsize=\figsize
\epsffile[130 380 510 735]{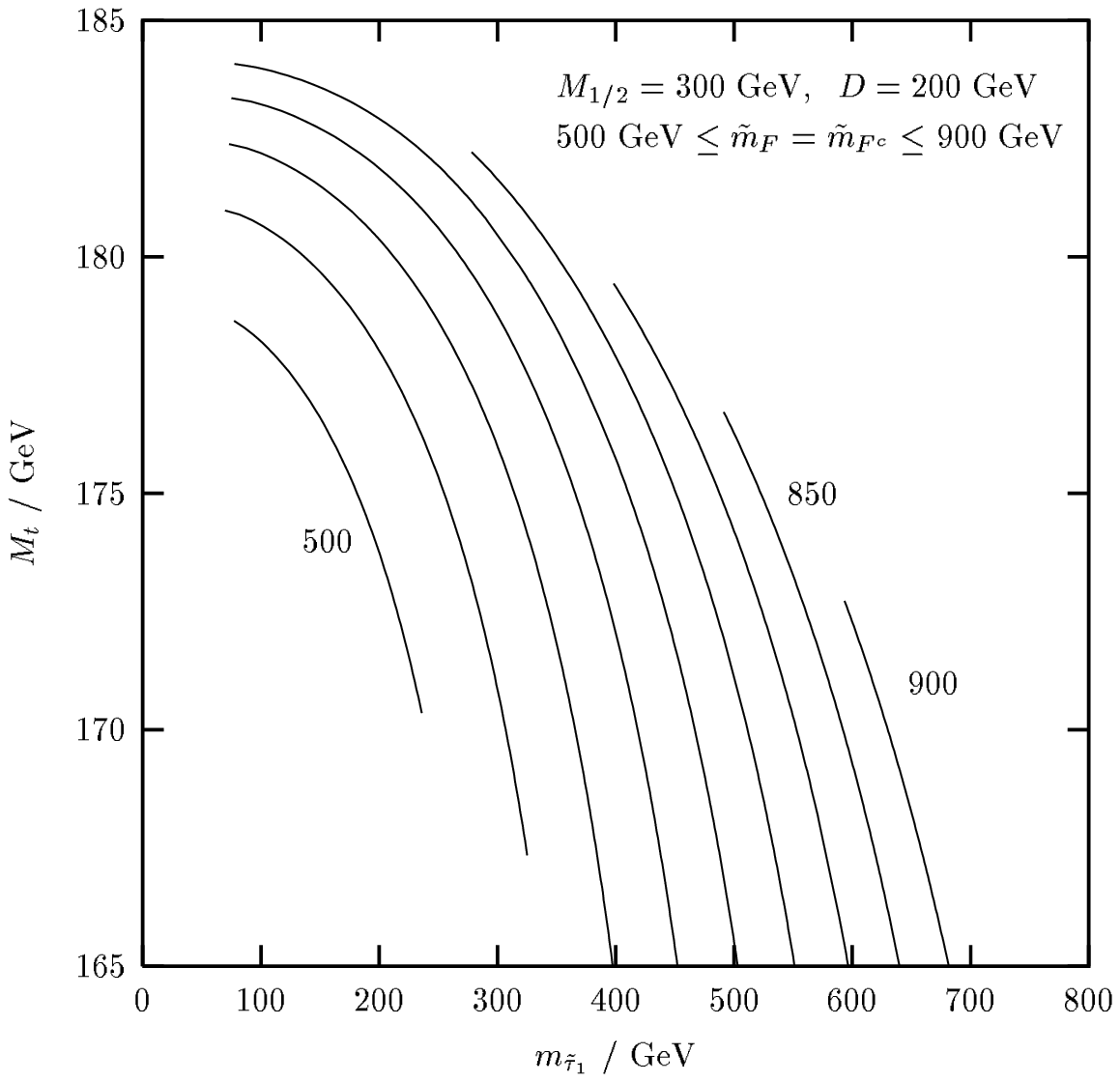}}

{\narrower\narrower\footnotesize\noindent
{FIG.~\FigMtVsStauScan}
Correlation between the top mass prediction $M_t$ and
the lightest charged slepton mass prediction $m_{\tilde\tau_1}$.
Each line corresponds to a fixed choice for the
soft mass of the scalar fields $\tilde m_F=\tilde m_{F^c}$
(labeled with a number)
and along it the soft mass for the unified Higgs bosons ${\tilde m}_h$
is varying
(increasing from the bottom to the top of the plot.)
\par}}}

\vbox{
\vbox{
\noindent
\hfil
\vbox{
\epsfxsize=\figsize
\epsffile[130 380 510 735]{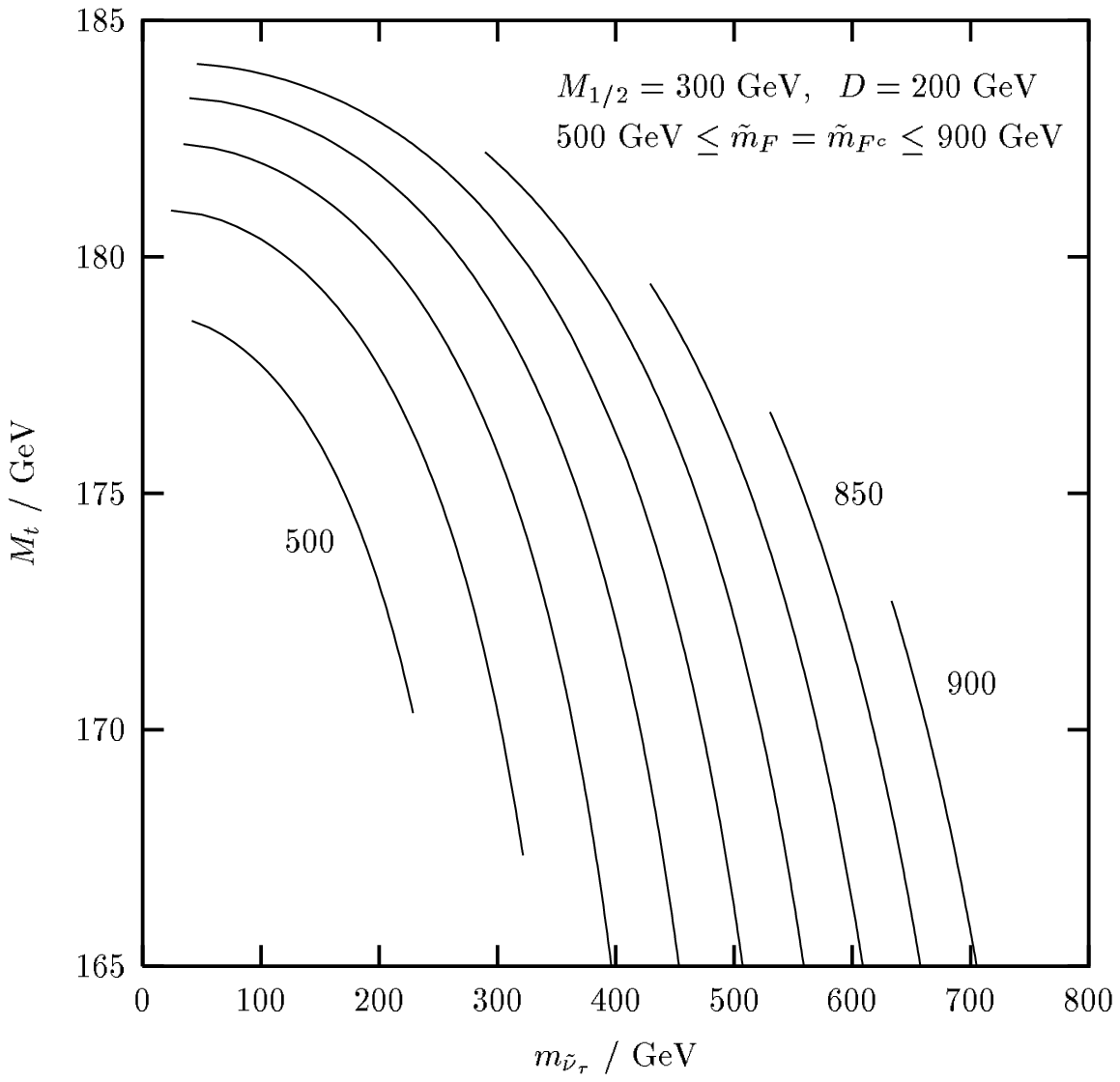}}

{\narrower\narrower\footnotesize\noindent
{FIG.~\FigMtVsSNeuScan}
Correlation between the top mass prediction $M_t$ and
the lightest sneutrino mass prediction $m_{\nu_3}$.
Each line corresponds to a fixed choice for the
soft mass of the scalar fields $\tilde m_F=\tilde m_{F^c}$
(labeled with a number)
and along it the soft mass for the unified Higgs bosons ${\tilde m}_h$
is varying
(increasing from the bottom to the top of the plot.)
\par}}

\vbox{
\noindent
\hfil
\vbox{
\epsfxsize=\figsize
\epsffile[130 380 510 735]{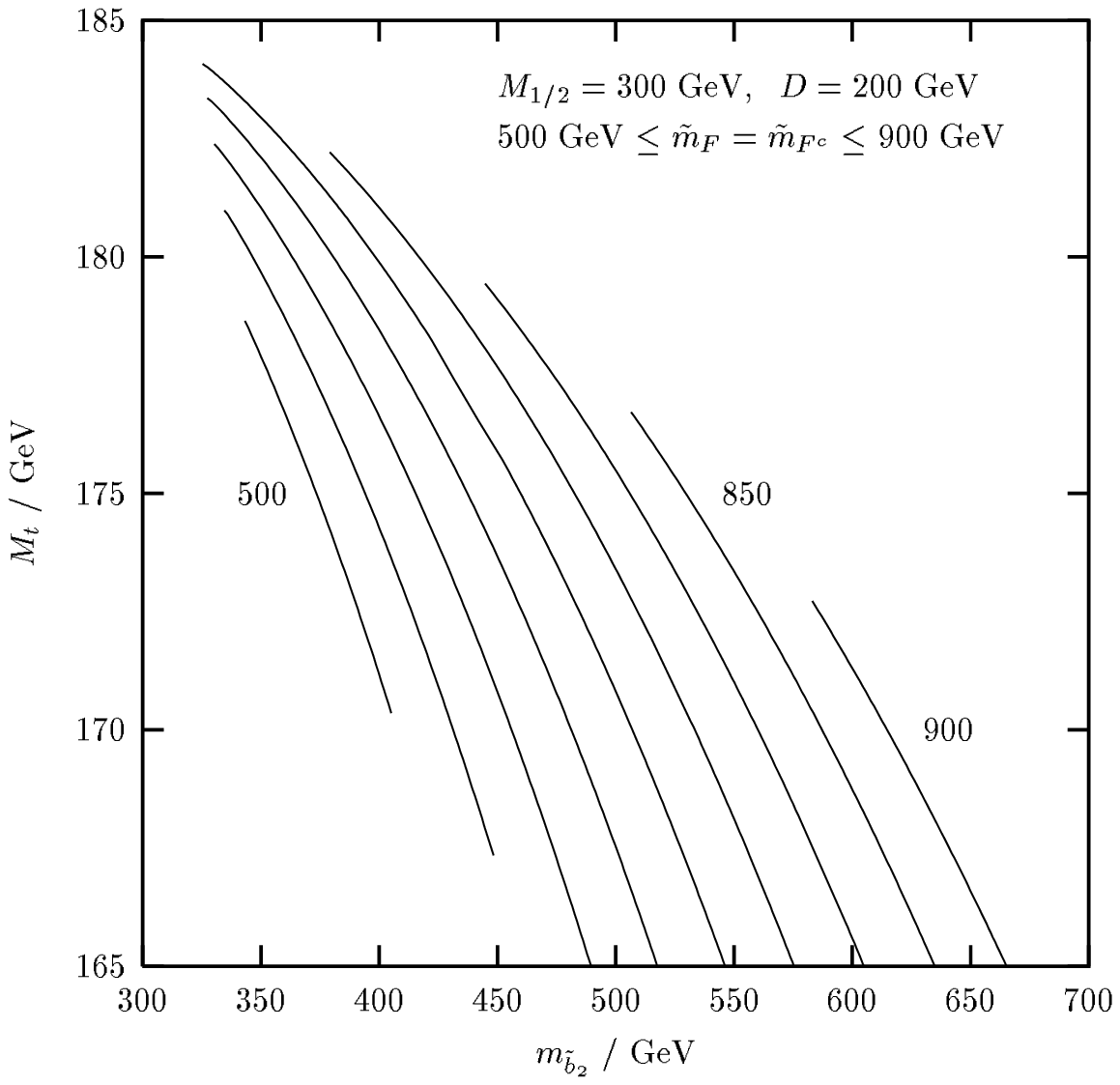}}

{\narrower\narrower\footnotesize\noindent
{FIG.~\FigMtVsSbottomScan}
Correlation between the top mass prediction $M_t$ and
the lightest sbottom mass prediction $m_{\tilde b_1}$.
Each line corresponds to a fixed choice for the
soft mass of the scalar fields $\tilde m_F=\tilde m_{F^c}$
(labeled with a number)
and along it the soft mass for the unified Higgs bosons ${\tilde m}_h$
is varying
(increasing from the bottom to the top of the plot.)
\par}}}

In Fig.~\FigMtVsSbottomScan\ the top mass prediction $M_t$ is plotted
against the lightest squark mass, that is, the right-handed sbottom
$\tilde b_2$. The reason why the right-handed down-type squarks are
lighter than the other squarks is because of the negative $D$-term
correction in Eq.~\refeqn{md-3gD}.
The prediction for $m_{\tilde b_2}$ is well above the experimental
lower bound $m^{exp}_{\tilde b_2} > 75 {\rm\ GeV}$ \cite{CCasoEtal}.

\vbox{
\noindent
\hfil
\vbox{
\epsfxsize=\figsize
\epsffile[130 380 510 735]{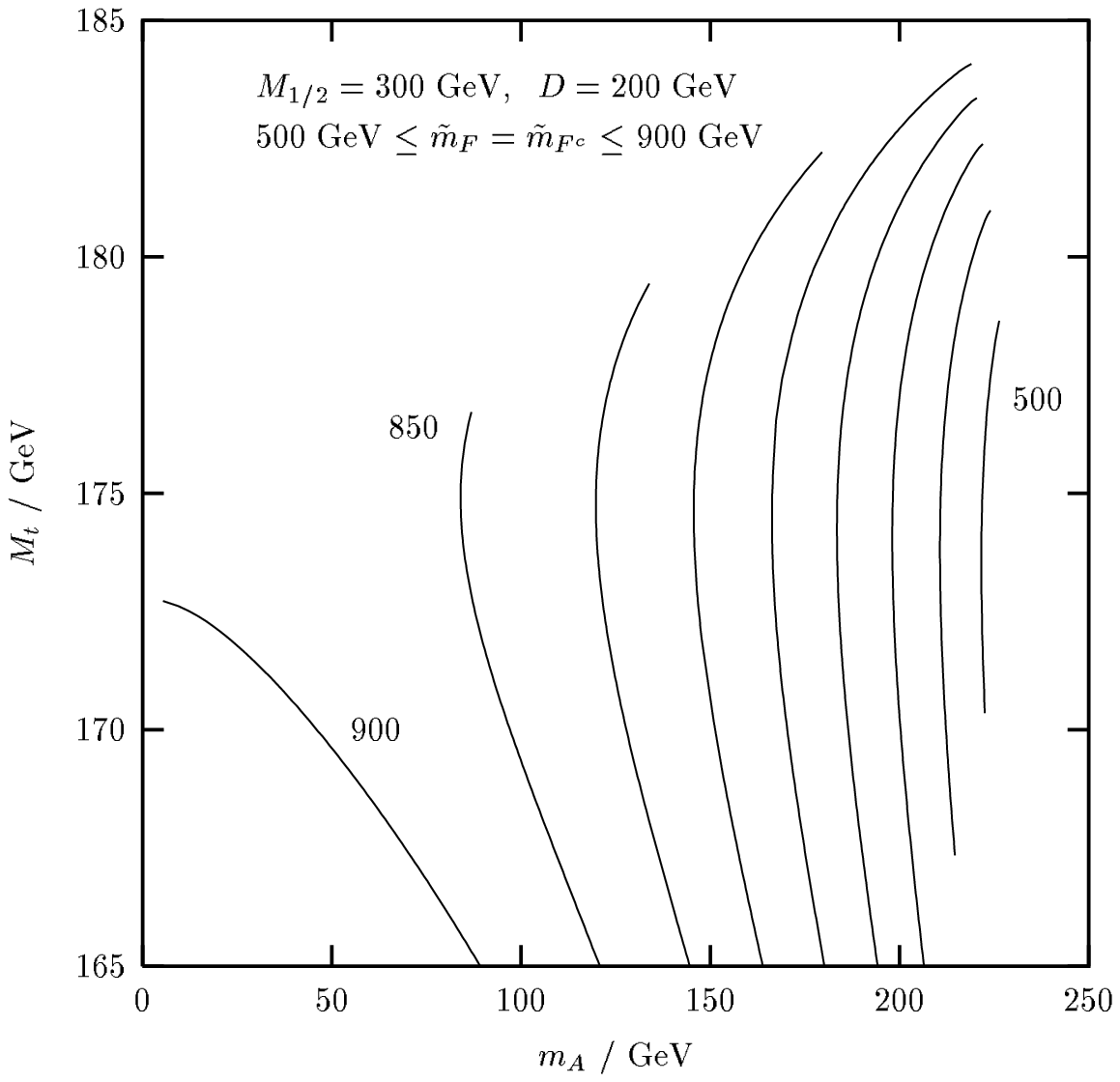}}

{\narrower\narrower\footnotesize\noindent
{FIG.~\FigMtVsMaScan}
Correlation between the top mass prediction $M_t$ and
the CP-odd Higgs boson mass prediction $m_A$.
Each line corresponds to a fixed choice for the
soft mass of the scalar fields $\tilde m_F=\tilde m_{F^c}$
(labeled with a number)
and along it the soft mass for the unified Higgs bosons ${\tilde m}_h$
is varying
(increasing from the bottom to the top of the plot.)
\par}}

Finally in Fig.~\FigMtVsMaScan\ we present the correlation between the
predictions for the top mass $M_t$ and
the mass of the CP-odd Higgs boson $m_A$.
We observe that $m_A$ decreases for increasing $\tilde m_{F,F^c}$
masses. In fact, for $\tilde m_{F,F^c}=900 {\rm\ GeV}$ we find that
$m_A$ is driven to zero.
This graph shows that only an upper bound
for the Higgs boson mass can be estimated : $m_A < 220 {\rm\ GeV}$
(the experimental lower bound is
$m^{exp}_A > 84 {\rm\ GeV}$ \cite{CCasoEtal}.)

We would like to end this section by remarking that,
as far as the top mass prediction is concerned,
the results in this section are similar to the $SO(10)$ model
\cite{BaDiFeTa,RaSa,PoPo2,LlMu,MuOlPo,MaNi}, since we have
taken the case of universal gaugino and scalar mass parameters,
with the D-term corrections reducing to those in $SO(10)$. 
However we have included the neutrino Yukawa coupling in our analysis,
and also the theory is different from $SO(10)$ above the GUT scale.
The reason we worked in this limit was to make contact with
the recent work on $SO(10)$, and to distinguish clearly
the effects of D-terms from the effect of explicit non-universality.
However we would like to emphasise that the results in
Figs.~\FigMtVsMhScan\--\FigMtVsMaScan\ are not only valuable by
themselves, but they also serve as a reference to which the
results in section VI.C should be compared with. Generally, the non-universal
422 model with $D$-term corrections will lead to a set of
predictions that is a combination of the results presented
in sections VI.C and VI.D.

\SECTION{E. $b\to s\gamma$ and $\tau\to\mu\gamma$ \\
            with non-universal A-terms}

It is well known that the decay $b\to s\gamma$ is a sensitive probe of
new physics. In the standard model the loop diagram involving the
$W$ boson and top quark give a theoretical prediction for
${\rm BR}(b\to s\gamma)=(3.28\pm0.33)\times 10^{-4}$ \cite{bsgSM}
\footnote{ This is the value of K. Chetyrkin \etal\ in Ref. \cite{bsgSM}.}
at the next-to-leading order in QCD.
This result turns out to be slightly larger than the official CLEO
measurement (see first Ref. in \cite{bsgCLEO}).
\footnote{
Note that an updated preliminary value by CLEO has been reported :
${\rm BR}(b\to s\gamma) = (3.15\pm 0.35 \pm 0.32)\times 10^{-4}$
(see second Ref. in \cite{bsgCLEO}.)}
However the recent ALEPH results
indicate a larger branching ratio \cite{bsgALEPH}. We quote~:
\begin{equation}
\matrix{
& & & \cr
{\rm BR}(b\to s\gamma) &=& (2.32\pm 0.57 \pm 0.35)\times 10^{-4} &
{\rm CLEO} \cr
& & & \cr
{\rm BR}(b\to s\gamma) &=& (3.11\pm 0.88 \pm 0.72)\times 10^{-4} &
{\rm ALEPH} \cr
& & & \cr}
\label{CLEOALEPH}
\end{equation}
where the first error is statistical and the second is systematic.
In view of the above large uncertainties we will take the conservative
range :
\begin{equation}
1.0\times 10^{-4} < {\rm BR}(b\to s\gamma) < 5.0\times 10^{-4}.
\label{bsgRange}
\end{equation}

In supersymmetric extensions of the SM the inclusion of
additional SUSY particles can spoil the above ``agreement'' \cite{bsgSUSY}.
The reason is because, generally, there is no guarantee that
the sparticle contributions are small or that they conspire to
cancel between themselves. Indeed, it is known that in two Higgs
extensions of the SM the contribution from the charged
Higgs--top quark loop $H^- t$ always interferes constructively with
the SM one.
Thus, the resulting branch ratio prediction is bigger
and dependent on the unknown charged Higgs mass $m_{H^-}$.
If $m_{H^-}$ is light then the prediction for the ${\rm BR}(b\to s\gamma)$
is larger than the experimental upper bound.
On the other hand, for a heavy Higgs mass $m_{H^-}\sim$ 1 TeV,
agreement with Eq.~\refeqn{CLEOALEPH} is still be possible.

The above situation can change drastically in SUSY models with third
family Yukawa unification. The reason is because the new
contributions arising from gluino-sbottom ($\tilde g \tilde b$),
neutralino-sbottom ($\chi^0 \tilde b$) and particularly from
chargino-stop ($\chi^- \tilde t$) loops are enhanced by large
large $\tan\beta\sim 40-50$ factors. In contrast to the charged
Higgs contribution, the SUSY amplitudes can either add constructively
or destructively with the $H^- t$ contribution or between themselves.
In order to recover the original agreement with experiment
a large part of the work in the literature \cite{bsgSUSY} explores the
idea of suppressing the $b\to s\gamma$ decay by canceling
the $\chi^- \tilde t$ against the $H^- t$ amplitude
(typically the gluino and neutralino amplitudes are small.)
Many of these results are obtained in the context of SUGRA/GUT models
with universal soft masses and trilinear terms at the Planck or GUT
scale with large gaugino masses $M_{1/2}\sim 1$ TeV.
Cancellation between $\chi^- \tilde t$ and $H^- t$ loops is possible
because of two reasons, firstly the large gaugino mass leads to
large stop masses thus suppressing the $\chi^- \tilde t$ amplitude
making it comparable in magnitude to the $H^- t$ term.
Secondly, the sign of $\mu$ is chosen to be positive.
This means that the $\chi^- \tilde t$ contribution has the opposite
sign of the $H^- t$ contribution thus allowing the cancellation to occur.

Clearly the above strategy is very attractive because it is based on
universal soft parameters which render it model independent.
However, it fails to address two issues. Firstly a large $M_{1/2}$
leads to fine-tuning of the $Z$ boson mass and secondly, a positive
$\mu$ is in conflict with the successful prediction for the top mass in
the context of GUT models with unified third family Yukawa couplings
which, as we have seen, require negative $\mu$.

In this section we explore a different solution to the enhanced
$b\to s\gamma$ decay in SUSY models with $t-b-\tau$ Yukawa unification.
The idea is to choose $\mu$ to be negative, for the reasons
stated earlier, and to suppress
$b\to s\gamma$ by allowing the trilinear soft $A$-terms to have
a non-universal family dependent structure. We will avoid 
unnatural tuning of electroweak symmetry breaking 
by taking a low value for the gaugino masses,
for example $M_{1/2}\sim 300 \> (250)$ GeV at the Planck (GUT) scale.
In this scheme we find that
$b\to s\gamma$ is dominated by the chargino-stop loop.

The purpose of this section is to investigate the possibility of
suppressing $b\to s\gamma$ by tuning the initial non-universal $A$-terms
at the Planck scale such that, at low energy, they are cancelled by
the flavour violating signals that naturally develop when the
parameters of our model evolve from high to low energy by the use of RGEs.
We will conclude by checking that the introduction of
non-universal $A$-terms is also compatible with the present upper bound on
${\rm BR}(\tau\to\mu\gamma)$.

\newpage

\SUBSECTION{$ b\to s\gamma$}

The standard model, charged Higgs, chargino and gluino diagrams
contributing to this decay are illustrated in Fig.~\FigBsgLoops.

\vbox{
\hfil
\vbox{
\hbox{
\hbox{
\epsfxsize=5cm
\epsffile[100 300 490 550]{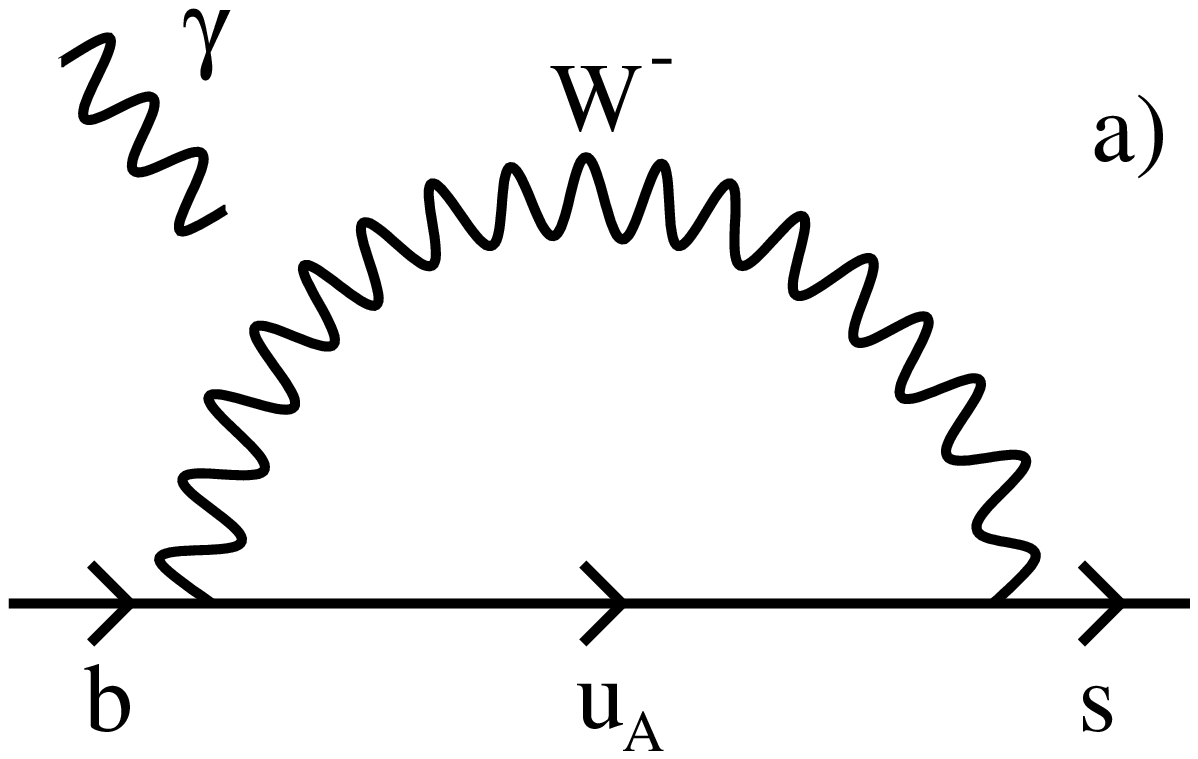}}
\hbox{
\epsfxsize=5cm
\epsffile[100 300 490 550]{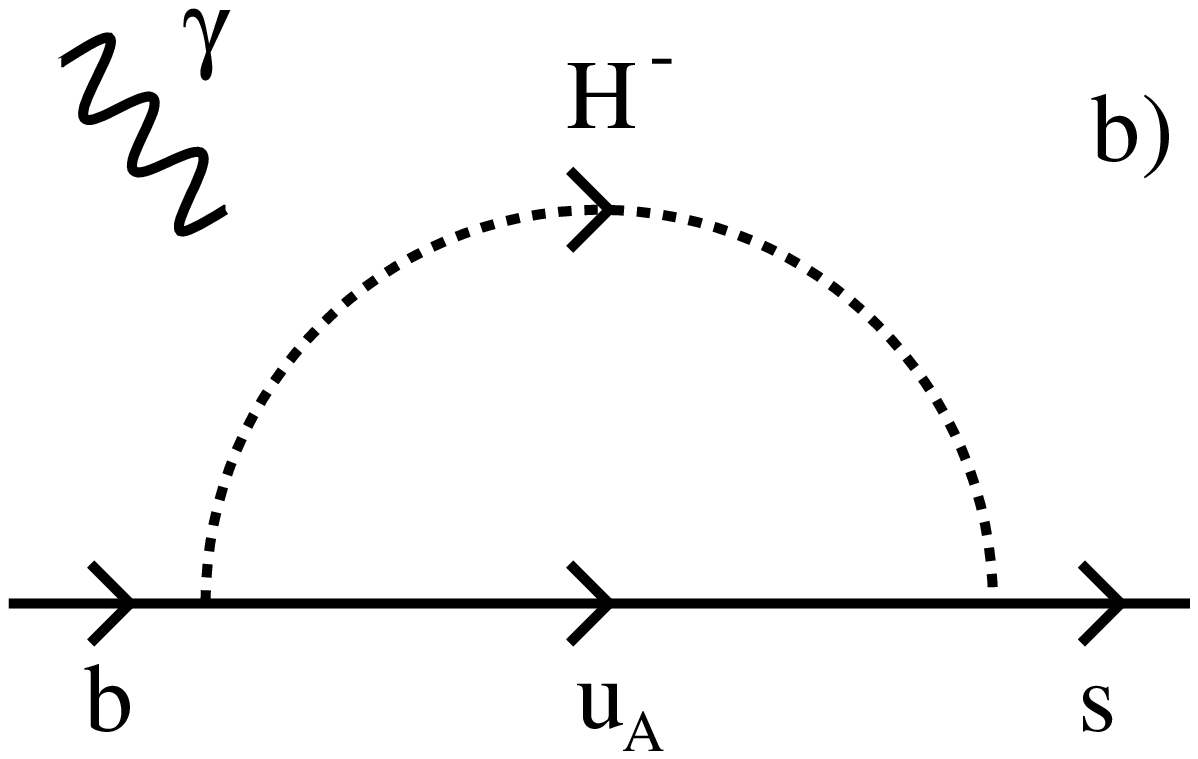}}
\hskip 0.5cm}}

\hfil
\vbox{
\hbox{
\hbox{
\epsfxsize=5cm
\epsffile[100 300 490 550]{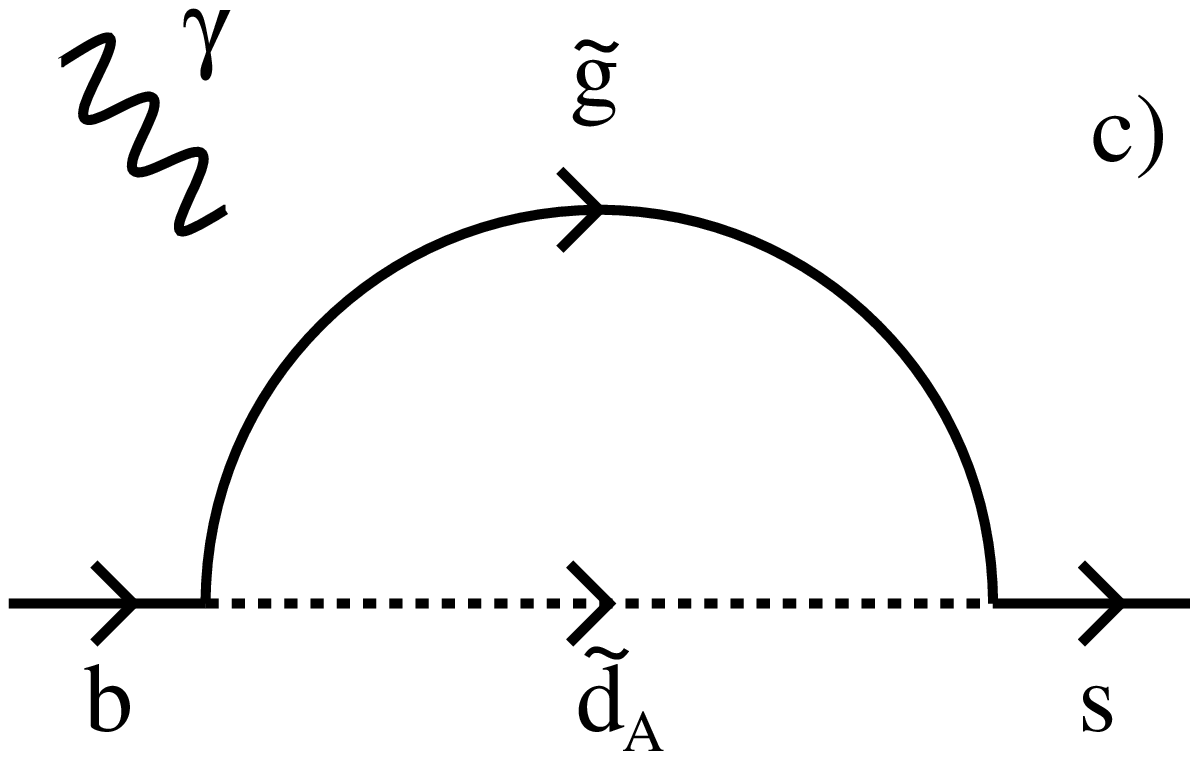}}
\hbox{
\epsfxsize=5cm
\epsffile[100 300 490 550]{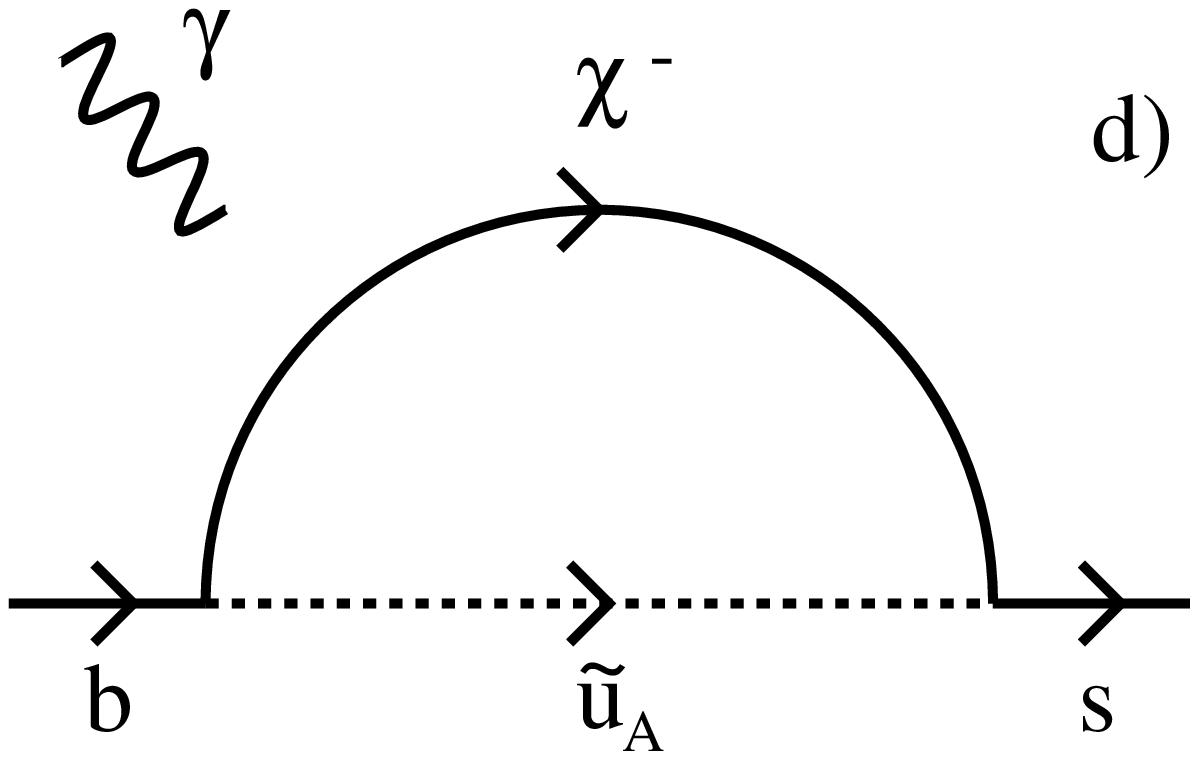}}
\hskip 0.5cm}}

\medskip

{\narrower\narrower\footnotesize\noindent
{FIG.~\FigBsgLoops.}
Diagrams responsible for the $b\to s\gamma$ decay considered
in this article (note that we did not compute the neutralino/sbottom
loop which is typically found to be small.)
The $W$ boson loop a) illustrates the standard model contribution.
The charged Higgs loop b) is present in two doublet Higgs extensions of the SM.
The gluino contribution c) was found to be small.
On the other hand, the chargino loop d) dominated the $b\to s\gamma$ decay.
\par\bigskip}}

\noindent
In models with large $\tan\beta$ the corresponding dominant amplitudes
are given by \cite{BeBoMaRi}~:
\begin{eqnarray}
A_{SM} &=& {\alpha_w \sqrt{\alpha_e} \over 4\sqrt{\pi}} {1\over M_W^2}
          V_{ts} V_{tb} \> 3 x_{tw}\> F_{12}(x_{tw})
\label{bsgSMAmp}\\
A_{H^-} &=& {\alpha_w \sqrt{\alpha_e} \over 4\sqrt{\pi}} {1\over M_W^2}
           V_{ts} V_{tb} \>  x_{th} \> F_{34}(x_{th}) \\
A_{\chi^-} &=& {\alpha_w \sqrt{\alpha_e} \over 2\sqrt{\pi}}
            \sum_{j=1}^2 \sum_{\alpha=1}^6 {1\over m_{\tilde U_\alpha}^2}
           H_{UL}^{j\alpha b}
          \left( G_{UL}^{j\alpha s} - H_{UR}^{j\alpha s} \right)
          {m_{\chi^-_j} \over m_b} F_{43}(x_{\chi^-_j \tilde U_\alpha})
\label{bsgChargAmp}\\
A_{\tilde g} &=& {\alpha_s \sqrt{\alpha_e} \over \sqrt{\pi}} C(R) e_D
          \sum_{\alpha=1}^6 {1\over m_{\tilde D_\alpha}^2}
          (V^{\tilde d d\dagger}_R)_{b \alpha} (V^{\tilde d d}_L)_{\alpha s}
          {m_{\tilde g} \over m_b} F_4(x_{\tilde g \tilde D_\alpha})
\label{bsgGluinoAmp}
\end{eqnarray}
where $\alpha_w = g^2/(4\pi)$, $\alpha_e = e^2/(4\pi)$,
$M_W$ is the $W$ boson mass,
$m_{\tilde U_\alpha}$ and $m_{\tilde D_\alpha}$
the sup/sdown type squark mass eigenstates, $V$ the CKM matrix,
$x_{tw} = m_t^2/M_W^2$, $x_{th} = m_t^2/m^2_{H^-}$,
$x_{\chi^-_j \tilde U_\alpha}= m^2_{\chi^-_j}/m^2_{\tilde U_\alpha}$,
$x_{\tilde g \tilde D_\alpha} = m^2_{\tilde g} / m^2_{\tilde D_\alpha}$,
$C(R) = 4/3$ and  $e_D=-1/3$.
The flavour matrices $V^{\tilde d d}_{L,R}$ describe the mismatch
between the transformations that diagonalize the
down-type squark and quark mass matrices
(See appendix \AppendixA\ for details.)
The functions $F$ are defined by $F_{mn}(x) = {2 \over 3} F_m(x) + F_n(x)$
with $F_{m,n}(x)$ as in Ref.~\cite{BeBoMaRi}.
The gaugino and higgsino vertex factors $G$ and $H$ are given by :
\begin{eqnarray}
G_{UL}^{j\alpha A} &=& T^c_{j1} (V^{\tilde ud}_L)_{\alpha A} \\
H_{UR}^{j\alpha A} &=& T^c_{j2} \sum_{B=1}^3
                     (V^{\tilde uu}_R)_{\alpha B}
                     (Y_u)_{BB} V_{BA} \label{HUR} \\
H_{UL}^{j\alpha A} &=& S^{c\dagger}_{2j} (Y_d)_{AA}
                      (V_L^{\tilde ud\dagger})_{A\alpha} \label{HUL}
\end{eqnarray}
where,
\begin{eqnarray}
Y_u &=& {\rm diag}(m_u,m_c,m_t)/
        (\sqrt{2} M_W \sin\beta) = \lambda'_u / g \\
Y_d &=& {\rm diag}(m_d,m_s,m_b)/
        (\sqrt{2} M_W \cos\beta) = \lambda'_d / g.
\end{eqnarray}
and $S^c$, $T^c$ diagonalize the chargino mass matrix.
Finally the precise definition of the
$V^{\tilde ud}_L$, $V^{\tilde uu}_R$ matrices that describe the mismatch
between the transformations required to diagonalize the up-type
squarks and quarks mass matrices can be found in appendix~\AppendixA.

\noindent
The branch ratio BR($b\to s\gamma$) can be computed from \cite{BeBoMaRi} :
\begin{equation}
{\rm BR}(b\to s\gamma) = {\Gamma(b\to s\gamma) \over
                           \Gamma(b\to ce\bar\nu)}
                          {\rm BR}(b\to ce\bar\nu)
\end{equation}
where the decay rate for $b\to s\gamma$ is given by :
\begin{equation}
\Gamma(b\to s\gamma) = {m^5_b \over 16\pi} |A^\gamma(m_b)|^2.
\end{equation}
The full expressions for $\Gamma(b\to ce\bar\nu)$ and $A^\gamma(m_b)$
(the QCD corrected amplitude at the scale of the process ($\sim m_b$)
-- obtained from the total sum of amplitudes
$A^\gamma(M_W)=A_{SM}+A_{H^-}+A_{\chi^-}+A_{\tilde g}$)
can be found in Ref.~\cite{BeBoMaRi}.

It is interesting to note that the chargino amplitude has two distinct
contributions. They are illustrated in Fig.~\FigBsgChargLoops.
In Fig.~\FigBsgChargLoops~a)
the helicity flip required in the decay is achieved at the
higgsino vertex. Thus, along the internal squark line
$\tilde t_L$--$\tilde c_L$, flavour
violation develops through a $(\Delta^u_{23})_{LL}$ mass insertion.
\footnote{The $(\Delta^u_{23})_{LL}$ parameter is the off-diagonal 23 entry of
the left-handed up-type squark mass matrix in a basis where the
up-type quark mass matrix is diagonal (see Ref.~\cite{GaMa}.)}
On the other hand, in Fig.~\FigBsgChargLoops~b)
the two higgsino vertices require that the
helicity flip must be accomplished through the
$\tilde t_L$--$\tilde c_R$ line via a $(\Delta^u_{23})_{LR}$ mass
insertion.

\vbox{
\hfil
\vbox{
\hbox{
\hbox{
\epsfxsize=5cm
\epsffile[100 300 490 550]{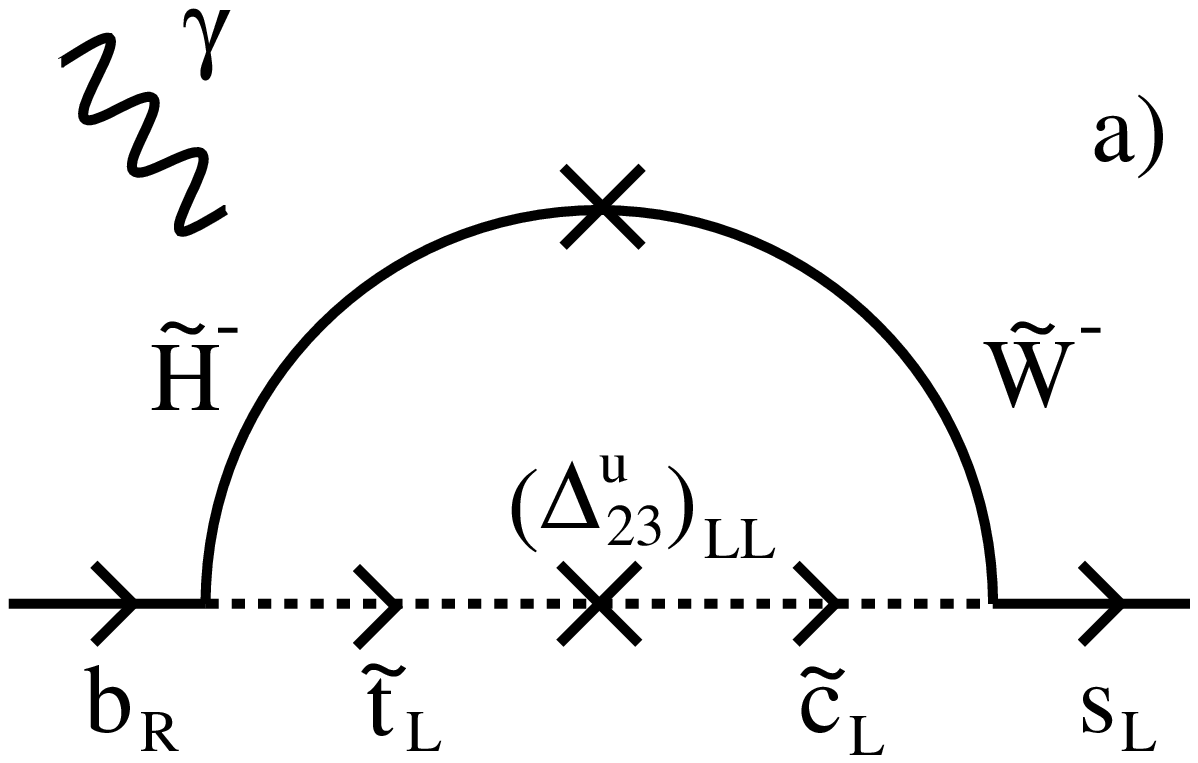}}
\hbox{
\epsfxsize=5cm
\epsffile[100 300 490 550]{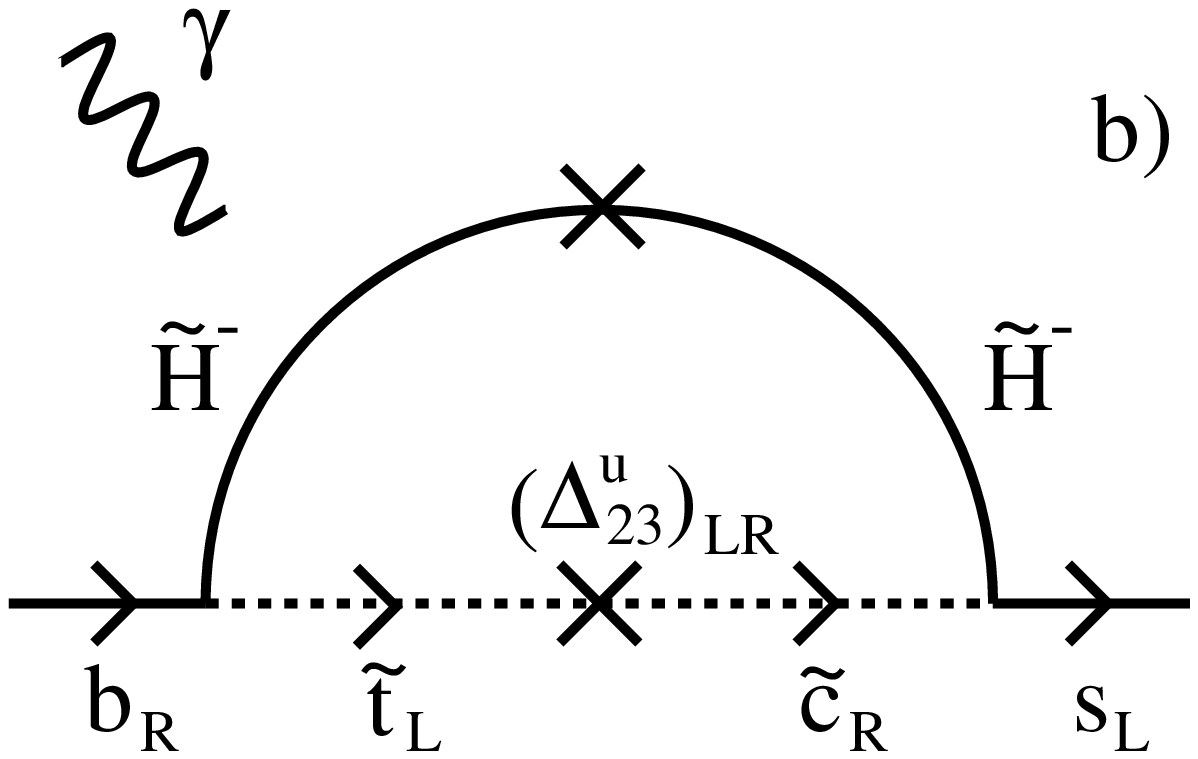}}
\hskip 0.5cm}}

\medskip

{\narrower\narrower\footnotesize\noindent
{FIG.~\FigBsgChargLoops}
Diagrams corresponding to the two contributions associated
with the chargino amplitude of Eq.~\refeqn{bsgChargAmp}.
In fig. a) flavour violation is introduced through
a $(\Delta^u_{23})_{LL}$ mass insertion along
the $\tilde t_L$--$\tilde c_L$ squark line.
In fig. b) a $(\Delta^u_{23})_{LR}$ mass insertion is introduced
along the $\tilde t_L$--$\tilde c_R$ chirality flipping squark line.
\par\bigskip}}

Remembering that we aim at reducing $A_{\chi^-}$ by introducing
tree-level explicit sources of flavour violation through
non-universal soft SUSY terms, the relevant question to address is
whether it is more appropriate to relax universality of the soft mass
terms for the fields in the $F$, $F^c$ multiplets,
thereby changing $(\Delta^u_{23})_{LL}$, or to modify the trilinear
$A$-terms, thereby changing $(\Delta^u_{23})_{LR}$.
For reasons that will became clear latter, when we study the
$\tau\to\mu\gamma$ decay, we will take the latter approach.

Traditionally the SUSY trilinear terms
are defined, at high energy, to be proportional to their
associated Yukawa matrix, \ie
$(\tilde A_u)_{AB} = A_0 (\lambda_u)_{AB}$ {\it etc}..
Although this assumption leads to obvious simplifications,
models inspired by string theory have been proposed \cite{GaKhTo}
in which the $A$-terms are not universal.
Motivated by these results we parametrise the $\tilde A$'s
by :
\begin{equation}
\matrix{
(\tilde A_u)_{AB} = A_0 x_{AB} (\lambda_u)_{AB} &
(\tilde A_d)_{AB} = A_0 x_{AB} (\lambda_d)_{AB} \cr
& \cr
(\tilde A_e)_{AB} = A_0 x_{AB} (\lambda_e)_{AB} &
(\tilde A_\nu)_{AB} = A_0 x_{AB} (\lambda_\nu)_{AB} \cr}
\label{trilinear}
\end{equation}
where $x_{AB}$ is a dimensionless matrix of order one that we
conveniently choose to have the following structure :
\begin{equation}
x_{AB} = \left(\matrix{
               1 & 1 & 1 \cr
               1 & 1 & x \cr
               1 & x & {1 \over 10}}\right)
\label{xAB}
\end{equation}

We now turn to study the phenomenological implications of
Eq.~\refeqn{xAB} to the ${\rm BR}(b\to s\gamma)$.
For illustrative purposes, and keeping in mind the successful
prediction for the top mass in the 422 model with $D$-terms,
we use, as in the previous section, the same
$\alpha_s = 0.120$, $M_b = 4.8 {\rm\ GeV}$ and
the same input at the Planck scale~:
$M_{1/2} = 300 {\rm\ GeV}$,
$\tilde m_{F} = \tilde m_{\bar F} = 500 {\rm\ GeV}$,
${\tilde m}_h = 750 {\rm\ GeV}$ and $D = 200 {\rm\ GeV}$.
Moreover we fixed  $A_0 = 2000 {\rm\ GeV}$
\footnote{$A_0(M_X) = 570 {\rm\ GeV}$}
and a negative sign for $\mu$.
At the GUT scale the Yukawa matrices were set as follows.
The eigenvalues of $\lambda_u$, $\lambda_d$, $\lambda_e$
were fixed by the requirement that the fermion masses at low energy
were reproduced.
The CKM mixing angles were assumed to be given by
$\theta_i = \theta^u_i+\theta^d_i$ with $\theta^u_i=\theta^d_i$
(the $\theta^{u,d}_i$ are the angles that parameterise the rotation
matrices that transform the left-handed up/down quarks
into their physical mass basis.)
The angles associated with the rotation required to transform the
left handed charged leptons into their physical mass basis
were given by $\theta^e_i=\theta^d_i$.
Finally $\lambda_\nu$ was set equal to $\lambda_u$.
For completeness we list in Table~\TabBsgYuk\
the values of the Yukawa matrices at the SUSY scale $Q=M_S=430 {\rm\ GeV}$.

\vbox{
\begin{center}
\begin{tabular}{c}
\multicolumn{1}{c}{\hskip 145pt TABLE \TabBsgYuk.} \cr
\noalign{\medskip}
\noalign{\hrule width 350pt}
\noalign{\smallskip}
\noalign{\hrule width 350pt}
\noalign{\medskip}
\end{tabular}
\end{center}
\vskip -15pt
$$
\lambda_u =
\left(\matrix{
\hfill  1.003 \times 10^{-5} &
\hfill  1.118 \times 10^{-6} &
\hfill \phantom{-}
        1.768 \times 10^{-8} \cr
\hfill -3.608 \times 10^{-4} &
\hfill  3.236 \times 10^{-3} &
\hfill  7.012 \times 10^{-5} \cr
\hfill  5.882 \times 10^{-4} &
\hfill -1.973 \times 10^{-2} &
\hfill  0.914 \hfill
}\right)
$$
$$
\lambda_d =
\left(\matrix{
\hfill  1.075 \times 10^{-3} &
\hfill \phantom{-}
\hfill  1.213 \times 10^{-4} &
\hfill  3.251 \times 10^{-6} \cr
\hfill  2.384 \times 10^{-3} &
\hfill  2.111 \times 10^{-2} &
\hfill -4.461 \times 10^{-4} \cr
\hfill -5.686 \times 10^{-4} &
\hfill  1.910 \times 10^{-2} &
\hfill  0.872 \hfill
}\right)
$$
$$
\lambda_e =
\left(\matrix{
\hfill \phantom{-}
        1.464 \times 10^{-4} &
\hfill -1.637 \times 10^{-5} &
\hfill  7.297 \times 10^{-7} \cr
\hfill  3.436 \times 10^{-3} &
\hfill  3.084 \times 10^{-2} &
\hfill -6.077 \times 10^{-4} \cr
\hfill  7.465 \times 10^{-4} &
\hfill  9.409 \times 10^{-3} &
\hfill  0.532 \hfill
}\right)
$$
$$
\lambda_\nu =
\left(\matrix{
\hfill  4.406 \times 10^{-6} &
\hfill  4.908 \times 10^{-7} &
\hfill \phantom{-}
        8.300 \times 10^{-9} \cr
\hfill -1.585 \times 10^{-4} &
\hfill  1.421 \times 10^{-3} &
\hfill  3.539 \times 10^{-5} \cr
\hfill  2.519 \times 10^{-4} &
\hfill -1.141 \times 10^{-2} &
\hfill  0.575 \hfill
}\right)
$$
\vskip -18pt
\vbox{
\begin{center}
\begin{tabular}{c}
\noalign{\hrule width 350pt}
\noalign{\smallskip}
\noalign{\hrule width 350pt}
\end{tabular}
\end{center}}

\vbox{
{\narrower\narrower\footnotesize\noindent
{TABLE \TabBsgYuk.}
Values of the Yukawa matrices at the effective SUSY scale
$Q=M_S=430 {\rm\ GeV}$.
The correct quark and charged lepton masses as well as the CKM matrix are
obtained after ``running-down'' these matrices firstly,
from $Q = M_S$ to $Q = M_Z$, using the SM RGEs, and secondly
from $Q=M_Z$ to $Q = 1 {\rm\ GeV}$,
using $SU(3)_c \times U(1)_{em}$ RGEs. (See section II.B for more details.)
\par\bigskip}}
}

\vbox{
\noindent
\hfil
\vbox{
\epsfxsize=\figsize
\epsffile[130 380 510 735]{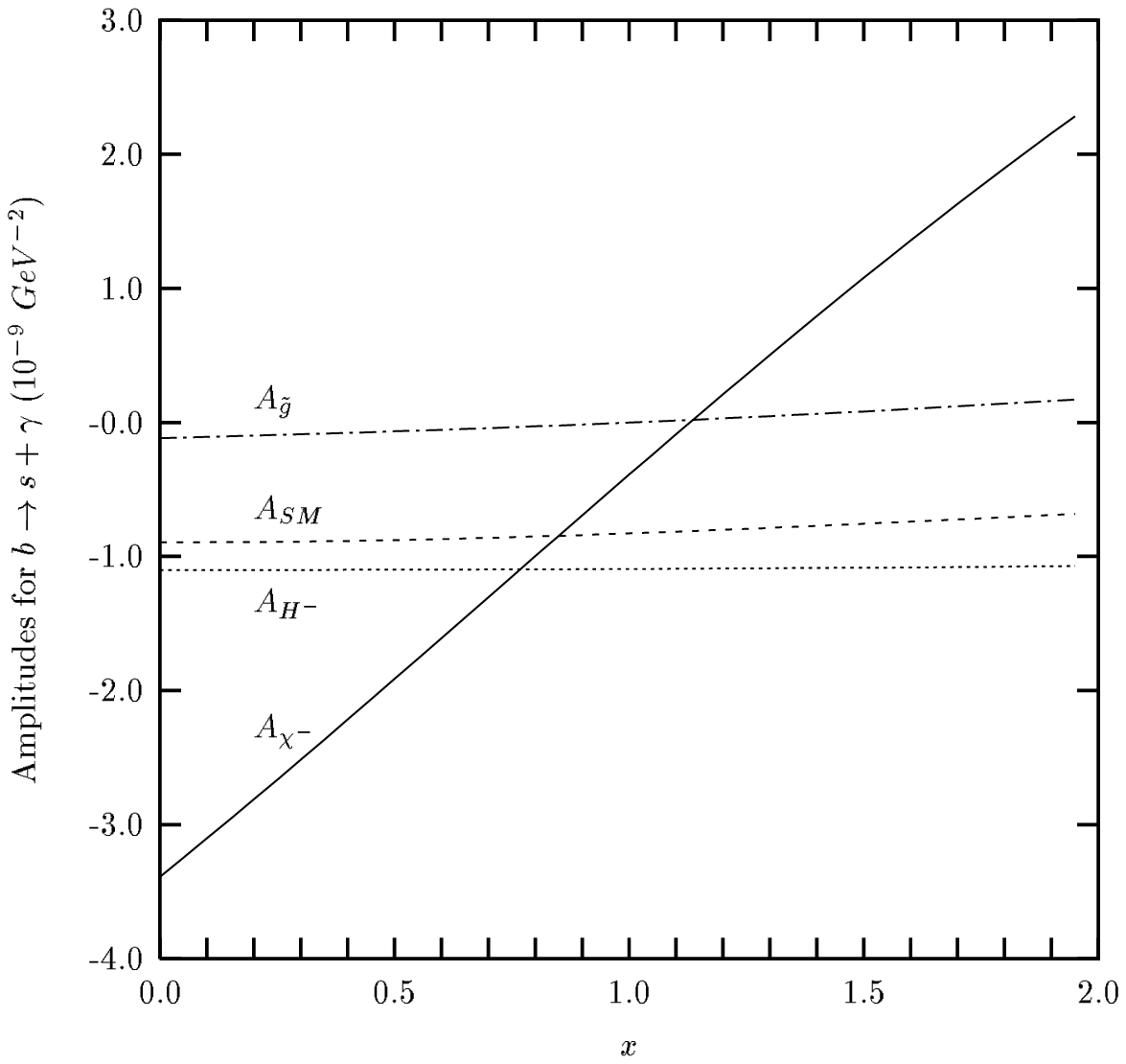}}

{\narrower\narrower\footnotesize\noindent
{FIG.~\FigBsgAmp}
Individual values for the amplitudes contributing
to the $b\to s\gamma$ decay against the
trilinear parameter $x$ of Eq.~\refeqn{xAB}.
One observes that the chargino amplitude $A_{\chi^-}$
is sensitive to $x$ and bigger than
$A_{SM}$, $A_{H^-}$ and $A_{\tilde g}$ for almost all the $x$ values.
Furthermore, the sign of $A_{\chi^-}$ changes as $x$ increases
from 0.0 to 2.0.
\par\bigskip}}

In Fig.~\FigBsgAmp\ we plot the SM, charged Higgs,
chargino and  gluino amplitudes in
Eqs.~\refeqn{bsgSMAmp}--\refeqn{bsgGluinoAmp}
against $x$. For $x\sim 0$ we approximately
recover the usual universal $A$-term model. This is because the first
and second family of the trilinear terms are small
(due to small Yukawa entries) and the third family $A$-terms at low
energy are not too sensitive to their initial value at high energy.
We see that because $\mu$ is negative, the chargino amplitude
is negative and interferes constructively with the other amplitudes.
However, as $x$ increases, the chargino amplitude becomes less
negative and the magnitude of $|A_{\chi^-}|$ steadily decreases.
At some point, around $x\sim 1.1$ the chargino
amplitude vanishes.
Beyond $x\sim 1.1$, $A_{\chi^-}$ is positive,
thus it interferes destructively with the SM and Higgs amplitudes.
Clearly, we see that by tuning $x$ it is easy to find a region where
$b\to s\gamma$ is suppressed.

\vbox{
\noindent
\hfil
\vbox{
\epsfxsize=\figsize
\epsffile[130 380 510 735]{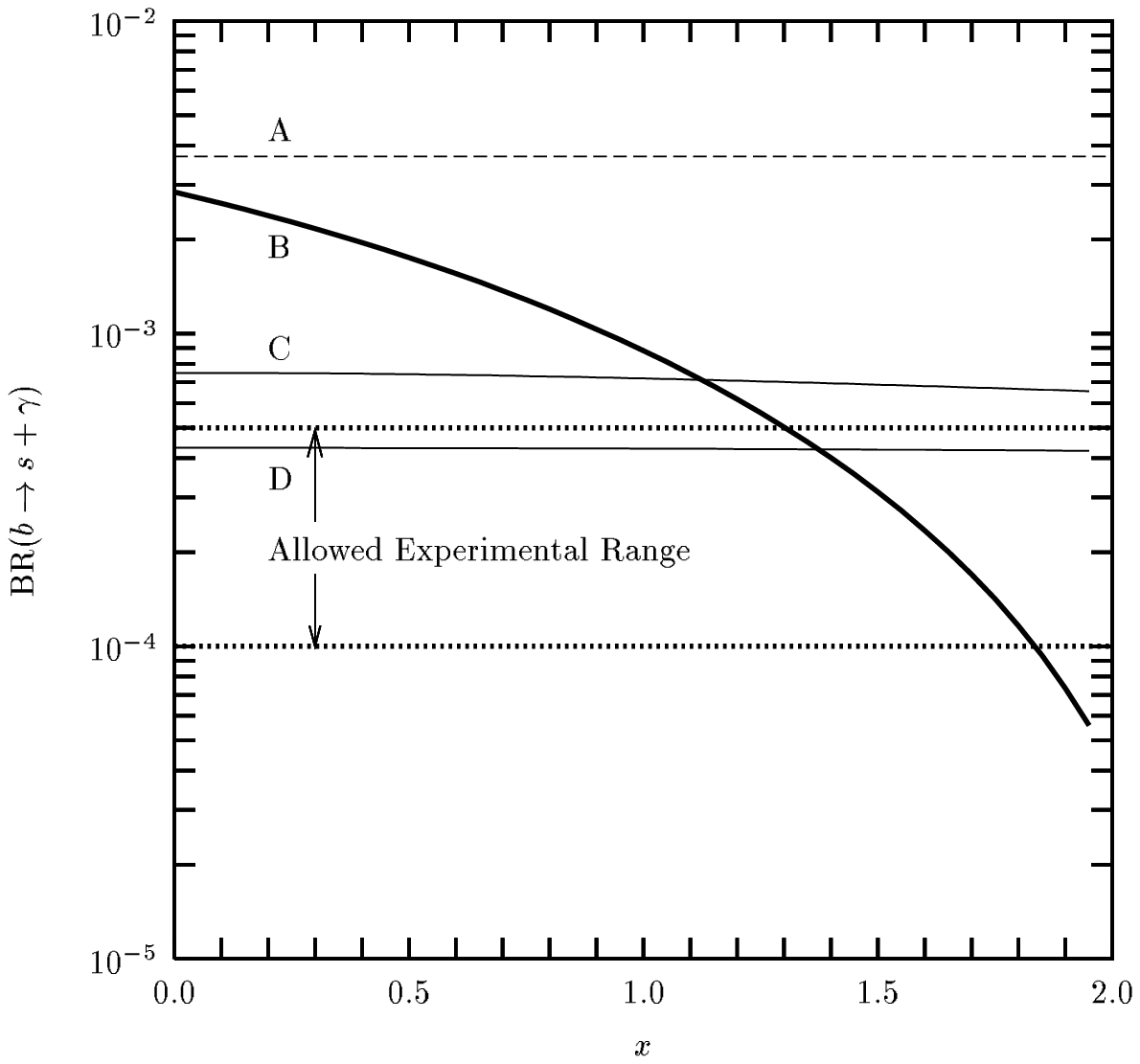}}

{\narrower\narrower\footnotesize\noindent
{FIG.~\FigBsgBR}
Branch ratio for the $b\to s\gamma$ decay against the
trilinear parameter $x$ in Eq.~\refeqn{xAB}.
The dashed line A corresponds to the result one obtains when $A_0 = 0$.
The solid line B was computed using the SM, charged Higgs,
gluino and higgsino contributions.
For the line C we used the SM and Higgs contributions only.
The line D was obtained using the SM amplitude alone.
The allowed experimental range is indicated by the area between
the two dotted lines.
We observe that for the line B an acceptable
prediction is obtained for $x\sim 1.60$.
\par\bigskip}}

In Fig.~\FigBsgBR\ we show the ${\rm BR}(b\to s\gamma)$ against $x$.
The area between the dashed lines indicates the experimental allowed
range in Eq.~\refeqn{bsgRange}.
The dashed line A corresponds to the result one obtains when $A_0=0$.
The solid lines D and C are plotted for illustrative purposes;
they correspond to the ${\rm BR}(b\to s\gamma)$ that is obtained
when only the SM amplitude, and the SM plus the charged Higgs amplitudes
are considered respectively.
The solid line B indicates the ${\rm BR}(b\to s\gamma)$ prediction
when all the amplitudes in Eqs.~\refeqn{bsgSMAmp}-\refeqn{bsgGluinoAmp}
are considered.
We observe that, due to a light charged Higgs mass $m_H^-\sim 130$ GeV,
the value of the ${\rm BR}(b\to s\gamma)$
when only the SM+$H^- t$ loops are considered lies above the
experimental range. On the other hand, the ${\rm BR}(b\to s\gamma)$ when
the chargino contribution is included (line B) starts at $x\sim 0$ very large
but is driven into compatibility with experiment at around
$x\sim 1.60 \pm 0.15$.

\SUBSECTION{$\tau\to\mu\gamma$}

In the previous section we showed that the parameterization of the
trilinear terms by Eq.~\refeqn{xAB}
lead to a suppressed ${\rm BR}(b\to s\gamma)$
compatible with the experimental data.
However, one should be careful about other possible implications
derived from this new source of flavour violation.
Clearly Eq.~\refeqn{xAB} also introduces lepton flavour violating
signals through the $\tilde A_{e,\nu}$ matrices in Eq.~\refeqn{trilinear}.
The purpose of this section is to check that the decay
$\tau\to\mu\gamma$
(for the successful value of $x$ fixed by $b\to s\gamma$)
is not in conflict with the experimental upper bound on
${\rm BR}(\tau\to\mu\gamma) < 3.0\times 10^{-6}$~\cite{tmgCLEO}.

\noindent
The effective lagrangian for the $\tau\to\mu\gamma$ decay is :
\begin{equation}
\textstyle
{\cal L}_{\tau\to\mu\gamma} =
{1 \over 2} {\bar\psi_\mu}(p-q)
\left( A^R P_R + A^L P_L \right) \sigma^{\alpha\beta}
\psi_\tau(p) F_{\alpha\beta}
\end{equation}
where $P_{R,L}$ are projection operators and
$F_{\alpha\beta}$ the electromagnetic field tensor.
The branch ratio can be computed using :
\begin{equation}
{\rm BR}(\tau\to\mu\gamma) = {12 \pi^2 \over G_F^2}
\left( |A^R|^2+|A^L|^2 \right)
\end{equation}
where $A^R = A^R_{\chi^-}+A^R_{\chi^0}$ and
$A^L = A^L_{\chi^0}$. Numerically we found that
the decay is dominated by the chargino-sneutrino loop diagram
illustrated in Fig.~\FigTmgChargLoop.

\vbox{
\hfil
\vbox{
\hbox{
\hbox{
\epsfxsize=5cm
\epsffile[100 300 490 550]{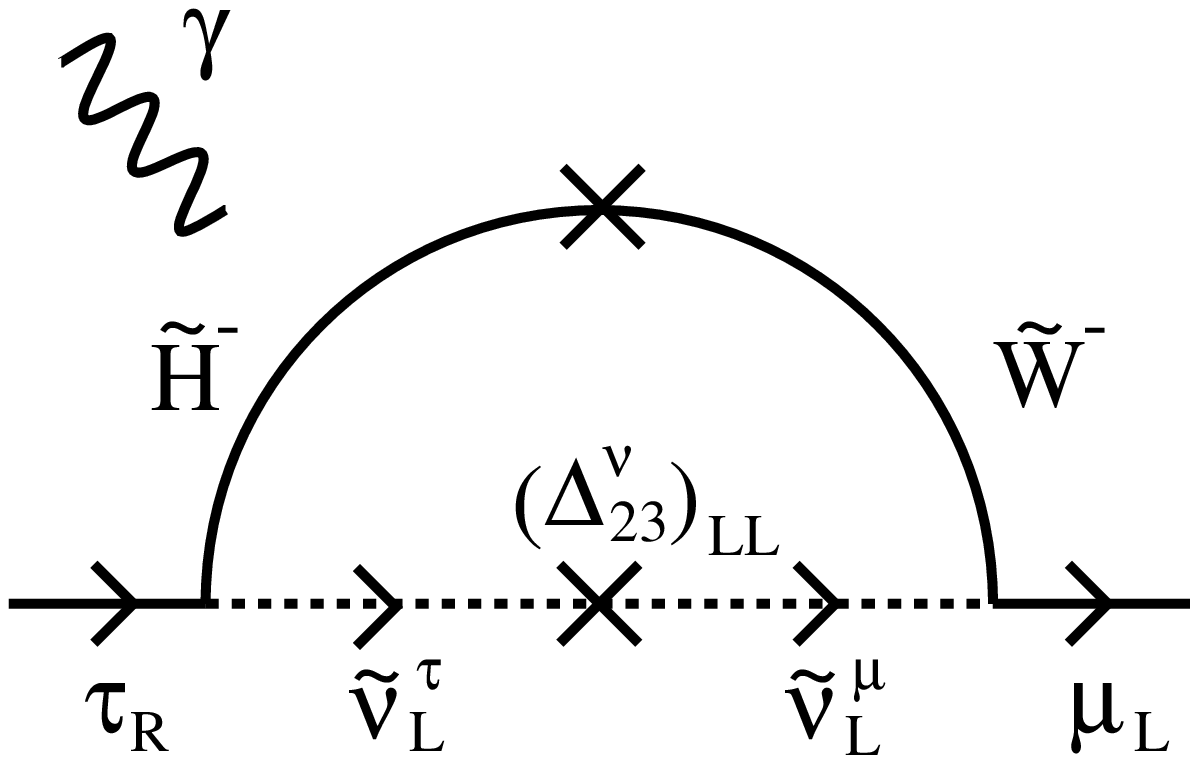}}
\hskip 0.5cm}}

\medskip

{\narrower\narrower\footnotesize\noindent
{FIG.~\FigTmgChargLoop}
Chargino diagram involved in the $\tau\to\mu\gamma$ decay
(the neutralino diagram was found to be subdominant.)
In this process lepton flavour violation develops
through the $(\Delta^\nu_{23})_{LL}$ mass insertion
along the $\tilde\nu^\tau_L$--$\tilde\nu^\mu_L$ sneutrino line.
\par\bigskip}}

\noindent
The corresponding amplitude is given by :
\begin{equation}
A^R_{\chi^-} = {\alpha_w \sqrt{\alpha_e} \over 2\sqrt{\pi}}
               \sum_{j=1}^2 \sum_{\alpha=1}^3
               {1\over m_{\tilde \nu_\alpha}^2}
               H_{\nu L}^{j\alpha \tau}
               G_{\nu L}^{j\alpha \mu}
               {m_{\chi^-_j} \over m_\tau}
               J(x_{\chi^-_j \tilde\nu_\alpha})
\end{equation}
where $m_{\tilde\nu_\alpha}$ is the physical mass of the sneutrinos,
$x_{\chi^-_j \tilde\nu_\alpha} = m^2_{\chi^-_j}/m^2_{\tilde\nu_\alpha}$
and $J$ a dimensionless function.
\footnote{Explicitly, $J(x) = [x^2-4x+3+2\ln(x)]/[2(x-1)^3]$.}
The gaugino and higgsino vertex factors are given by :
\begin{eqnarray}
G_{\nu L}^{j\alpha A} &=& T^c_{j1} (V^{\tilde\nu e}_{LL})_{\alpha A} \cr
H_{\nu L}^{j\alpha A} &=& {S^{c\dagger}_{2j}} (Y_e)_{AA}
                          (V^{\tilde\nu e\dagger}_{LL})_{A\alpha}
\end{eqnarray}
where
$Y_e = {\rm diag}(m_e,m_\mu,m_\tau)/
       (\sqrt{2} M_W \cos\beta) = \lambda'_e / g$.
Finally the sneutrino/charged lepton flavour matrix is defined
by $V^{\tilde\nu e}_L = S^{\tilde\nu}_{LL} T^{e\dagger}$
where $S^{\tilde\nu}_{LL}$ diagonalizes the sneutrino mass matrix and
$T^e$ is the matrix that rotates the left-handed charged leptons
into their physical mass eigenstates
(see appendix \AppendixA\ for details.)

We see from Fig.~\FigTmgChargLoop\
that the $\tau$ decay develops through the
sneutrino $\tilde\nu^\tau_L$--$\tilde\nu^\mu_L$ scalar line via the
$(\Delta^\nu_{23})_{LL}$ mass insertion.
Comparing the bottom with the tau decay, we note that while in the
former decay the presence of right-handed up-type squarks allows
the diagram in Fig.~\FigBsgChargLoops\ b) to exist,
in the latter decay, the corresponding lepton analogue of such
diagram does not exist since the effective low energy theory
does not include right-handed sneutrinos ($\tilde\nu_R$).

We can now justify why we preferred to consider non-universal
$A$-terms rather than non-universal soft SUSY masses.
If we had chosen to introduce non-universal soft squark masses
then unification of quarks and leptons in the $F$ and $F^c$
multiplets would demand similar non-universal slepton masses.
In this scenario it would be difficult to simultaneously
suppress $b\to s +\gamma$ and keep ${\rm BR}(\tau\to\mu\gamma)$
below the experimental bound.
The reason is because both decays can proceed via a
$(\Delta_{23})_{LL}$ mass insertion.
In contrast, when non-universal $A$-terms are introduced, we can control
$b\to s\gamma$ via the $(\Delta^u_{23})_{LR}$ insertion, while
leaving the prediction for ${\rm BR}(\tau\to\mu\gamma)$ approximately
unchanged because the leptonic decay proceeds via the
$(\Delta^\nu_{23})_{LL}$ insertion.

\vbox{
\vbox{
\noindent
\hfil
\vbox{
\epsfxsize=\figsize
\epsffile[130 380 510 735]{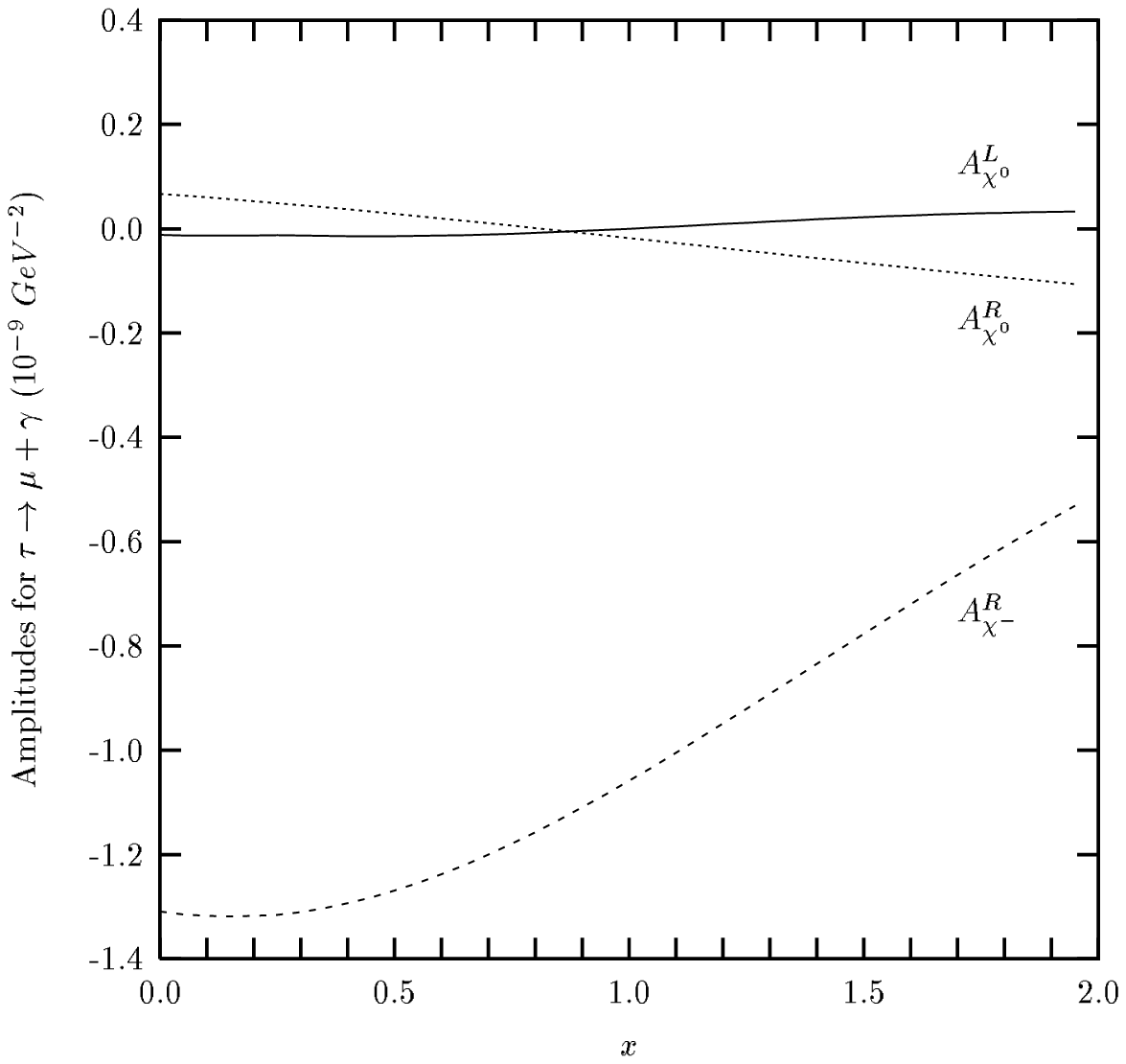}}

{\narrower\narrower\footnotesize\noindent
{FIG.~\FigTmgAmp}
Values for the chargino $A^R_\chi$ and neutralino
amplitudes $A^{L,R}_{\chi^0}$ contributing to the $\tau\to\mu\gamma$
decay against the trilinear parameter $x$ in Eq.~\refeqn{xAB}.
We observe that $A^{L,R}_{\chi^0}$ are small compared with $A^R_\chi$.
Moreover, the magnitude of $A^R_\chi$ slowly decreases as $x$ increases.
\par}}

\vbox{
\noindent
\hfil
\vbox{
\epsfxsize=\figsize
\epsffile[130 380 510 735]{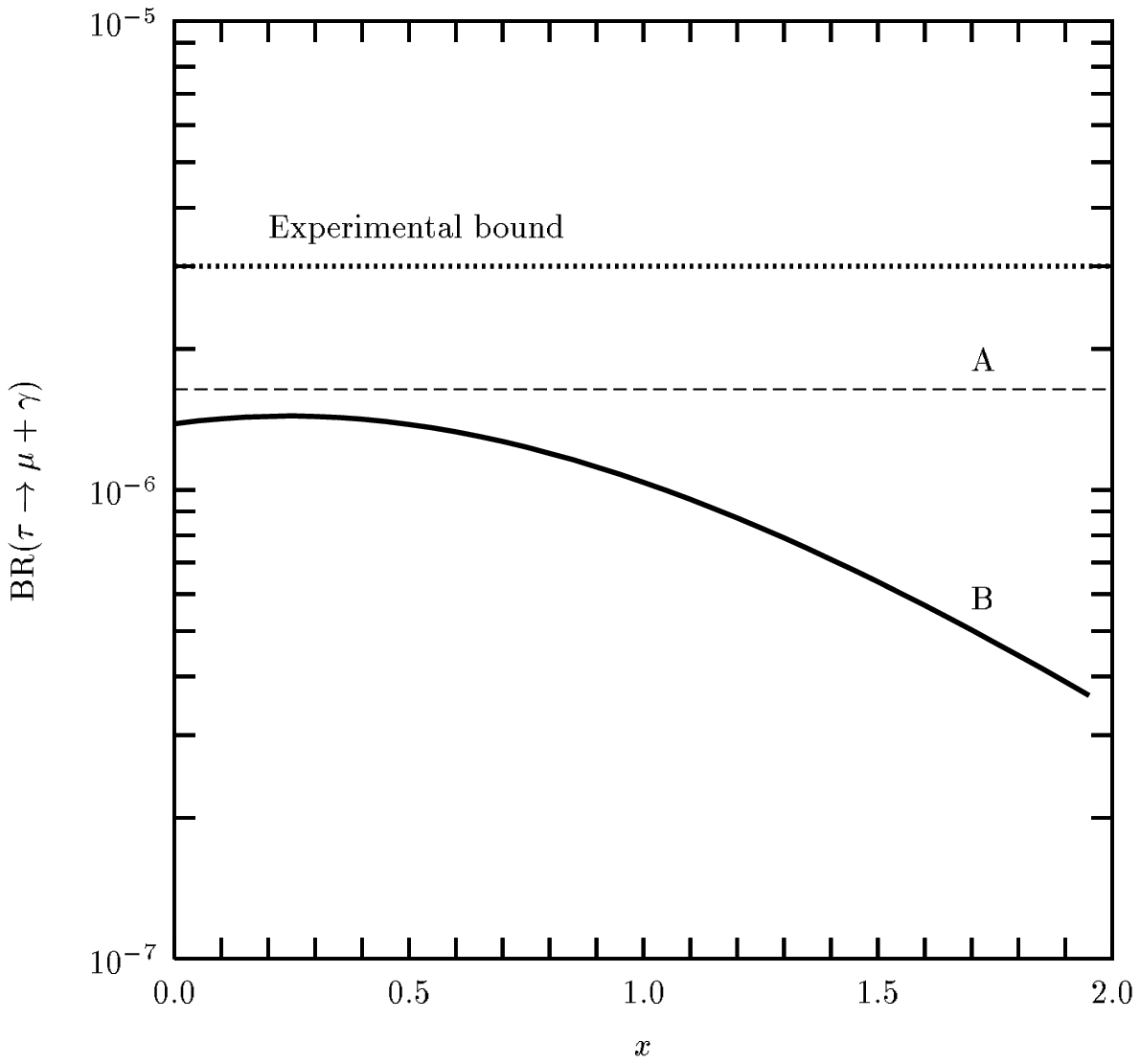}}

{\narrower\narrower\footnotesize\noindent
{FIG.~\FigTmgBR}
Branch ratio for the $\tau\to\mu\gamma$ decay against the
trilinear parameter $x$ in Eq.~\refeqn{xAB}.
The dashed line A corresponds to the result when $A_0=0$.
We observe that the solid line B is driven to smaller values
as $x$ increases. In fact, for large $x\sim 2.0$,
the predicted branch ratio is roughly one order of magnitude
smaller than the present experimental bound (here indicated
by the dotted line.)
\par}}}

In Fig.~\FigTmgAmp\ and Fig.~\FigTmgBR\ we show how the amplitudes and the
branch ratio of the $\tau\to\mu\gamma$ decay depend
on the trilinear parameter $x$ in Eq.~\refeqn{xAB}.
We see that, for $x\sim 0$, the prediction
for the tau decay is close to the experimental limit
${\rm BR}(\tau\to\mu\gamma) < 3.0 \times 10^{-6}$.
However as $x$ increases the magnitude of the chargino amplitude
$|A_{\chi^-}|$ slowly decreases thus driving the prediction for
${\rm BR}(\tau\to\mu\gamma)$ to smaller values.
For $x\sim 1.60$ the predicted  ${\rm BR}(\tau\to\mu\gamma)$
is about one order of magnitude below the experimental upper bound.

In summary, we have verified that by introducing non-universal $A$-terms
one can simultaneously satisfy the present experimental constraints
on $b\to s\gamma$ and $\tau\to\mu\gamma$ even when $\mu$ is negative.

Finally we point that the success in suppressing the
$b\to s\gamma$ decay is not without some tuning.
Indeed we found that the suppression works well only
for a restricted range of $x\sim 1.60\pm 0.15$.
Thus taking as a measure of tuning $x$ the ratio
$\delta=\delta x / x$ we find $\delta\sim 19$ \%.
Nevertheless, it is unclear if it is more natural to expect
$b\to s\gamma$ to be suppressed by flavour physics, implying tuning
in the flavour parameters, or by suppressing the charged Higgs and
chargino amplitudes by demanding large $m_{H^-}$ and stop masses,
implying tuning in setting $M_Z$.

\newpage

\SECTION{VII. CONCLUSION}

We have studied Yukawa unification,
including the effects of a physical neutrino mass consistent
with the Superkamiokande observations.
We began our study by reviewing the usual mSUGRA scenario
with universal soft mass parameters, but including the
effect of the neutrino Yukawa coupling.
Assuming hierarchical neutrino masses, and mixing angles
arising mainly from the neutrino sector (so that the charged lepton
Yukawa matrix has no large off-diagonal entries) 
we saw that the
usual predictions are not much affected. 
\footnote{Large off-diagonal entries in the neutrino Yukawa
matrix will not affect our results very much since the
right-handed neutrinos decouple from the
right-hand sides of the RGEs at high energy.}
For example the
usual result that 
positive $\mu$ is not allowed, and negative $\mu$ leads to 
top quark masses which are too small is still valid.

We then analysed a string/$D$-brane inspired \422 model since this
allows the most general non-universal scalar and gaugino masses,
and is therefore the perfect laboratory for studying the effects
of non-universality.
We explored the sensitivity of the predictions to variations
in all these non-universal soft masses, and showed that the usual
results can change considerably as a function of the
degree of non-universality of the soft parameters.
We then switched on the usual $D$-term contributions which arise in
$SO(10)$, and which by themselves are enough to
allow successful electroweak symmetry breaking, and
permit small corrections to the $b$-quark mass which implies
an acceptable large top quark mass.
We studied the effect of the $D$-terms in the \422 model,
for the sake of clarity working in the (approximate) $SO(10)$ limit of the
model, in order to distinguish clearly the effect of the D-terms
from that of explicit non-universality. Including D-terms
as the only source of non-universality, we studied the
sparticle spectrum in some detail, although these results should be
considered in conjunction with the effect of the explicit non-universal
scalar masses and non-universal gaugino masses considered previously,
since in a realistic \422 model both effects will simulaneously be present.

We then turned to rare decays such as $b \rightarrow s \gamma$
and $\tau \rightarrow \mu \gamma$
which severely constrain Yukawa unification.
Assuming that the only source of non-universality is 
due to $D$-terms, we showed how
non-universal trilinear parameters 
can lead to cancellations of the important effects
and so provide information about the family-dependent supersymmetry
breaking soft Lagrangian. Again, the effect of more general non-universal
scalar masses and non-universal gaugino masses 
associated with the \422 model will modify the results,
but the main message remains clear: family-dependent non-universality
will play an important role in $b \rightarrow s \gamma$, and such
effects should be considered in conjunction with the effects
of family-independent non-universality.
We have shown that successful Yukawa unification can be achieved
if both types of non-universality are simultaneously present.

In summary Yukawa unification is well motivated from
both a theoretical point of view, and from the point of view
of predicting large $\tan \beta$ which helps to raise the Higgs mass.
We have found that Yukawa unification is perfectly viable providing the
soft masses are non-universal in both family-independent
and family-dependent ways, and we have explored the specific correlations 
between the soft parameters required in order to satisfy
all the constraints simultaneously. 
In our view the sensitivity of Yukawa unification to soft SUSY
breaking parameters is to be welcomed, since it provides
a window into the soft supersymmetry breaking Lagrangian.

\SECTION{ACKNOWLEDGMENTS}
The work of M.O. was supported by JNICT under contract grant :
PRAXIS XXI/BD/ 5536/95.

\newpage

\SECTION{APPENDIX \AppendixA \\
         Notation and Conventions}

In this appendix we briefly summarize our conventions and
explain in detail the notation concerning the diagonalization of
the mass matrices.
The superpotential of the MSSM+$\nu^c$ model is given by :
\begin{eqnarray}
{\cal W} =
\epsilon_{\alpha\beta}
(&& \!\!\!\!\!\!\!\!\!\!
     \phantom{+}
      u^c_A (\lambda_u)_{AB} q^\alpha_B h^\beta_u
     -d^c_A (\lambda_d)_{AB} q^\alpha_B h^\beta_d \nonumber \\
 && \!\!\!\!\!\!\!\!\!\!
     +\nu^c_A (\lambda_\nu)_{AB} l^\alpha_B h^\beta_u
     -e^c_A (\lambda_e)_{AB} l^\alpha_B h^\beta_d
     +\mu h_u^\alpha h_d^\beta) +
{\textstyle {1\over 2}} (M_\nu)_{AB} \nu^c_A \nu^c_B
\end{eqnarray}
where $\epsilon_{12} = - \epsilon_{21} = 1$,
$A,B = 1,..,3$ are flavour indices and
$\alpha,\beta=1,2$ are $SU(2)_L$ indices.
The Yukawa matrices are diagonalized by the following transformations~:
\begin{equation}
S^u \lambda_u T^{u\dagger} = \lambda^\prime_u\quad
S^d \lambda_d T^{d\dagger} = \lambda^\prime_d\quad
S^e \lambda_e T^{e\dagger} = \lambda^\prime_e\quad
S^\nu \lambda_\nu T^{\nu\dagger} = \lambda^\prime_\nu
\end{equation}
where the primed $\lambda'$s are diagonal.
In this notation the CKM matrix is given by $V=T^u T^{d\dagger}$.
The full lagrangian of the model also includes
trilinear, soft scalar masses and gaugino masses given by~:
\begin{eqnarray}
{\cal V} =
\epsilon_{\alpha\beta} (
&& \!\!\!\!\!\!\!\!\!\!
\phantom{+}
 \tilde u^c_A (\tilde A_u)_{AB} \tilde q^\alpha_B h^\beta_u
-\tilde d^c_A (\tilde A_d)_{AB} \tilde q^\alpha_B h^\beta_d \nonumber \\
&& \!\!\!\!\!\!\!\!\!\!
+\tilde \nu^c_A (\tilde A_\nu)_{AB} \tilde l^\alpha_B h^\beta_u
-\tilde e^c_A (\tilde A_e)_{AB} \tilde l^\alpha_B h^\beta_d
+m_3 h_u^\alpha h_d^\beta)+{\rm h.c.}
\end{eqnarray}
\begin{eqnarray}
{\cal L} = {\textstyle {1 \over 2}} (
&& \!\!\!\!\!\!\!\!\!\!
m_2^2 |h_u|^2+
m_1^2 |h_d|^2+
\tilde q^*_A (\tilde m^2_q)_{AB} \tilde q_B+
\tilde l^*_A (\tilde m^2_l)_{AB} \tilde l_B+ \nonumber \\
&& \!\!\!\!\!\!\!\!\!\!
\tilde u_A^c (\tilde m^2_{u^c})_{AB} \tilde u^{c*}_B+
\tilde d_A^c (\tilde m^2_{d^c})_{AB} \tilde d^{c*}_B+
\tilde e_A^c (\tilde m^2_{e^c})_{AB} \tilde e^{c*}_B+
\tilde \nu_A^c (\tilde m^2_{\nu^c})_{AB} \tilde \nu^{c*}_B) + {\rm h.c.}
\end{eqnarray}
\begin{equation}
{\cal L} =
{\textstyle {1 \over 2}} M_1 \lambda_1 \lambda_1 +
{\textstyle {1 \over 2}} M_2 \lambda^a_2 \lambda^a_2 +
{\textstyle {1 \over 2}} M_3 \lambda^m_3 \lambda^m_3 + {\rm h.c.}
\end{equation}
The chargino and neutralino mass matrices can be
conveniently written in the basis of the following 4-component
gaugino and higgsino fields :
\footnote{$\lambda^\pm = (\lambda^1_2\mp i\lambda^2_2)/ \sqrt{2}$}
\begin{equation}
\tilde W^- = \left(\matrix{ -i\lambda^- \cr
                             i{\bar\lambda}^+}\right)\quad
\tilde H^- = \left(\matrix{ \tilde H^2_d \cr
                             i{\bar{\tilde H}}^1_u}\right)\quad
\end{equation}
\begin{equation}
\tilde B   = \left(\matrix{ -i\lambda_1 \cr
                             i{\bar\lambda}_1}\right)\quad
\tilde W^0 = \left(\matrix{ -i\lambda^3_2 \cr
                             i{\bar\lambda}^3_2}\right)\quad
\tilde H^0_d = \left(\matrix{ \tilde H^1_d \cr
                              {\bar{\tilde H}}^1_d}\right)\quad
\tilde H^0_u = \left(\matrix{ \tilde H^2_u \cr
                              {\bar{\tilde H}}^2_u}\right).
\end{equation}
Thus,
\begin{equation}
{\cal L} = -
\left(\bar{\tilde W}^-_L \> \bar{\tilde H}^-_L \right)
\left(\matrix{
M_2 & \sqrt{2} M_W s_\beta \cr
\sqrt{2} M_W c_\beta & \mu  \cr} \right)
\left( \matrix{ \tilde W^-_R \cr \tilde H^-_R} \right) + {\rm h.c.}
\label{ChargMassMatrix}
\end{equation}
and,
\begin{equation}
{\cal L} = - {1 \over 2}
\left(\bar{\tilde B} \>
      \bar{\tilde W}^0 \>
      \bar{\tilde H}^1_d \>
      \bar{\tilde H}^2_u\right)
\left(\matrix{ M_1 & 0 & -M_Z c_\beta s_\theta &  M_Z s_\beta s_\theta \cr
               0 & M_2 &  M_Z c_\beta c_\theta & -M_Z s_\beta c_\theta \cr
               -M_Z c_\beta s_\theta &  M_Z c_\beta c_\theta & 0 & -\mu \cr
                M_Z s_\beta s_\theta & -M_Z s_\beta c_\theta & -\mu & 0 \cr}
      \right)
\left(\matrix{ \tilde B \cr
               \tilde W^0 \cr
               \tilde H^1_d \cr
               \tilde H^2_u }\right)
\label{NeutMassMatrix}
\end{equation}
where $s_\beta = \sin\beta$, $c_\beta = \cos\beta$
($\tan\beta = v_2/v_1$ = the ratio of the up/down Higgs VEVs)
and $\theta$ the weak mixing angle.
The chargino mass matrix $M^C$ in Eq.~\refeqn{ChargMassMatrix}
and the neutralino matrix $M^N$ in Eq.~\refeqn{NeutMassMatrix}
are diagonalized by :
\begin{eqnarray}
S^C M^C T^{C\dagger} &=& {\rm Diag}(m_{\chi^-_1},
                                    m_{\chi^-_2}) \\
S^N M^N S^{N\dagger} &=& {\rm Diag}(m_{\chi^0_1},
                                    m_{\chi^0_2},
                                    m_{\chi^0_3},
                                    m_{\chi^0_4})
\end{eqnarray}
\noindent
The mass matrices for the charged scalar sparticles are written
in the following basis~:
\begin{equation}
\left(\matrix{ \tilde u_L \cr \tilde u_R }\right) =
\left(\matrix{ \tilde u \hfill \cr \tilde u^{c*}}\right)
\quad\quad
\left(\matrix{ \tilde d_L \cr \tilde d_R }\right) =
\left(\matrix{ \tilde d \hfill \cr \tilde d^{c*}}\right)
\quad\quad
\left(\matrix{ \tilde e_L \cr \tilde e_R }\right) =
\left(\matrix{ \tilde e \hfill \cr \tilde e^{c*}}\right)
\end{equation}
Explicitly we find :
\begin{equation}
\left(\tilde u^\dagger_L \> \tilde u^\dagger_R \right)
\left(
\matrix{
\tilde m^2_q + m_u^\dagger m_u + M_Z^2 Z_{u} c_{2\beta} &
-\mu\lambda^\dagger_u v_1+\tilde A^\dagger_u v_2  \cr
-\mu\lambda_u v_1+\tilde A_u v_2 &
\tilde m^2_{u^c} + m_u m_u^\dagger + M_Z^2 Z_{u^c} c_{2\beta}
}\right)
\left(\matrix{\tilde u_L \cr \tilde u_R}\right)
\label{SupMassMatrix}
\end{equation}
\begin{equation}
\left(\tilde d^\dagger_L \> \tilde d^\dagger_R \right)
\left(
\matrix{
\tilde m^2_q + m_d^\dagger m_d + M_Z^2 Z_{d} c_{2\beta} &
-\mu\lambda^\dagger_d v_2+\tilde A^\dagger_d v_1  \cr
-\mu\lambda_d v_2+\tilde A_d v_1 &
\tilde m^2_{d^c} + m_d m_d^\dagger + M_Z^2 Z_{d^c} c_{2\beta}
}\right)
\left(\matrix{\tilde d_L \cr \tilde d_R}\right)
\label{SdownMassMatrix}
\end{equation}
\begin{equation}
\left(\tilde e^\dagger_L \> \tilde e^\dagger_R \right)
\left(
\matrix{
\tilde m^2_l + m_e^\dagger m_e + M_Z^2 Z_{e} c_{2\beta} &
-\mu\lambda^\dagger_e v_2+\tilde A^\dagger_e v_1  \cr
-\mu\lambda_e v_2+\tilde A_e v_1 &
\tilde m^2_{e^c} + m_e m_e^\dagger + M_Z^2 Z_{e^c} c_{2\beta}
}\right)
\left(\matrix{\tilde e_L \cr \tilde e_R}\right)
\label{SleptonMassMatrix}
\end{equation}
\noindent
The light sneutrino mass matrix in the $\tilde \nu_L =\tilde\nu$
basis, after the heavy right-handed sneutrinos are
integrated out, is given by :
\begin{equation}
\tilde \nu_L^\dagger
\left( \tilde m^2_l+M_Z^2 Z_{\nu_L} c_{2\beta} \right)
\tilde \nu_L
\label{SneutrinoMassMatrix}
\end{equation}
\noindent
The Z factors in Eqs.\refeqn{SupMassMatrix}-\refeqn{SneutrinoMassMatrix}
are defined by $Z_f = I_f-Q_f s^2_\theta$
where $I_f$ is the isospin and
      $Q_f$ the electric charge of the $f$ field :
\begin{equation}
\matrix{
Z_u     = ({\textstyle +{1 \over 2}})-({\textstyle +{2 \over 3}}) s^2_\theta &
Z_{u^c} = ({\textstyle  {    0    }})-({\textstyle -{2 \over 3}}) s^2_\theta \cr
& \cr
Z_d     = ({\textstyle -{1 \over 2}})-({\textstyle -{1 \over 3}}) s^2_\theta &
Z_{d^c} = ({\textstyle  {    0    }})-({\textstyle +{1 \over 3}}) s^2_\theta \cr
& \cr
Z_e     = ({\textstyle -{1 \over 2}})-({\textstyle -{    1    }}) s^2_\theta &
Z_{e^c} = ({\textstyle  {    0    }})-({\textstyle +{    1    }}) s^2_\theta \cr
& \cr
Z_\nu   = ({\textstyle +{1 \over 2}})-({\textstyle  {\phantom{+}0}}) s^2_\theta &
\cr
}
\end{equation}

\noindent
The diagonalization of the up-type squark squared mass matrix
$M^{\tilde u2}$ of Eq.~\refeqn{SupMassMatrix},
down-type squark matrix $M^{\tilde d2}$ of Eq.~\refeqn{SdownMassMatrix},
charged slepton matrix $M^{\tilde e2}$ of Eq.~\refeqn{SleptonMassMatrix}
and sneutrino matrix of Eq.~\refeqn{SneutrinoMassMatrix}
is achieved in the following way :
\begin{eqnarray}
S^{\tilde u} M^{\tilde u 2} S^{\tilde u\dagger} &=&
{\rm Diag}(m_{\tilde U_1},..,m_{\tilde U_6}) \\
&=&
{\rm Diag}(m_{\tilde u_1},m_{\tilde c_1},m_{\tilde t_1},
           m_{\tilde u_2},m_{\tilde c_2},m_{\tilde t_2}) \\
S^{\tilde d} M^{\tilde d 2} S^{\tilde d\dagger} &=&
{\rm Diag}(m_{\tilde D_1},..,m_{\tilde D_6}) \\
&=&
{\rm Diag}(m_{\tilde d_1},m_{\tilde s_1},m_{\tilde b_1},
           m_{\tilde d_2},m_{\tilde s_2},m_{\tilde b_2}) \\
S^{\tilde e} M^{\tilde e 2} S^{\tilde e\dagger} &=&
{\rm Diag}(m_{\tilde E_1},..,m_{\tilde E_6}) \\
&=&
{\rm Diag}(m_{\tilde e_1},m_{\tilde \mu_1},m_{\tilde \tau_1},
           m_{\tilde e_2},m_{\tilde \mu_2},m_{\tilde \tau_2}) \\
S^{\tilde\nu}_{LL} M^{\tilde\nu 2}_{LL} S^{\tilde\nu \dagger}_{LL} &=&
{\rm Diag}(m_{\tilde\nu_1},m_{\tilde\nu_2},m_{\tilde\nu_3})
\end{eqnarray}.

Finally it is convenient to define the following matrices that
are, in a way, supersymmetric generalizations of the CKM matrix,
\ie they describe the flavour properties of vertices that involve
the interaction between a SUSY scalar particle and
a standard model fermion :
\begin{equation}
\matrix{
(V^{\tilde u u}_R)_{\alpha A} = \sum_{B=1}^3
S^{\tilde u}_{\alpha B+3} S^{u\dagger}_{BA} \hfill &
(V^{\tilde u u\dagger}_R)_{A \alpha} = \sum_{B=1}^3
S^u_{AB} S^{\tilde u\dagger}_{B+3 \alpha} \hfill \cr
& \cr
(V^{\tilde d d}_L)_{\alpha A} = \sum_{B=1}^3
S^{\tilde d}_{\alpha B} T^{d\dagger}_{BA} \hfill &
(V^{\tilde d d\dagger}_L)_{A \alpha} = \sum_{B=1}^3
T^d_{AB} S^{\tilde d\dagger}_{B\alpha} \hfill \cr
& \cr
(V^{\tilde d d}_R)_{\alpha A} = \sum_{B=1}^3
S^{\tilde d}_{\alpha B+3} S^{d\dagger}_{BA} \hfill &
(V^{\tilde d d\dagger}_R)_{A \alpha} = \sum_{B=1}^3
S^d_{AB} S^{\tilde d\dagger}_{B+3 \alpha} \hfill \cr
& \cr
(V^{\tilde u d}_L)_{\alpha A} = \sum_{B=1}^3
S^{\tilde u}_{\alpha B} T^{d\dagger}_{BA} \hfill &
(V^{\tilde u d\dagger}_L)_{A\alpha} = \sum_{B=1}^3
T^d_{AB} S^{\tilde u\dagger}_{B\alpha} \hfill \cr
& \cr
(V^{\tilde \nu e}_{LL})_{\alpha A} = \sum_{B=1}^3
(S^{\tilde\nu}_{LL})_{\alpha B} T^{e\dagger}_{BA} \hfill &
(V^{\tilde \nu e\dagger}_{LL})_{A \alpha} = \sum_{B=1}^3
T^e_{AB}(S^{\tilde\nu\dagger}_{LL})_{B\alpha} \hfill \cr}
\end{equation}

\newpage

\SECTION{APPENDIX \AppendixB \\
         The supersymmetry mass scale}

In this section we will define the effective SUSY scale $M_S$ which we
used to represent the average sparticle masses. Above $M_S$ the theory
was described by the MSSM+$\nu^c$ and below it we had the Standard Model.
Traditionally it is convenient to introduce three $T_i$ scales
that describe the effect of the decoupling of the SUSY particles on
the gauge couplings \cite{BDWright} :
\begin{eqnarray}
T_1 &=& m_{\tilde q}^{11/25}
        m_{\tilde l}^{9/25}
        m_{\tilde H}^{4/25}
        m_H^{1/25}
        \label{T1} \\
T_2 &=& m_{\tilde q}^{9/25}
        m_{\tilde W}^{8/25}
        m_{\tilde H}^{4/25}
        m_{\tilde l}^{3/25}
        m_H^{1/25}
        \label{T2} \\
T_3 &=& m_{\tilde q}^{1/2}
        m_{\tilde g}^{1/2}
        \label{T3}
\end{eqnarray}
where $m_{\tilde q}$, $m_{\tilde l}$, $m_{\tilde H}$, $m_H$,
$m_{\tilde W}$ and $m_{\tilde g}$ are the squark, slepton, higgsino,
heavy CP-even Higgs boson, wino and gluino masses respectively.
In models with gauge unification at $M_X$ the
prediction for $\alpha_s(M_{\rm Z})$ depends on $M_{SUSY}$
given by \cite{LaPo2} :
\begin{equation}
M_{SUSY} = T_1^{-25/19}\>
           T_2^{100/19}\>
           T_3^{-56/19}
\label{MSUSY}
\end{equation}
Combining the equations above we obtain \cite{CaPoWa} :
\begin{equation}
M_{SUSY} = m_{\tilde H}
              \left( \frac{m_{\tilde W}}{m_{\tilde g}} \right)^{28/19}
              \left[
              \left( \frac{m_{\tilde l}}{m_{\tilde q}} \right)^{3/19}
              \left( \frac{m_H}         {m_{\tilde H}} \right)^{3/19}
              \left( \frac{m_{\tilde W}}{m_{\tilde H}} \right)^{4/19}
              \right]
\label{msusy}
\end{equation}

In this article we did not use the above scale to describe the average
sparticle masses since as was emphasized in
Refs.~\cite{LaPo2,CaPoWa}, $M_{SUSY}$ is only related to the overall
sparticle masses in the unlikely case of degenerate SUSY spectrum.
Instead, we introduced a new scale $M_S$ which was defined such that
the sum of the squares of the threshold corrections induced
by the decoupling of the SUSY spectrum on the gauge couplings is
minimal :
\begin{equation}
{\partial \over \partial M_S}
\left\{
\sum_{i=1}^3
\frac{B_i}{2 \pi} \ln \left(M_S \over T_i \right)
\right\}^2 = 0
\qquad
{\rm where}
\qquad
\textstyle
B_i=({5\over 2},{25\over 6},4)
\end{equation}
Explicitly we find that $M_S$ is given by :
\begin{eqnarray}
M_{\rm S} &=& T_1^{225/1426}\>
              T_2^{625/1426}\>
              T_3^{576/1426} \label{MS} \\
          &=& m_{\tilde q}^{612/1426} \>
              m_{\tilde g}^{228/1426} \>
              m_{\tilde W}^{200/1426} \>
              m_{\tilde l}^{156/1426} \>
              m_{\tilde H}^{136/1426} \>
              m_H         ^{ 34/1426} \label{ms}
\end{eqnarray}
where the exponent values of the $m_{\tilde q}$,..,$m_H$ terms are :
0.43, 0.20, 0.14, 0.11, 0.10 and 0.02 respectively.
It is interesting to note that the unphysical nature of
$M_{SUSY}$
can be immediately identified through
the appearance of negative exponents (see Eq.~\refeqn{MSUSY})
whereas the $T_i$'s in Eqs.~\refeqn{T1}-\refeqn{T3}
and $M_S$ in Eq.~\refeqn{MS} and Eq.~\refeqn{ms}
are weighed by positive numbers.

In Table~\TabMsPred\ we list the low energy values of the masses
of the gauginos, the effective SUSY scales in
Eqs.~\refeqn{T1}-\refeqn{T3}, and the value of the new supersymmetry
scale $M_S$ defined in Eq.~\refeqn{ms} against the often used
$M_{SUSY}$ of Eq.~\refeqn{msusy}.
We observe that while $M_S$ is a good average of the $T_i$'s,
$M_{SUSY}$ fails to represent a meaningful effective SUSY mass.

\vbox{
\begin{center}
\begin{tabular}{crrrrrrrr}
\multicolumn{9}{c}{TABLE \TabMsPred.} \cr
\noalign{\medskip}
\noalign{\hrule height\rulerheight}
\noalign{\smallskip}
\noalign{\hrule height\rulerheight}
\noalign{\medskip}
Case &
$m_{\tilde B}$ &
$m_{\tilde W}$ &
$m_{\tilde g}$ &
$T_1$ &
$T_2$ &
$T_3$ &
$M_{S}$ &
$M_{SUSY}$ \\
\noalign{\medskip}
\noalign{\hrule height\rulerheight}
\noalign{\medskip}
A & 173 & 334 &  942 & 440 & 454 &  874 &  588 &  68 $\>$ \null \\
B & 356 & 674 & 1787 & 851 & 887 & 1650 & 1132 & 150 $\>$ \null \\
C & 174 & 334 &  942 & 438 & 433 &  879 &  577 &  53 $\>$ \null \\
D & 356 & 674 & 1789 & 818 & 820 & 1662 & 1091 & 103 $\>$ \null \\
E & 173 & 334 &  976 & 449 & 474 &  903 &  610 &  76 $\>$ \null \\
F & 355 & 673 & 1845 & 889 & 935 & 1700 & 1181 & 172 $\>$ \null \\
G & 173 & 334 &  976 & 451 & 466 &  907 &  607 &  68 $\>$ \null \\
H & 356 & 673 & 1845 & 876 & 911 & 1707 & 1166 & 150 $\>$ \null \\
I & 264 & 503 & 1415 & 718 & 714 & 1319 &  915 & 116 $\>$ \null \\
\noalign{\medskip}
\noalign{\hrule height\rulerheight}
\noalign{\smallskip}
\noalign{\hrule height\rulerheight}
\end{tabular}
\end{center}

{\narrower\narrower\footnotesize\noindent
{TABLE \TabMsPred.}
Predicted values for the bino ($m_{\tilde B}$), wino ($m_{\tilde W}$)
and gluino ($m_{\tilde g}$) masses at low energies and for
the three SUSY scales $T_i$ and the effective
$M_{S}$, $M_{SUSY}$ scales defined in Eqs.~\refeqn{ms},\refeqn{msusy}
(masses given in GeV units.).
The first column indicates the input for each model as defined by
list of values in Table \TabCasesDef.
\par\bigskip}}

\newpage

\SECTION{APPENDIX C \\
         Tables for the SUSY corrections}

In this appendix we present some examples which
illustrate the magnitude of the SUSY corrections
to the bottom and tau mass.

In Table~\TabGluinoCorr\ we systematically list the values of all
the parameters, evaluated at low energy,
that are needed to compute the gluino correction.
The columns refer to~:
the input taken according to Table~\TabCasesDef;
the value of the strong gauge coupling;
the ratio of the MSSM VEVs;
the mass of the gluino; the value of the Higgs mixing parameter $\mu$;
the masses of the physical bottom squarks;
the value of the dimensionless function
$I_{\tilde g}= m_{\tilde g} \mu
 I(m_{\tilde g}^2, m_{\tilde b_1}^2,m_{\tilde b_2}^2 )$;
\footnote{
$
I(x,y,z) = - {xy\ln(x/y)+yz\ln(y/z)+zx\ln(z/x) \over (x-y)(y-z)(z-x)}
$
}
and finally, in the last column, the value of
the gluino correction (in percentage \%.)
In Table~\TabHiggsinoCorr\ we list similar
values appropriate for the evaluation of the higgsino correction.
For example, $\alpha_t = \lambda_t^2 / 4\pi$,
$m_{\tilde t_{1,2}}$ are stop masses and
$I_{\tilde H}= A_t \mu I(\mu^2, m_{\tilde t_1}^2,m_{\tilde t_2}^2 )$.
Comparing Table~\TabGluinoCorr\ with Table~\TabHiggsinoCorr\ we read that
$\Delta^{\tilde g} m_b \sim -20 \%$ dominates
over $\Delta^{\tilde H} m_b \sim 6~\%$.

In Table~\TabBinoCorr\ we present, in the last column,
the values for the bino corrections to the tau mass $\Delta^{\tilde B} m_\tau$.
In this table we also give the values for $\alpha' = g'^2/4\pi$,
for the bino mass $m_{\tilde B}$, for the stau masses and for
$I_{\tilde B}= m_{\tilde B} \mu I(m_{\tilde B}^2,
 m_{\tilde\tau_1}^2,m_{\tilde\tau_2}^2)$.
We read that, on average, $\Delta^{\tilde B} m_\tau$ is -2.45 \%.

Finally we computed
$\Delta m_d$ and $\Delta m_s$ which we
show in Table \TabSusyCorr. Two comments deserve attention.
Firstly, $\Delta^{\tilde H} m_{d,s} \simeq 0$
due to small $\lambda_{d,s}$ Yukawa couplings.
Secondly, we see that the gluino contribution is not universal,
{\it i.e.} $\Delta^{\tilde g} m_d \simeq \Delta^{\tilde g}m_s
\not\simeq \Delta^{\tilde g}m_b$ due to
$m_{\tilde d, \tilde s} \not\simeq m_{\tilde b}$.
In conjunction, they lead to non-universal SUSY corrections
$\Delta m_{d,s} \not\simeq \Delta m_b$, thus
slightly affecting the ratio $\lambda_{d,s}(M_X)/\lambda_b(M_X)$
\cite{BlRaPo}.

\vbox{
\begin{center}
\begin{tabular}{cccrrrrcc}
\multicolumn{9}{c}{TABLE \TabGluinoCorr.} \cr
\noalign{\medskip}
\noalign{\hrule height\rulerheight}
\noalign{\smallskip}
\noalign{\hrule height\rulerheight}
\noalign{\medskip}
Case &
$\alpha_3$ &
$\tan \beta$ &
$m_{\tilde g}$ &
$\mu$ &
$m_{\tilde b_1}$ &
$m_{\tilde b_2}$ &
$I_{\tilde g}$ &
$\Delta^{\tilde g} m_b$ \cr
\noalign{\medskip}
\noalign{\hrule height\rulerheight}
\noalign{\medskip}
A & 0.0904 & 42.16 &  942 & -417 &  775 &  691 & -0.305 & -24.65 \cr
B & 0.0848 & 44.16 & 1787 & -795 & 1431 & 1316 & -0.311 & -24.72 \cr
C & 0.0906 & 35.94 &  942 & -339 &  790 &  725 & -0.238 & -16.43 \cr
D & 0.0850 & 37.51 & 1789 & -636 & 1475 & 1373 & -0.238 & -16.11 \cr
E & 0.0960 & 47.15 &  976 & -494 &  794 &  692 & -0.358 & -34.36 \cr
F & 0.0896 & 49.40 & 1845 & -942 & 1452 & 1323 & -0.367 & -34.44 \cr
G & 0.0960 & 41.99 &  976 & -445 &  804 &  719 & -0.313 & -26.81 \cr
H & 0.0897 & 44.06 & 1845 & -847 & 1480 & 1364 & -0.320 & -26.82 \cr
I & 0.0920 & 47.34 & 1415 & -703 & 1144 & 1035 & -0.347 & -32.03 \cr
\noalign{\medskip}
\noalign{\hrule height\rulerheight}
\noalign{\smallskip}
\noalign{\hrule height\rulerheight}
\end{tabular}
\end{center}

{\narrower\narrower\footnotesize\noindent
{TABLE \TabGluinoCorr.}
Values of the parameters required to compute the gluino correction to the
bottom quark mass $\Delta^{\tilde g} m_b \sim -20\%$
(all evaluated at low energy.)
The gluino mass is given by $m_{\tilde g}$,
the sbottom masses by $m_{\tilde b_{1,2}}$ and the dimensionless
function $I_{\tilde g}$ is given by
$I_{\tilde g} = m_{\tilde g} \mu
                I(m_{\tilde g}^2, m_{\tilde b_1}^2,m_{\tilde b_2}^2 )$.
\par\bigskip}}

\vbox{
\begin{center}
\begin{tabular}{cccrrrrcc}
\multicolumn{9}{c}{TABLE \TabHiggsinoCorr.} \cr
\noalign{\medskip}
\noalign{\hrule height\rulerheight}
\noalign{\smallskip}
\noalign{\hrule height\rulerheight}
\noalign{\medskip}
Case &
$\alpha_t$ &
$\tan \beta$ &
$A_t$ &
$\mu$ &
$m_{\tilde t_1}$ &
$m_{\tilde t_2}$ &
$I_{\tilde H}$ &
$\Delta^{\tilde H} m_b$ \cr
\noalign{\medskip}
\noalign{\hrule height\rulerheight}
\noalign{\medskip}
A & 0.0476 & 42.16 &  -780 & -417 &  812 &  665 & 0.408 & 6.51 \cr
B & 0.0482 & 44.16 & -1410 & -795 & 1457 & 1291 & 0.401 & 6.80 \cr
C & 0.0363 & 35.94 &  -879 & -339 &  831 &  690 & 0.385 & 4.00 \cr
D & 0.0364 & 37.51 & -1605 & -636 & 1504 & 1345 & 0.376 & 4.09 \cr
E & 0.0597 & 47.15 &  -728 & -494 &  826 &  676 & 0.406 & 9.10 \cr
F & 0.0602 & 49.40 & -1302 & -942 & 1472 & 1303 & 0.399 & 9.45 \cr
G & 0.0499 & 41.99 &  -814 & -445 &  842 &  692 & 0.416 & 6.94 \cr
H & 0.0504 & 44.06 & -1465 & -847 & 1506 & 1338 & 0.409 & 7.23 \cr
I & 0.0573 & 47.34 & -1055 & -703 & 1170 & 1013 & 0.400 & 8.64 \cr
\noalign{\medskip}
\noalign{\hrule height\rulerheight}
\noalign{\smallskip}
\noalign{\hrule height\rulerheight}
\end{tabular}
\end{center}

{\narrower\narrower\footnotesize\noindent
{TABLE \TabHiggsinoCorr.}
Values of the parameters required to compute the higgsino correction to the
bottom quark mass $\Delta^{\tilde H} m_b \sim 6~\%$
(all evaluated at low energy.)
The higgsino mass is approximately given by $|\mu|$,
the stop masses are given by $m_{\tilde t_{1,2}}$ and the dimensionless
function $I_{\tilde H}$ is given by
$I_{\tilde H} = A_t \mu
                I(\mu^2, m_{\tilde t_1}^2,m_{\tilde t_2}^2 )$.
\par\bigskip}}

\vbox{
\begin{center}
\begin{tabular}{cccrrrrcc}
\multicolumn{9}{c}{TABLE \TabBinoCorr.} \cr
\noalign{\medskip}
\noalign{\hrule height\rulerheight}
\noalign{\smallskip}
\noalign{\hrule height\rulerheight}
\noalign{\medskip}
Case &
$\alpha'$ &
$\tan \beta$ &
$m_{\tilde B}$ &
$\mu$ &
$m_{\tilde \tau_1}$ &
$m_{\tilde \tau_2}$ &
$I_{\tilde B}$ &
$\Delta^{\tilde B} m_\tau$ \cr
\noalign{\medskip}
\noalign{\hrule height\rulerheight}
\noalign{\medskip}
A & 0.0104 & 42.16 & 173 & -417 &  348 &  178 & -0.669 & -2.33 \cr
B & 0.0105 & 44.16 & 356 & -795 &  653 &  387 & -0.656 & -2.42 \cr
C & 0.0104 & 35.94 & 174 & -339 &  343 &  207 & -0.507 & -1.51 \cr
D & 0.0105 & 37.51 & 356 & -636 &  657 &  424 & -0.494 & -1.55 \cr
E & 0.0104 & 47.15 & 173 & -494 &  355 &  143 & -0.860 & -3.36 \cr
F & 0.0105 & 49.40 & 355 & -942 &  652 &  352 & -0.818 & -3.37 \cr
G & 0.0104 & 41.99 & 173 & -445 &  350 &  175 & -0.715 & -2.49 \cr
H & 0.0105 & 44.06 & 356 & -847 &  655 &  387 & -0.696 & -2.56 \cr
I & 0.0104 & 47.34 & 264 & -703 &  549 &  329 & -0.641 & -2.52 \cr
\noalign{\medskip}
\noalign{\hrule height\rulerheight}
\noalign{\smallskip}
\noalign{\hrule height\rulerheight}
\end{tabular}
\end{center}

{\narrower\narrower\footnotesize\noindent
{TABLE \TabBinoCorr.}
Values of the parameters required to compute the bino correction to the
tau lepton mass $\Delta^{\tilde B} m_\tau \sim -2.45 \%$.
(all evaluated at low energy.)
The bino mass is given by $m_{\tilde B}$,
the stau masses by $m_{\tilde\tau_{1,2}}$ and the dimensionless
function $I_{\tilde B}$ is given by
$I_{\tilde B}= m_{\tilde B} \mu
               I(m_{\tilde B}^2, m_{\tilde\tau_1}^2, m_{\tilde\tau_2}^2)$
\par\bigskip}}

\vbox{
\begin{center}
\begin{tabular}{cccccccc}
\multicolumn{8}{c}{TABLE \TabSusyCorr.} \cr
\noalign{\medskip}
\noalign{\hrule height\rulerheight}
\noalign{\smallskip}
\noalign{\hrule height\rulerheight}
\noalign{\medskip}
Case &
$\Delta^{\tilde g} m_d$ &
$\Delta^{\tilde H} m_d$ &
$\Delta^{\tilde g} m_s$ &
$\Delta^{\tilde H} m_s$ &
$\Delta m_d$ &
$\Delta m_s$ &
$\Delta m_b$ \cr
\noalign{\medskip}
\noalign{\hrule height\rulerheight}
\noalign{\medskip}
A & -19.24 & 0.00 & -19.25 & 0.01 & -19.24 & -19.23 & -17.20 \cr
B & -19.28 & 0.00 & -19.28 & 0.01 & -19.28 & -19.27 & -17.04 \cr
C & -12.95 & 0.00 & -12.95 & 0.01 & -12.95 & -12.94 & -11.35 \cr
D & -12.72 & 0.00 & -12.72 & 0.01 & -12.72 & -12.71 & -11.16 \cr
E & -26.38 & 0.00 & -26.39 & 0.02 & -26.38 & -26.37 & -24.40 \cr
F & -26.38 & 0.00 & -26.38 & 0.02 & -26.38 & -26.36 & -24.19 \cr
G & -20.92 & 0.00 & -20.92 & 0.01 & -20.92 & -20.91 & -18.87 \cr
H & -20.91 & 0.00 & -20.92 & 0.01 & -20.91 & -20.90 & -18.67 \cr
I & -24.40 & 0.00 & -24.40 & 0.02 & -24.40 & -24.38 & -22.55 \cr

\noalign{\medskip}
\noalign{\hrule height\rulerheight}
\noalign{\smallskip}
\noalign{\hrule height\rulerheight}
\end{tabular}
\end{center}

{\narrower\narrower\footnotesize\noindent
{TABLE \TabSusyCorr.}
Gluino and higgsino SUSY corrections to the down and strange masses
$\Delta^{\tilde g,\tilde H} m_{d,s}$ and the total correction to the
down, strange and bottom quark masses $\Delta m_{d,s,b}$ (in percentage values.)
\par\bigskip}}

\newpage

\SECTION{APPENDIX D \\
         Boundary conditions for the gauge and gaugino masses}

In this appendix we review the origin of Eq.~\refeqn{Gauge422}
that relates the gauge and the gaugino masses of the
$SU(3)_c \otimes SU(2)_L \otimes U(1)_Y$ MSSM
and of the \422\ Pati-Salam Model.

We start by considering the constraint on the
$U(1)_Y$, $SU(2)_R$ and $SU(4)$ gauge couplings
$g'$, $g_{2R}$ and $g_4$.
The covariant derivative for the heavy
Higgs boson $H$ that breaks the GUT symmetry is :
\begin{equation}
D_\mu H = \partial_\mu H +
          i  g_{2R} \tau_{R}^a W^a_{R\mu} H +
          i  g_4 T^m G^m_\mu H
\end{equation}
where $\tau_{R}^a =  {1 \over 2}  \sigma^a$ and $T^m$ are the $SU(2)_R$
and $SU(4)$ group generators with associated gauge bosons $W^a_R$
and $G^m$ ($a=1,..,3$ and $m=1,..,15$).
When $H$ develops a non-vanishing VEV
$\langle H_\nu \rangle = \sqrt{2} V_\nu$
along the neutrino direction the
quadratic interaction $|D_\mu H|^2$ generates the following mass terms~:
\begin{equation}
2 g^2_{2R} V_\nu^2 (\tau_R^a \tau_R^b)_{22} W^{a\mu}_R W^b_{R\mu}
\end{equation}
\begin{equation}
4 g_{2R} g_4 V_\nu^2 (\tau_R^3)_{22} (T^{15})_{44} W^{3\mu}_R G^{15}_\mu
\end{equation}
\begin{equation}
2 g_4^2 V_\nu^2 (T^m T^n)_{44} G^{m\mu} G^n_\mu.
\end{equation}
where $\tau_R^3 = {\rm diag}({1 \over 2},-{1 \over 2})$ and
$T^{15} = \sqrt{3 \over 2} \> {\rm Diag}
                              ({1\over 6},{1\over 6},{1\over 6},-{1\over 2})$
 are diagonal matrices.
Upon explicit substitution of the group generators
(and after adding to the above expressions similar terms associated with
the $\bar H$ Higgs field, with a VEV
$\langle \bar H_\nu \rangle = \sqrt{2} \bar V_\nu$)
the mixing between the $W^3_R$, $G^{15}$
gauge bosons can be written as :
\begin{equation}
{1 \over 2} V^2
\left(\matrix{W^{3\mu}_R & G^{15\mu}}\right)
\left(\matrix{g^2_{2R} &  - \sqrt{3 \over 2} g_4 g_{2R} \cr
              - \sqrt{3 \over 2} g_4 g_{2R} &
              {3 \over 2} g_4^2 \cr}\right)
\left(\matrix{W^3_{\mu_R} \cr
              G^{15}_\mu  \cr}\right)
\label{GaugeMatrix}
\end{equation}
where $V^2 = V_\nu^2 + \bar V_\nu^2$.
The matrix above can be diagonalized by a unitary matrix parameterised
by the rotation angle $\alpha$ given by :
\begin{equation}
\sin\alpha =
{g_{2R} \over \sqrt{ g^2_{2R} + {3 \over 2} g_4^2}}
\quad\quad
\cos\alpha =
{\sqrt{3\over 2} g_4 \over \sqrt{ g^2_{2R} + {3 \over 2} g_4^2}}
\label{RotAngle}
\end{equation}
to yield a massless $B$ state (the MSSM bino) and a heavy gauge boson
$X$ with a mass $m_X^2 = V^2 ( g^2_{2R} + {3 \over 2} g^2_4 )$.
These are related to the $W^3_R$, $G^{15}$ bosons through :
\begin{equation}
\left(\matrix{ B \cr X \cr}\right) =
\left(\matrix{\phantom{-} \cos\alpha & \sin\alpha \cr
             -\sin\alpha & \cos\alpha}\right)
\left(\matrix{ W^3_R \cr G^{15} \cr}\right)
\quad,\quad
\left(\matrix{ W^3_R \cr G^{15} \cr}\right) =
\left(\matrix{\cos\alpha & -\sin\alpha \cr
              \sin\alpha & \phantom{-}\cos\alpha}\right)
\left(\matrix{ B \cr X \cr}\right)
\end{equation}
Finally, in order to obtain the first equation in Eq.~\refeqn{Gauge422}
we need to relate $g'$ to $g_{2R}$
and $g_4$. This can be achieved by examining, for example,
the kinetic term for the fermions in the $F$ representation~:
\begin{eqnarray}
{\cal L} = i {\bar F} \gamma^\mu D_\mu F
\sim - {\bar F} \gamma^\mu \left[ g_4 T^{15} G^{15}_\mu \right] F
\sim - {\bar u} \gamma^\mu \left[ \textstyle \sqrt{3 \over 2} g_4
\left(1 \over 6 \right)
(\sin\alpha B_\mu) \right] u
\end{eqnarray}
and comparing it to the $U(1)_Y$ neutral current Lagrangian :
\begin{equation}
{\cal L} = - {\bar u} \gamma^\mu \left[ \textstyle g' \left( 1 \over 6 \right)
B_\mu \right] u
\end{equation}
We get :
\begin{equation}
\textstyle
g' = \sqrt{3 \over 2} g_4 \sin\alpha
\label{gprime}
\end{equation}
Thus combining Eq.~\refeqn{RotAngle} and \refeqn{gprime} we find :
\begin{equation}
{1 \over g^{\prime 2}} = {1 \over g^2_{2R}}+{1 \over {3 \over 2} g_4^2}
\label{gprime2}
\end{equation}
which agrees with Eq.~\refeqn{Gauge422} after replacing
$g^{\prime 2} = {3 \over 5} g_1^2$.

In the second part of this appendix we will justify
the second equation in Eq.~\refeqn{Gauge422}.
We start by considering the following Higgs-gaugino-higgsino
interaction :
\begin{equation}
{\cal L} = i \sqrt{2} H^\dagger
\left( g_{2R} \tau^a_R \lambda^a_R +
       g_4    T^m \lambda^m_4 \right) \tilde H + {\rm h.c.}
\end{equation}
where $H$ is the heavy Higgs boson, $\tilde H$ the heavy higgsino field,
and $\lambda^a_R$, $\lambda^m_4$ the $SU(2)_R$, $SU(4)$ gauginos
(a similar expression applies to $\bar H$). When the Higgs bosons $H$,
$\bar H$ develop their VEVs, we obtain (after appropriately
substituting the generator matrices $\tau^3_R$, $T^{15}$) :
\begin{equation}
\textstyle
{\cal L} = - g_{2R} (-i\lambda^3_R)
\left[ V_\nu {\tilde H}_\nu +
       {\bar V}_\nu {\tilde{\bar H}}_\nu \right]+
\sqrt{3 \over 2} g_4 (-i\lambda^{15}_4)
\left[ V_\nu {\tilde H}_\nu +
       {\bar V}_\nu {\tilde{\bar H}}_\nu \right]+{\rm h.c.}
\label{GauginoHiggsino}
\end{equation}
From Eq.~\refeqn{GauginoHiggsino}
it is clear that the gauginos only couple to the
$ V_\nu {\tilde H}_\nu+
 {\bar V}_\nu {\tilde{\bar H}}_\nu$
combination of higgsinos.
Thus, it is convenient to define new fields $\tilde N_1$
and $\tilde N_2$ through~:
\begin{equation}
\left(\matrix{ \tilde N_1 \cr \tilde N_2 \cr}\right) =
\left(\matrix{\cos\theta_N & \sin\theta_N \cr
             -\sin\theta_N & \cos\theta_N}\right)
\left(\matrix{ \tilde H_\nu \cr \tilde{\bar H}_\nu }\right)
\end{equation}
where $\tan\theta_N = \bar V_\nu / V_\nu$.
The bosonic partners of $\tilde N_1$, $\tilde N_2$ have VEVs given by
$\langle N_1 \rangle = \sqrt{2} V_1 =
  \sqrt{2} V_\nu \cos\theta_N + \sqrt{2} \bar V_\nu \sin\theta_N =
  \sqrt{2} ( V_\nu^2 + \bar V_\nu^2)^{1/2}$ and
$\langle N_2 \rangle = 0$.
\noindent
The gaugino mass matrix in the
$(-i \lambda^3_R,-i\lambda^{15}_4,\tilde N_1)$ basis can be written as~:
\begin{equation}
M_\lambda = \left(\matrix{
M_{2R} & 0 & g_{2R} V_1 \cr
0 & M_4 &
-\sqrt{3 \over 2} g_4 V_1 \cr
g_{2R} V_1 &
-\sqrt{3 \over 2} g_4 V_1 & 0 \cr}
\right)
\label{gauginomatrix}
\end{equation}
where the 13, 23 entries derive from Eq.~\refeqn{GauginoHiggsino}
and $M_{2R}$, $M_4$
are explicit light soft masses for the $SU(2)_{2R}$, $SU(4)$ gauginos.
The two heavy eigenvalues of the above matrix are approximately given
by $M_{\lambda_{2,3}} \sim \pm (g^2_{2R}+{3 \over 2} g^2_4)^{1/2} V_1$.
The lightest eigenvalue (the bino mass) can easily be computed from
$M_1 = {\rm Det} M_\lambda / ( M_{\lambda_1} M_{\lambda_2} )$.
We find :
\begin{equation}
M_1 =
{{3 \over 2} g^2_4 \over g^2_{2R} + {3 \over 2} g^2_4} M_{2R}+
{ g^2_{2R} \over g^2_{2R} + {3 \over 2} g^2_4 } M_4
\label{BINOMASS}
\end{equation}
Finally, replacing the result of Eq.~\refeqn{gprime2}
into Eq.~\refeqn{BINOMASS} gives us the
expression we wanted to prove :
\begin{equation}
{M_1 \over g^{\prime 2}} =
{M_{2R} \over g^2_{2R}}+
{M_4 \over {3 \over 2} g^2_4}
\end{equation}

We would like to conclude this appendix with the following observation.
The rotation
that diagonalizes the $W^3_R$--$G^{15}$ gauge bosons mass matrix
(see Eq.~\refeqn{GaugeMatrix})
also simultaneously block diagonalizes the gaugino mass matrix
in Eq.~\refeqn{gauginomatrix}. Indeed we note that :
\begin{equation}
\left(\matrix{ c_\alpha & s_\alpha & 0 \cr
              -s_\alpha & c_\alpha & 0 \cr
               0 & 0 & 1 \cr}\right)
M_\lambda
\left(\matrix{ c_\alpha &-s_\alpha & 0 \cr
               s_\alpha & c_\alpha & 0 \cr
               0 & 0 & 1 \cr}\right) \sim
\left(\matrix{ M_{\lambda_1} & {\cal O}(M) & 0 \cr
               {\cal O}(M) & {\cal O}(M) & {\cal O}(V_1) \cr
               0 & {\cal O}(V_1) & 1 \cr}\right)
\label{apprgaugino}
\end{equation}
where $s_\alpha = \sin\alpha$, $c_\alpha = \cos\alpha$ and
${\cal O}(M)$ is a number of order $M_{2R}$ and/or $M_4$.
Thus we find from Eq.~\refeqn{apprgaugino} that
$M_{\lambda_1} = c^2_\alpha M_{2R}+s^2_\alpha M_4$ which is nothing
less than the bino mass of Eq.~\refeqn{BINOMASS} re-written in terms of the
gauge mixing angle $\alpha$ defined in Eq.~\refeqn{RotAngle}.

\newpage

\SECTION{APPENDIX E \\
         Expressions for the $D$-terms}

The $D$-term corrections to the soft masses in
Eqs.~\refeqn{mQgD}-\refeqn{m1gD} arise when the
rank of the Pati-Salam group is reduced from five to four due to gauge
symmetry breaking.
In this appendix we show that the coefficients of the corrections are
related to the charge carried by the fields under the $U(1)$ broken
generator and that their magnitude depends on the difference between
the soft masses of the heavy Higgs that break the GUT symmetry.

We start by carefully reporting our index conventions.
The matrix elements of the fields of the
$SU(4)\otimes SU(2)_L\otimes SU(2)_R$ model in
Eqs.~\refeqn{FcFields}-\refeqn{RHiggs} are
indicated by~:
\begin{equation}
\matrix { F^c_{\dot\alpha i} : (\bar 4,1,\bar 2) \quad &
          \bar H_{\dot\theta p} : (\bar 4,1,\bar 2) \quad & \cr
          & & h_\rho^{\phantom{\rho}\dot\sigma} : (1,\bar 2,2) \quad \cr
          F^{\gamma k} : (4,2,1) \quad &
          H^{\dot\eta r} : (4,1,2) \quad & }
\label{Index422}
\end{equation}
where $i,k,p,r=1,..,4$ are $SU(4)$ indices,
$\dot\alpha,\dot\theta,\dot\eta,\dot\sigma=1,2$ are $SU(2)_R$ indices and
$\gamma,\rho=1,2$ are $SU(2)_L$ indices.
The position of the indices is the following, the first/second index
refers to the line/column of the matrices in
Eqs.~\refeqn{FcFields}-\refeqn{RHiggs}.
Moreover up/down indices
are related to the representation of the multiplet.
For example, the lower $\rho$ index of $h$ indicates that $h$
transforms in the $\bar 2$ (anti-matter) representation of $SU(2)$
under the $SU(2)_L$ symmetry,
whereas the upper $\dot\sigma$ index indicates that $h$ transforms
in the fundamental representation of $SU(2)$ under the $SU(2)_R$ symmetry.

The $D$-term contributions from the $SU(2)_{2R}$ and $SU(4)$ groups
are given by~:
\begin{equation}
{\cal V} = {\textstyle {1\over 2}} g^2_{2R}\sum_{a=1}^3    D^a_{2R} D^a_{2R}+
           {\textstyle {1\over 2}} g^2_{4} \sum_{m=1}^{15} D^m_4 D^m_4.
\label{VDterms}
\end{equation}
We focus on the $a=3$ and $m=15$ contributions of Eq.~\refeqn{VDterms}
which involve the
\begin{equation}
\textstyle
\tau_R^3 = {\rm diag}({1 \over 2},-{1 \over 2}) \qquad\qquad
T^{15} = \sqrt{3 \over 2} \> {\rm diag}
                             ({1\over 6},{1\over 6},{1\over 6},-{1\over 2})
\end{equation}
diagonal generators of the $SU(2)_{2R}$ and $SU(4)$ groups.
Using the notation of Eq.~\refeqn{Index422}
we find that $D^3_{2R}$, $D^{15}_4$ are given by :

\begin{eqnarray}
   \!\!\!\!\!\!\!\!\!
D^2_{2R}
& \!\!\! = & \!\!\!
           \bar H^{\dagger}_{\dot\theta p}
           (-\tau^{3*}_R)_{\dot\theta \dot\omega}
           \bar H_{\dot\omega p}+
           H^{\dagger\dot\eta r}
           (\tau^3_R)_{\dot\eta \dot\xi}
           H^{\dot\xi r}+
           \tilde F^{c\dagger}_{\dot\alpha i}
           (-\tau^{3*}_R)_{\dot\alpha \dot\beta}
           \tilde F^{c}_{\dot\beta i}+
           h_\rho^{\dagger\phantom{\rho}\dot\sigma}
           (\tau^{3}_R)_{\dot\sigma \dot\pi}
           h_\rho^{\phantom{\rho}\dot\pi} \\
   \!\!\!\!\!\!\!\!\!
D^{15}_4
& \!\!\! = & \!\!\!
           \bar H^{\dagger}_{\dot\theta p}
           (-T^{15*})_{pq}
           \bar H_{\dot\theta q}+
           H^{\dagger\dot\eta r}
           (T^{15})_{rs}
           H^{\dot\eta s}+
           \tilde F^{c\dagger}_{\dot\alpha i}
           (-T^{15*})_{ij}
           \tilde F^{c}_{\dot\alpha j}+
           \tilde F^{\dagger\gamma k}
           (T^{15})_{kl}
           \tilde F^{\gamma l}
\end{eqnarray}
\noindent
Assuming that the heavy Higgs develop VEVs along the neutrino
directions~:
\begin{equation}
\langle\bar H \rangle = \bar H_{\dot 2 4} = \bar H_\nu \qquad
\langle H \rangle = H^{\dot 2 4} = H_\nu
\end{equation}
we can expand $D^3_{2R}$, $D^{15}_4$ :
\begin{eqnarray}
   \!\!\!\!\!\!\!\!\!
D^3_{2R}
 & \!\!\! = & \!\!\!
\bar H^{\dagger}_\nu (-\tau^{3*}_R)_{\dot 2 \dot 2} \bar H_\nu+
H^\dagger_\nu (\tau^3_R)_{\dot 2 \dot 2} H_\nu + \nonumber \\
 & \!\!\!   & \!\!\!
\tilde d^{c\dagger}(-\tau^{3*}_R)_{\dot 1 \dot 1} \tilde d^c+
\tilde e^{c\dagger}(-\tau^{3*}_R)_{\dot 1 \dot 1} \tilde e^c+
\tilde u^{c\dagger}(-\tau^{3*}_R)_{\dot 2 \dot 2} \tilde u^c+
\tilde \nu^{c\dagger}(-\tau^{3*}_R)_{\dot 2 \dot 2} \tilde \nu^c+ \nonumber \\
 & \!\!\!   & \!\!\!
h^\dagger_d(\tau^3_R)_{\dot 1 \dot 1} h_d+
h^\dagger_u(\tau^3_R)_{\dot 2 \dot 2} h_u \nonumber \\
 & \!\!\! = & \!\!\!
+{\textstyle {1\over 2}} |\bar H_\nu|^2
-{\textstyle {1\over 2}} |H_\nu|^2
-{\textstyle {1\over 2}} |\tilde d^c|^2
-{\textstyle {1\over 2}} |\tilde e^c|^2
+{\textstyle {1\over 2}} |\tilde u^c|^2
+{\textstyle {1\over 2}} |\tilde \nu^c|^2
+{\textstyle {1\over 2}} |h_d|^2
-{\textstyle {1\over 2}} |h_u|^2 \label{D32R} \\
& & \nonumber \\
   \!\!\!\!\!\!\!\!\!
D^{15}_4
 & \!\!\! = & \!\!\!
\bar H^{\dagger}_\nu (-T^{15*})_{44} \bar H_\nu+
H^\dagger_\nu (T^{15})_{44} H_\nu + \nonumber \\
 & \!\!\!   & \!\!\!
\tilde d^{c\dagger}(-T^{15*})_{11} \tilde d^c+
\tilde e^{c\dagger}(-T^{15*})_{44} \tilde e^c+
\tilde u^{c\dagger}(-T^{15*})_{11} \tilde u^c+
\tilde \nu^{c\dagger}(-T^{15*})_{44} \tilde \nu^c+ \nonumber \\
 & \!\!\!   & \!\!\!
\tilde q^{\dagger}(T^{15})_{11} \tilde q  +
\tilde l^{\dagger}(T^{15})_{44} \tilde l \nonumber \\
 & \!\!\! = & \!\!\!
{\textstyle \sqrt{3\over 2}} \, [
+{\textstyle {1\over 2}} |\bar H_\nu|^2
-{\textstyle {1\over 2}} |H_\nu|^2
-{\textstyle {1\over 6}} |\tilde d^c|^2
+{\textstyle {1\over 2}} |\tilde e^c|^2
-{\textstyle {1\over 6}} |\tilde u^c|^2
+{\textstyle {1\over 2}} |\tilde \nu^c|^2
+{\textstyle {1\over 6}} |\tilde q|^2
-{\textstyle {1\over 2}} |\tilde l|^2 ] \label{D154}
\end{eqnarray}
One can summarize the results of Eqs.~\refeqn{D32R}-\refeqn{D154} by writing :
\begin{equation}
D^3_{2R} = D_{H}+\sum_\phi I_\phi |\phi|^2 \hfill \qquad
D^{15}_4 = {\textstyle \sqrt{3\over 2}} D_H +
             {\textstyle \sqrt{3\over 2}} \sum_\phi
             {\textstyle \left( B-L \over 2 \right)}_\phi |\phi|^2
\label{D32RD154}
\end{equation}
where $D_H = {\textstyle {1\over 2}} (|\bar H_\nu|^2-|H_\nu|^2)$
and $\phi$ denotes any of the light fields
$\tilde u^c$, $\tilde d^c$, $\tilde e^c$, $\tilde \nu^c$,
$\tilde q$, $\tilde l$, $h_u$, $h_d$.
The factor $I_\phi$ refers to the charge carried by $\phi$ with
respect to the $SU(2)_R$ group and $(B-L)/2$ to the
semi-difference between the baryon and lepton numbers of $\phi$.
These are can be read from the coefficients of the terms in
Eqs.~\refeqn{D32R}-\refeqn{D154}
and are collected in Table~\TabDtermCharges.

\vbox{
\begin{center}
\begin{tabular}{ccccccccc}
\multicolumn{9}{c}{TABLE \TabDtermCharges.} \cr
\noalign{\medskip}
\noalign{\hrule height\rulerheight}
\noalign{\smallskip}
\noalign{\hrule height\rulerheight}
\noalign{\medskip}
&
$\tilde u^c$ &
$\tilde d^c$ &
$\tilde e^c$ &
$\tilde \nu^c$ &
$\tilde q$ &
$\tilde l$ &
$h_u$ &
$h_d$ \cr
\noalign{\medskip}
\noalign{\hrule height\rulerheight}
\noalign{\medskip}
$I_\phi$ &
$+{1 \over 2}$ &
$-{1 \over 2}$ &
$-{1 \over 2}$ &
$+{1 \over 2}$ &
\hfill 0 &
\hfill 0 &
$-{1 \over 2}$ &
$+{1 \over 2}$ \cr
$\left( B-L \over 2 \right)_\phi$ &
$-{1 \over 6}$ &
$-{1 \over 6}$ &
$+{1 \over 2}$ &
$+{1 \over 2}$ &
$+{1 \over 6}$ &
$-{1 \over 2}$ &
\hfill 0 &
\hfill 0 \cr
\noalign{\medskip}
\noalign{\hrule height\rulerheight}
\noalign{\smallskip}
\noalign{\hrule height\rulerheight}
\end{tabular}
\end{center}

{\narrower\narrower\footnotesize\noindent
{TABLE \TabDtermCharges.}
Charges carried by the light scalar fields under the
$SU(2)_{2R}$ and (Baryon-Lepton)/2 symmetries.
\par
\bigskip}}
\noindent
Finally, using Eq.~\refeqn{D32RD154} into Eq.~\refeqn{VDterms} gives :
\begin{eqnarray}
{\cal V}
&=&
{\textstyle {1\over 2}} g^2_{2R} D^3_{2R} D^2_{2R} +
{\textstyle {1\over 2}} g^2_4    D^{15}_4 D^{15}_4 \\
&=&
{\textstyle {1\over 2}}
g^2_{2R} \,
[ \, 2 D_H \sum_\phi I_\phi |\phi|^2 ]+
{\textstyle {1\over 2}}
g^2_4 \,
[ \, 2 \, {\textstyle {3\over 2}}
  D_H \sum_\phi {\textstyle \left( B-L \over 2 \right)_\phi} |\phi|^2 ] \\
&=&
{\textstyle {1\over 2}} ( |\bar H_\nu|^2 - |H_\nu|^2 )
\sum_\phi
\left\{ \, g^2_{2R} I_\phi + {\textstyle {3\over 2}}
  g^2_4 {\textstyle \left( B-L \over 2 \right)_\phi} \,\right\} |\phi|^2
\label{VDterms2}
\end{eqnarray}
The equation above deserves two comments.
Firstly, we see that the broken $U(1)$ generator $X$ resulting from the GUT
symmetry breaking :
\begin{equation}
SU(4) \otimes SU(2)_{2R} \to SU(3)_c\otimes U(1)_Y\otimes U(1)_X
\end{equation}
is given by $X=I+(B-L)/2$ whereas the unbroken hypercharge
is $Y= -I+(B-L)/2$.
Secondly, comparing Eq.~\refeqn{VDterms2} with
Eqs.~\refeqn{mQgD}-\refeqn{m1gD} we find :
\begin{equation}
D^2 = {\textstyle {1\over 8}} ( |\bar H_\nu|^2 - |H_\nu|^2)
\label{D2HH}
\end{equation}
From Eq.~\refeqn{D2HH} alone, one might be tempted to conclude that the natural
scale for the $D$-term is of the order of the mass of the heavy Higgs.
However this is not true.

The scale of the $D$-term in Eq.~\refeqn{D2HH} can be estimated upon the
minimization of the heavy Higgs potential given by~:
\begin{equation}
{\cal V} =
{\textstyle {1\over 8}}
(g^2_{2R}+{\textstyle {3\over 2}} g^2_4)
(\bar H_\nu^2-H_\nu^2)^2+
\lambda_S^2 (\bar H_\nu H_\nu - M_H^2)^2+
m^2_{\bar H} \bar H_\nu^2 +
m^2_H H_\nu^2
\label{VHH}
\end{equation}
where the first term is a $D$-term,
the second an $F_S$-term where $S$ is the gauge singlet of
Eq.~\refeqn{W3}
and $M_H \sim 10^{16}$ GeV,
and $m_{\bar H}$, $m_H$ soft scalar masses.

The minimization conditions for ${\cal V}$, re-written as combinations of
$\partial {\cal V} / \partial\bar H_\nu \pm
 \partial {\cal V} / \partial     H_\nu = 0$, are given by~:
\begin{equation}
[ \,g^2_H (\bar H^2_\nu - H^2_\nu)^2 +
     \lambda_S^2(\bar H_\nu H_\nu - M_H^2) \, ]
(\bar H_\nu + H_\nu) = -m^2_{\bar H} \bar H_\nu-m^2_H H_\nu
\label{MinH1}
\end{equation}
\begin{equation}
[ \,g^2_H (\bar H^2_\nu + H^2_\nu)^2 -
     \lambda_S^2(\bar H_\nu H_\nu - M_H^2) \, ]
(\bar H_\nu - H_\nu) = -m^2_{\bar H} \bar H_\nu~+m^2_H H_\nu
\label{MinH2}
\end{equation}
where $g^2_H = {1\over 4} ( g^2_{2R} + {3\over 2} g^2_4 )$.
In the limit of negligible soft masses the right-hand side of
Eqs.~\refeqn{MinH1}-\refeqn{MinH2} vanishes.
Thus, we find that ${\bar H}_\nu = H_\nu = M_H$, implying that $D=0$.
This result could had been anticipated since,
if $m_{\bar H} = m_H$ then the Higgs potential is invariant
under the exchange of the $\bar H_\nu \leftrightarrow H_\nu$ fields.
\noindent
When the soft masses are not zero, but still much smaller than $M_H$
a perturbative solution to Eqs.~\refeqn{MinH1}-\refeqn{MinH2} is appropriate.
In terms of the new parameters $m$, $\bar m$ defined by :
\begin{equation}
\bar H_\nu = M_H - \bar m \qquad
     H_\nu = M_H -      m
\end{equation}
the minimization conditions are :
\begin{equation}
2\lambda_S^2 (\bar m+ m) M = m^2_{\bar H}+m^2_H \qquad
4 g^2_H      (\bar m- m) M = m^2_{\bar H}-m^2_H
\end{equation}
Thus we find :
\begin{equation}
\bar m = {1 \over \lambda^2 + 2 g_H^2}
         \left( m^2_{\bar H} \over M \right) \qquad
     m = {1 \over \lambda^2 + 2 g_H^2}
         \left( m^2_{H} \over M \right)
\end{equation}
and finally :
\begin{equation}
D^2 = {m^2_H - m^2_{\bar H} \over 4 \lambda^2_S + 2 g^2_{2R}+3 g^2_4}
\end{equation}
We see that, in spite of $D$ being the difference of two GUT scale
masses, it actually scales with the difference between the heavy soft Higgs
masses. The reason is because $D^2 \sim \epsilon M_H^2$ where
$\epsilon = (m^2_H-m^2_{\bar H}) / M^2_H$ is a very small parameter
which measures the amount of the $\bar H_\nu \leftrightarrow H_\nu$
symmetry breaking of the potential $\cal V$ in Eq.~\refeqn{VHH}.

\newpage


\end{document}